%% file: main.tex
\documentclass{ieeeaccess}

\usepackage{cite}
\usepackage{amsmath,amssymb,amsfonts}
\usepackage{graphicx}
\usepackage{textcomp}
\usepackage{url}
\usepackage[nolist]{acronym}
\usepackage{subfigure}
\usepackage{float}
\usepackage{dsfont}
\usepackage{algpseudocode}
\usepackage{makecell}
\usepackage{multirow}
\usepackage{multicol}
\usepackage{booktabs}
\usepackage[ruled,vlined]{algorithm2e}
\usepackage{soul}
\usepackage{tabularx}
\usepackage{rotating}
\usepackage{afterpage}
\usepackage{pdflscape}
\usepackage{color}
\usepackage{lscape}
\usepackage{longtable}
\usepackage{array}
\usepackage[flushleft]{threeparttable}
\usepackage{colortbl}
\usepackage[utf8]{inputenc}

\def\BibTeX{{\rm B\kern-.05em{\sc i\kern-.025em b}\kern-.08em
    T\kern-.1667em\lower.7ex\hbox{E}\kern-.125emX}}

\input{./Definitions.tex}

\newcommand{\blue}[1]{#1}
\newcommand{\red}[1]{#1}

\usepackage{fancyhdr}
\fancypagestyle{mahmood}{%
   \fancyhf{}% clear all fields
   \fancyhead[C]{\footnotesize
     © 2025 IEEE. Personal use of this material is permitted. Permission from IEEE must be obtained for all other uses, in any current or future media, including reprinting/republishing this material for advertising or promotional purposes, creating new collective works, for resale or redistribution to servers or lists, or reuse of any copyrighted component of this work in other works.
   }
}

\begin{document}
% \history{Date of publication xxxx 00, 0000, date of current version xxxx 00, 0000.}
\doi{10.1109/ACCESS.2025.DOI}

\title{Service Registration, Indexing, Discovery \& Selection; An Architectural Survey Toward a GenAI‑Driven Future}

\author{
\uppercase{Mohammad Farhoudi}\authorrefmark{1}, \IEEEmembership{Student Member, IEEE}, 
\uppercase{Masoud Shokrnezhad}\authorrefmark{2}, 
and \uppercase{Tarik Taleb}\authorrefmark{3}, \IEEEmembership{Senior Member, IEEE}
}

\address[1]{University of Oulu, Oulu, Finland 
(e-mail: mohammad.farhoudi@oulu.fi)}
\address[2]{ICTFICIAL Oy, Espoo, Finland (e-mail: masoud.shokrnezhad@ictficial.com)}
\address[3]{Ruhr University Bochum (RUB), Bochum, Germany (e-mail: tarik.taleb@rub.de)}

\corresp{Corresponding author: Mohammad Farhoudi (e-mail: mohammad.farhoudi@oulu.fi).}

\begin{abstract}
The emergence of sixth‑generation (6G) networks marks a paradigm shift: by unifying an edge‑to‑cloud computing continuum with ultra‑high‑performance networking, 6G will enable capabilities far beyond today’s boundaries. As use‑case diversity grows exponentially and user adoption drives traffic to unprecedented and highly dynamic levels, novel service orchestration mechanisms are indispensable. In this paper, we adopt an architectural viewpoint, examining Service Registration, Indexing, Discovery, and Selection (SRIDS) as fundamental elements of 6G service provision. We first establish the theoretical foundations of SRIDS in 6G by defining its core concepts, detailing its end‑to‑end workflow, reviewing current standardization efforts, and projecting its future design objectives, including reliability, scalability, automaticity and adaptability, determinism, efficiency, sustainability, semantic‑awareness, security, privacy, and trust. We then perform a comprehensive literature review and gap analysis \blue{encompassing both existing surveys and recent research efforts, identifying conceptual and methodological gaps that hinder unified SRIDS in 6G}. Next, we introduce a taxonomy that classifies SRIDS mechanisms into centralized, distributed, decentralized, and hybrid architectures, and systematically examine the relevant studies within each category. Each work is evaluated against the extracted design objectives. Building on these findings, we propose a hybrid architectural framework, combining centralized data management to ensure consistency and agility with distributed coordination to enhance scalability in emerging 6G use cases. The framework incorporates innovative technologies, such as Generative Artificial Intelligence (GenAI). We conclude by highlighting open challenges and suggesting directions for future research.
\end{abstract}


\titlepgskip=-15pt

\maketitle

\thispagestyle{mahmood}

\input{sections/sec1}
\input{sections/sec2}
\input{sections/sec3}
\input{sections/sec4}
\input{sections/sec5}
\input{sections/sec6}

\section{Conclusion}\label{sec:conclusion}
In this paper, we adopted an architectural viewpoint, examining service registration, indexing, discovery, and selection as fundamental elements of 6G service provision. We first established the theoretical foundations of SRIDS in 6G by defining core concepts, detailing the end-to-end workflow, reviewing current standardization efforts, and projecting future design objectives that included reliability, scalability, automaticity and adaptability, determinism, efficiency, sustainability, semantic awareness, security, privacy, and trust. A comprehensive literature review and gap analysis identified weaknesses in existing SRIDS frameworks, leading to the introduction of a taxonomy classifying SRIDS mechanisms into centralized, distributed, decentralized, and hybrid architectures. Based on these findings, we proposed a hybrid architectural framework that combines centralized data management, ensuring consistency and agility, with distributed coordination, thereby enhancing scalability for emerging 6G use cases. Additionally, we highlighted potential research directions that emphasized the need for adaptive \ac{ML} techniques, predictive analytics, smart security enhancements, addressing network challenges, and global and local optimization techniques. These avenues emerged as critical to overcoming the challenges posed by the dynamic and multifaceted nature of service provisioning in the future 6G landscape. The insights presented in this paper serve as a foundation for ongoing research, anticipating significant advancements in SRIDS methodologies that cater to the complexities of next-generation networks.

\section*{Acknowledgment}
This research work is partially supported by the Business Finland 6Bridge 6Core project under Grant No. 8410/31/2022, the European Union’s Horizon Europe research and innovation programme under the 6G-SANDBOX project with Grant Agreement No. 101096328, the 6G-Path project with Grant No. 101139172, and the Research Council of Finland 6G Flagship Programme under Grant No. 369116. The paper reflects only the authors' views.
% The European Commission and the Spanish Ministry are not responsible for any use that may be made of the information it contains.  

\bibliographystyle{ieeetran}
\bibliography{references/IEEEabrv, references/conf_short, references/bibliography}

\begin{IEEEbiography}[{\includegraphics[width=1in,height=1.25in,clip,keepaspectratio]{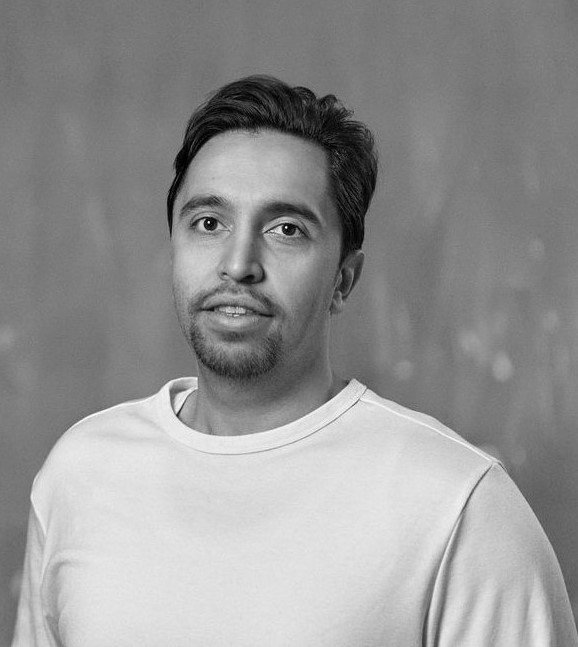}}]{Mohammad Farhoudi} (Student Member, IEEE)
received his B.Sc. degree in Information Technology and M.Sc. degree in Computer Networking from Amirkabir University of Technology, Tehran, Iran. He is currently pursuing a Ph.D. in Communication Engineering at the University of Oulu, Finland, within the Centre for Wireless Communications unit. He has served as a Senior Research Assistant at Amirkabir University of Technology, utilizing SDN and NFV in large-scale data and transport networks. He has contributed to multiple European research projects and collaborates with academic and industrial partners.
\end{IEEEbiography}

\begin{IEEEbiography}[{\includegraphics[width=1in,height=1.25in,clip,keepaspectratio]{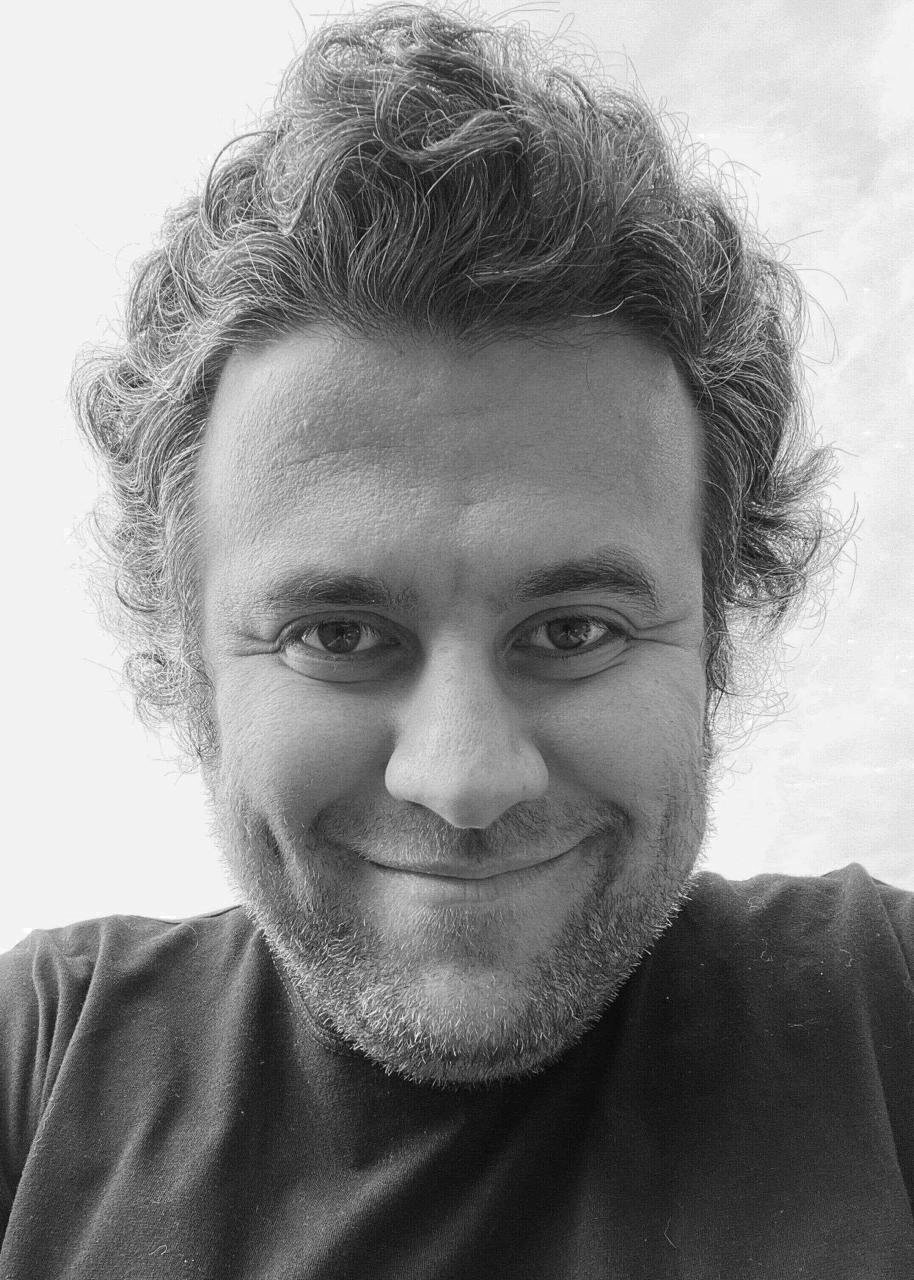}}]{Masoud Shokrnezhad} received his Ph.D. degree (recognized as a bright talent) in computer networks from Amirkabir University of Technology (Tehran Polytechnic), Tehran, Iran, in 2019. He is currently a postdoctoral researcher with the Center of Wireless Communications at the University of Oulu, Finland. Between June 2021 and December 2021, he was a postdoctoral researcher at the School of Electrical Engineering, Aalto University, Espoo, Finland. Prior to that, he worked as a senior system designer and engineer with FavaPars and Pouya Cloud Technology in Tehran, Iran, since 2013. Throughout his career, Dr. Shokrnezhad has been involved in numerous national, international, and European projects focused on designing and developing computing and networking frameworks. He has also co-managed a startup that develops SDWAN solutions for B2B use cases. 
\end{IEEEbiography}

\begin{IEEEbiography}[{\includegraphics[width=1in,height=1.25in,clip,keepaspectratio]{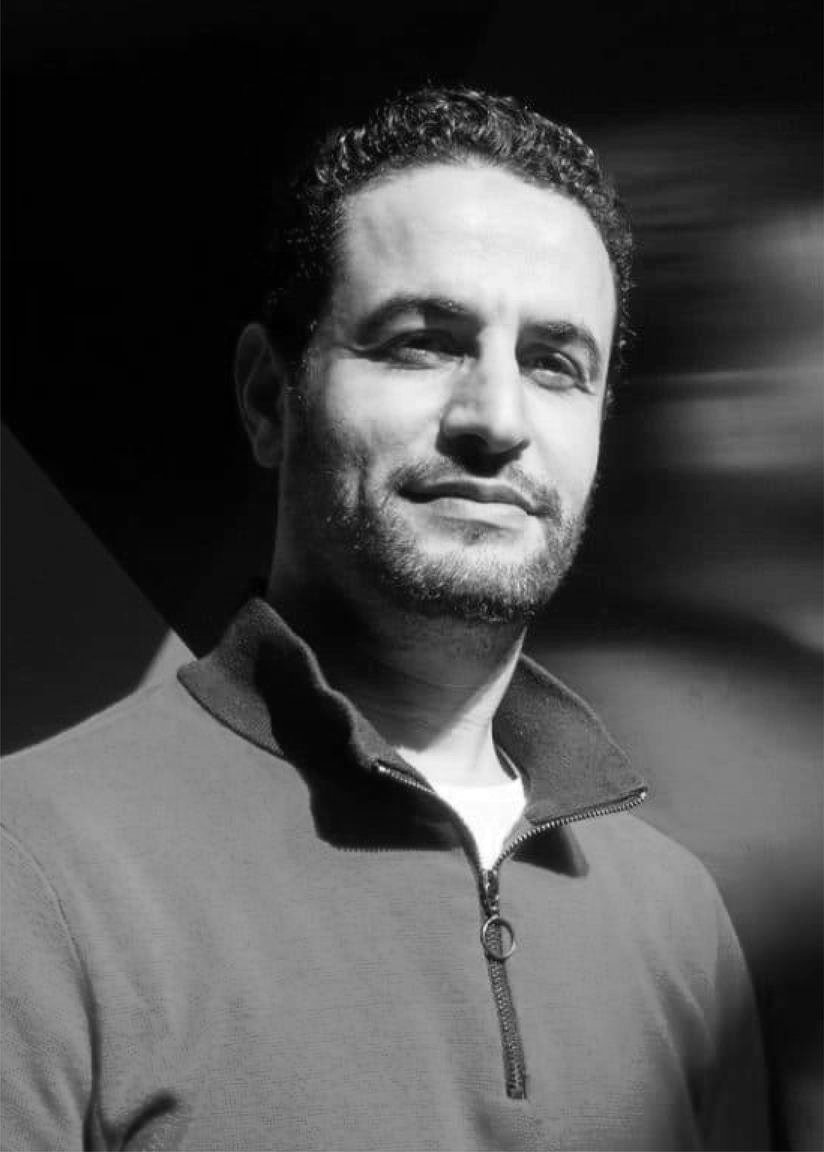}}]{Tarik Taleb (Senior Member, IEEE)}received the B.E. degree (with distinction) in information engineering and the M.Sc. and Ph.D. degrees in information sciences from Tohoku University, Sendai, Japan, in 2001, 2003, and 2005, respectively. He is currently a Full Professor at Ruhr University Bochum, Germany. He was a Professor with the Centre of Wireless Communications, University of Oulu, Oulu, Finland. He is the founder of ICTFICIAL Oy, and the founder and the Director of the MOSA!C Lab. From October 2014 to December 2021, he was an Associate Professor with the School of Electrical Engineering, Aalto University, Espoo, Finland. Prior to that, he was working as a Senior Researcher and a 3GPP Standards Expert with NEC Europe Ltd., Heidelberg, Germany. Before joining NEC and till March 2009, he worked as an Assistant Professor with the Graduate School of Information Sciences, Tohoku University, in a lab fully funded by KDDI. From 2005 to 2006, he was a Research Fellow with the Intelligent Cosmos Research Institute, Sendai. Prof. Taleb has been directly engaged in the development and standardization of the Evolved Packet System as a member of the 3GPP System Architecture Working Group. His current research interests include AI-based network management, architectural enhancements to mobile core networks, network softwarization and slicing, mobile cloud networking, network function virtualization, software-defined networking, software-defined security, and mobile multimedia streaming.
\end{IEEEbiography}

\EOD

\end{document}

%% file: Definitions.tex
\begin{acronym}
\acro{GenAI}{Generative Artificial Intelligence}
\acro{LLM}{Large Language Model}
\acro{BS}{Base Station}
\acro{UAV}{Unmanned Aerial Vehicle}
\acro{RAG}{Retrieval-Augmented generation}
\acro{RSU}{Roadside Unit}
\acro{MAC}{Multiple-Access Control}
\acro{UE}{User Equipment}
\acro{LoS}{Line-of-Sight}
\acro{PoA}{Point of Attachment}
\acro{MINLP}{Mixed Integer Non-Linear Programming}
\acro{DRL}{Deep Reinforcement Learning}
\acro{GNN}{Graph Neural Network}
\acro{NFV}{Network Function Virtualization}
\acro{VNF}{Virtual Network Function}
\acro{SDN}{Software-Defined Network}
\acro{CFN}{Compute First Network}
\acro{TSN}{Time-Sensitive Networking}
\acro{AR}{Augmented Reality}
\acro{6G}{Sixth-Generation}
\acro{Wi-Fi}{Wireless Fidelity}

\acro{SDP}{Service Discovery Protocol}
\acro{SRIDS}{Service Registration, Indexing, Discovery, and Selection}
\acro{NACK}{Negative ACknowledgment}
\acro{ID}{identifier}
\acro{DAG}{Directed Acyclic Graph}
\acro{OASIS}{Organization for the Advancement of Structured Information Standards}
\acro{UDDI}{Universal Description, Discovery and Integration}
\acro{XML}{Extensible Markup Language}
\acro{SOAP}{Simple Object Access Protocol}
\acro{SPML}{Service Provisioning Markup Language}
\acro{3GPP}{Third Generation Partnership Project}
\acro{CAPIF}{Common API Framework}
\acro{SA5}{Service and System Aspects}
\acro{CFTF}{Common Functions Task Force}
\acro{ETSI}{European Telecommunications Standards Institute}
\acro{ZSM}{Zero-touch network and Service Management}
\acro{GSMA}{Global System for Mobile Communications Association}
\acro{ONAP}{Open Network Automation Platform}
\acro{TM}{Telemanagement}

\acro{TLS}{Transport Layer Security}
\acro{mTLS}{mutual \ac{TLS}}
\acro{REST}{Representational State Transfer}
\acro{RPC}{Remote Procedure Call}
\acro{TCP}{Transmission Control Protocol}
\acro{UDP}{User Datagram Protocol}
\acro{MEC}{Multi-access Edge Computing}
\acro{SAP}{Service Advertisement Protocol}
\acro{API}{Application programming interface}
\acro{SHA}{Secure Hash Algorithm}

\acro{MANET}{Mobile Ad-Hoc Network}
\acro{MSN}{Mobile Social Network}
\acro{IP}{Internet Protocol}
\acro{IoT}{Internet of Things}
\acro{DHT}{Distributed Hash Table}

\acro{SLP}{Service Location Protocol}
\acro{DNS}{Domain Name Server}
\acro{ONS}{Object Name Service}
\acro{DDDS}{Dynamic Delegation Discovery System}
\acro{GS1}{Global Standard1}
\acro{UPnP}{Universal Plug and Play}
\acro{CAN}{Content Addressable Network}
\acro{CoAP}{Constrained Application Protocol}
\acro{HTTP}{HyperText Transfer Protocol}
\acro{IOTA}{Internet of Things Application}
\acro{IBSD}{IOTA‐Based Services Discovery}
\acro{IRI}{IOTA Reference Implementation}
\acro{SD‐AMC}{Service Discovery Ad‐hoc Mobile Cloud}

\acro{NoSQL}{No Structured Query Language}
\acro{MTD}{Moving Target Defense}
\acro{KAN}{Kolmogorov Arnold Network}
\acro{ML}{Machine Learning}
\acro{OWL}{Web Ontology Language}
\acro{RDF}{Resource Description Framework}

\acro{PKI}{Public Key Infrastructure}
\acro{CA}{Certificate Authority}
\acro{BFT}{Byzantine-Fault-Tolerant}

\acro{LoRA}{Low-Rank Adaptation}
\acro{TD3}{Twin Delayed Deterministic Policy Gradient}
\acro{GPS}{Global Positioning System}
\acro{NFV-MANO}{\ac{NFV} Management and Orchestration}

\end{acronym}

%% file: sections/sec1.tex
\section{Introduction}\label{sec:introduction}
\begin{figure}[t!]
    \centering
    \includegraphics[width=3.09in]{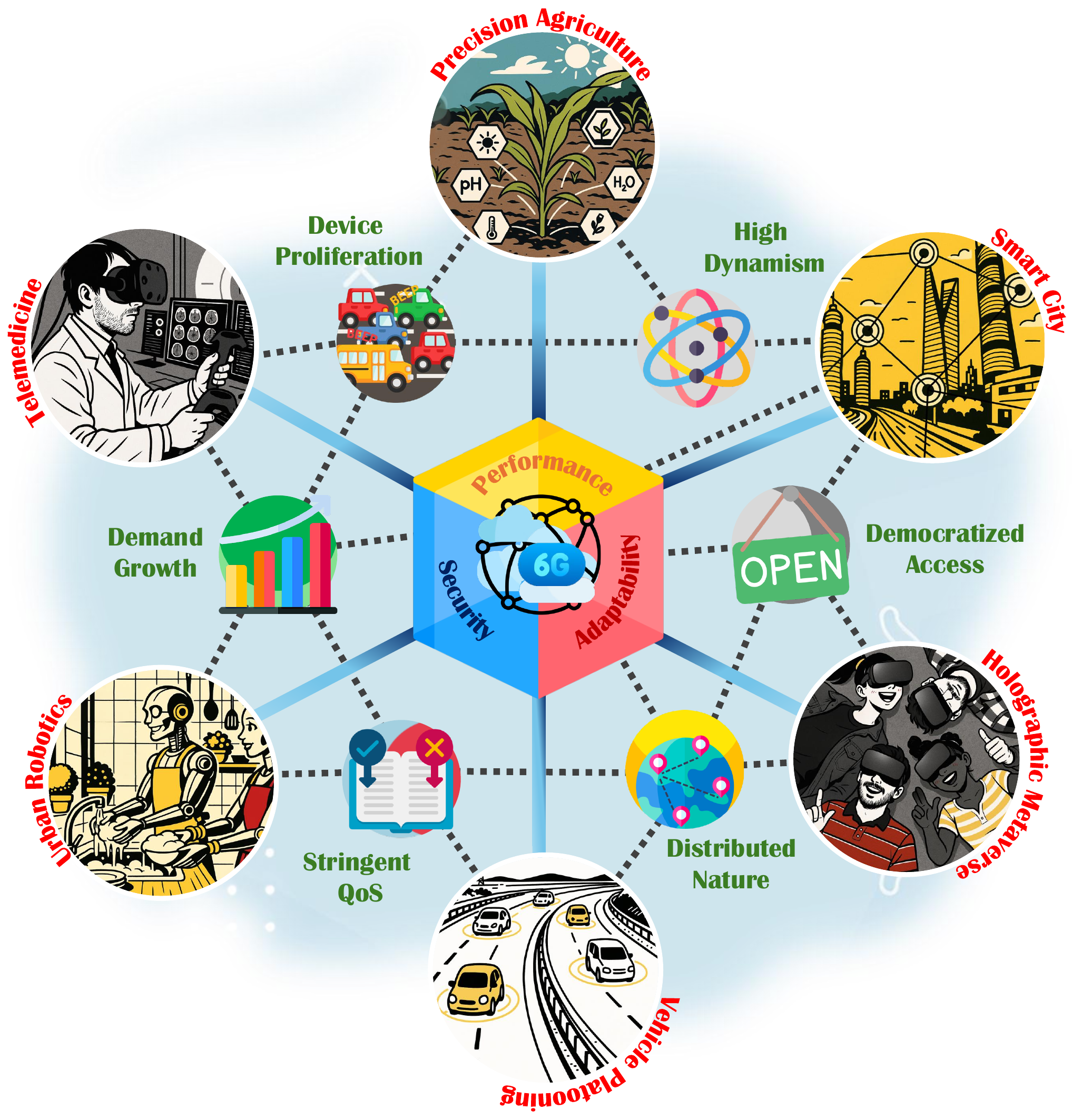}
    \vspace{-8pt}
    \caption{Emerging innovative use cases in the future, facilitated by the proliferation of 6G technologies.}
    \label{fig:novel_services_usecases}
    \vspace{-6pt}
\end{figure}

% In
\PARstart{T}{he}
pace of technological transformation is accelerating rapidly, driven by breakthroughs such as quantum communications, neuromorphic processors, and \ac{GenAI} that continually extend the boundaries of performance and functionality \cite{Quantum2023, Klusch2024, Sengar2024}. Emerging paradigms, from AI‑driven microelectromechanical systems to \acp{LLM}, promise to redefine human–technology interactions and reshape societal infrastructures. This evolution has already given rise to a myriad of novel use cases, including vehicle platooning, smart cities based on digital twins, urban robotics, biosensor-driven precision agriculture, telemedicine, and metasurface holographic metaverse, as depicted in Fig.~\ref{fig:novel_services_usecases}. Next‑generation systems are expected to provide the infrastructure necessary to support these use cases. On the networking side, \red{Sixth-Generation (6G)} network promises ultra‑high data rates, ultra‑low latency, and massive device connectivity \cite{6garchitecture_tarik}. On computing, 6G envisions an edge-to-cloud continuum where distributed resources span user-proximated edge nodes to centralized cloud data centers. The integrated orchestration of networking and computing across this continuum transcends conventional paradigms, unlocking synergistic performance gains across the entire system \cite{farhoudi2023qosaware}.

The effective realization of such advanced use cases on 6G relies on a robust service provisioning framework, with \textit{\ac{SRIDS}} forming its backbone elements \cite{DiscoveryAccess2024}. \textit{Service registration} allows providers to formally advertise services, including their function, component, and requirement descriptions. \textit{Service indexing} organizes and annotates these services and the deployed instances of their components in a semantically rich registry, ensuring that discovery queries can be answered both efficiently and accurately. Through \textit{Service discovery}, users or orchestration agents identify candidate services and their instances that satisfy functional and contextual requirements. \textit{Service selection} then evaluates these candidates using multi-criteria decision algorithms that consider real-time telemetry, such as resource latency, system load, energy status, and user mobility, to select the most suitable instances. Continuous performance monitoring feeds back into \ac{SRIDS}, enabling automatic re-registration when service properties change, dynamic re-indexing upon instance updates, and proactive re-selection to maintain optimal quality as system states evolve.

The implementation of \ac{SRIDS} for future use cases presents significant challenges, which arise from the unprecedented scale of connected devices and the generated demand; the stringent quality requirements; the high dynamism stemming from both mobile users and fluctuating distributed infrastructure resources; and the necessity for universality to democratize access (Fig.~\ref{fig:novel_services_usecases}). Considering these challenges, effective \ac{SRIDS} mechanisms must excel in three critical dimensions that determine system viability in 6G systems. \textit{Performance} aspects necessitate supporting millions of concurrent connections across extensive geographical regions while maintaining stringent quality-of-service parameters despite perpetual state fluctuations, with energy efficiency throughout the service lifecycle preventing economic and environmental constraints \cite{uusitalo2024, Ssemakula2024}. \textit{Security} considerations address architecture vulnerabilities through comprehensive protection against unauthorized access, malicious actors, non-compliant participants, and compromised infrastructure elements \cite{securityOrchestration, Rajesh2025}. Simultaneously, \textit{flexibility} features require autonomous real-time reconfiguration capabilities without human intervention, complemented by semantic comprehension functionality that enables operations based on meanings and intentions rather than syntax-driven processing and rigid protocol adherence, while accommodating previously unanticipated functionalities \cite{farhoudi2024, semCom2024}. \blue{In this context, {\ac{GenAI}} serves as the cognitive backbone of AI orchestration, enabling {\ac{SRIDS}} to interpret user intents and generate adaptive delivery. Beyond conventional discriminative AI, it autonomously produces data, code, and policies while constructing task-specific representations on demand, facilitating autonomous decision-making, service adaptation, and intent comprehension \mbox{\cite{genAISurvey}}. It underpins the concept of \textit{Trusted Distributed AI} by operating collaboratively across cross-domains, wherein generative models act as verifiable agents that enhance {\ac{SRIDS}} without compromising data sovereignty.}

To pave the way for addressing the aforementioned design objectives, we aim to investigate the solutions put forth in the literature concerning \ac{SRIDS} \blue{towards a {\ac{GenAI}}-driven future}, with a specific emphasis on their architectural aspects. A focus on architecture is crucial, as the manner in which the core elements of \ac{SRIDS} are modularized, orchestrated, and interconnected directly determines its ability to efficiently and reliably support the scale, stringency, dynamism, and universality intrinsic to 6G systems. The placement of control logic, the delineation of communication interfaces, and the degree of decentralization all influence not only performance outcomes, but also systemic properties such as security. A thoughtful architectural design also underpins the seamless integration of advanced features such as flexibility, ensuring that the system remains both user-intuitive and technically robust even as it evolves to serve novel use cases. The contributions of this paper encompass the following aspects:
\begin{itemize}
    \item We establish a precise definition of the \ac{SRIDS} key concepts and standardized workflow, followed by a detailed explanation of a specific future use case (incorporating advanced technologies such as \acp{LLM}), alongside their attendant design objectives.
    
    \item We introduce a novel taxonomic framework that classifies existing \ac{SRIDS} architectures into four distinct paradigms: centralized, distributed, decentralized, and hybrid. For each classification, we conduct a rigorous comparative analysis of inherent advantages and limitations, accompanied by a systematic review and interpretation of state-of-the-art methods within each category.
    
    \item We present a methodical assessment of how current literature addresses the design objectives of future use cases for \ac{SRIDS}, culminating in a formal gap analysis that quantifies the disparity between existing solutions and required capabilities.
    
    \item To address identified deficiencies, we propose a novel architectural framework that synthesizes the strengths of existing approaches while mitigating their limitations through strategic integration of emergent 6G capabilities (particularly embedded \ac{GenAI}). Furthermore, we delineate critical research directions necessary for the practical realization of this proposed architecture.
\end{itemize}

The remainder of this paper is organized as follows. Section \ref{sec:basic_concepts} elaborates on the fundamental concepts underpinning \ac{SRIDS} and delineates their anticipated future design objectives. Section \ref{sec:literatureReview} presents a critical review of existing survey literature focused on \ac{SRIDS}, detailing their specific scopes and research methodologies. In Section \ref{sec:architecture}, we examine the pertinent \ac{SRIDS} \blue{survey literature} concerning architectural implementations. Section \ref{sec:discussion} conducts a gap analysis comparing the capabilities of existing solutions documented in the literature with the identified future design objectives. To address these gaps, Section \ref{sec:future-directions} introduces a proposed future \ac{SRIDS} architecture and delineates necessary directions for subsequent research. Finally, Section \ref{sec:conclusion} synthesizes the principal findings of this study and offers concluding remarks.

%% file: sections/sec2.tex
\section{Background}\label{sec:basic_concepts}
In this section, we establish the foundational principles of \ac{SRIDS} by elucidating its core concepts within the 6G architecture and adopting consistent terminology for the entirety of this work. Subsequently, we examine the \ac{SRIDS} workflow, focusing on collaboration among participating entities to reveal the underlying mechanisms enabling efficient service provisioning. We then derive a \ac{LLM}-based use case, highlighting the significant challenges and primary design objectives for \ac{SRIDS} in the context of next-generation systems. Finally, we discuss ongoing standardization efforts pertinent to the advancement of \ac{SRIDS}. For improved clarity and reference, Table \ref{table:acronym-list} provides a comprehensive list of abbreviations and acronyms employed throughout the paper.

\subsection{6G as the Enabling Substrate}
\begin{figure}[t!]
    \centering
    \includegraphics[width=3.2in]{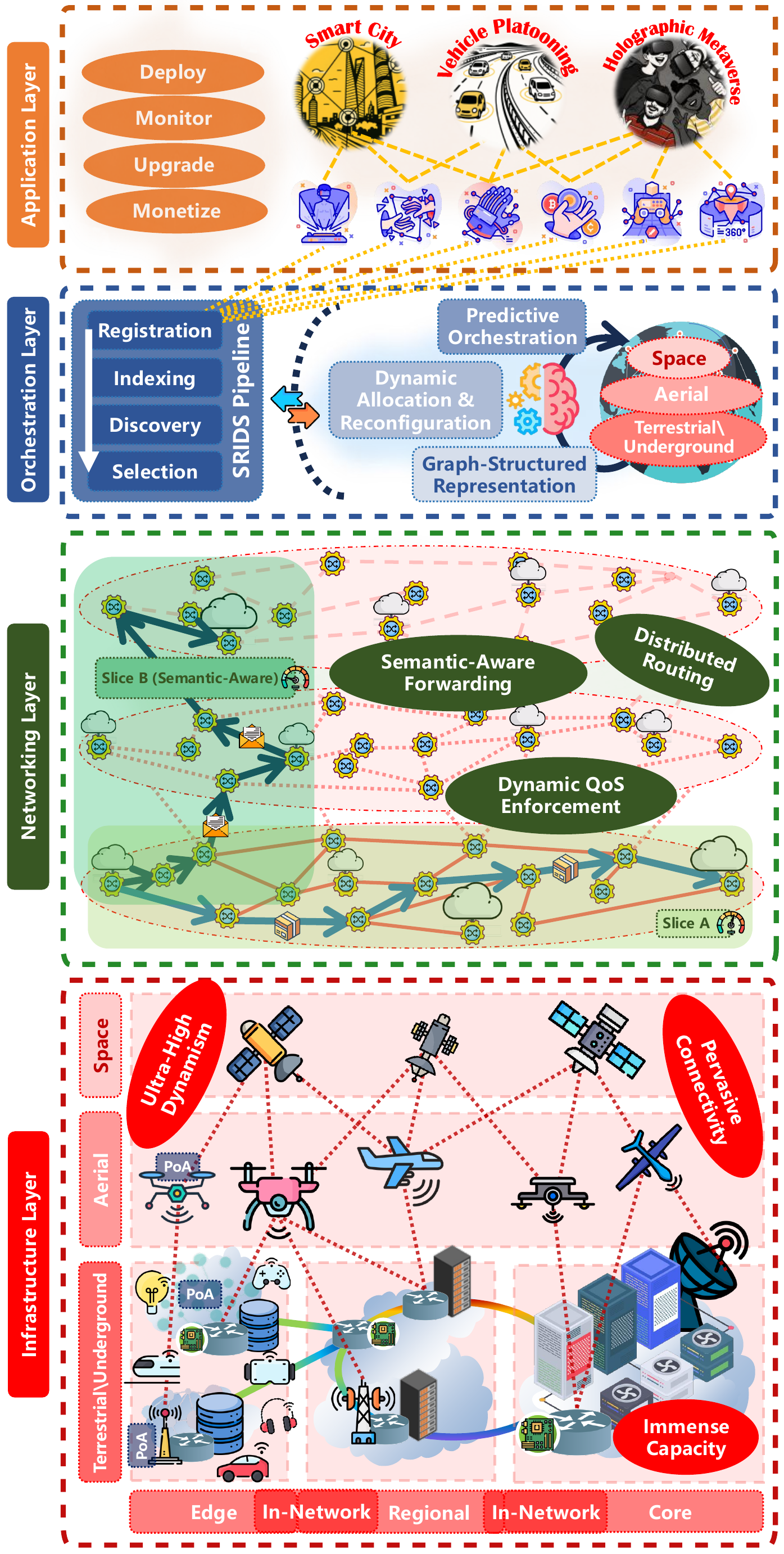}
    \vspace{-2pt}
    \caption{Mapping the future network: a breakdown of the 6G architecture and its core modules in the four layers: Infrastructure, Networking, Orchestration, and Application.}
    \label{fig:6g_architecture}
    % \vspace{-5pt}
\end{figure}

As depicted in Fig.~\ref{fig:6g_architecture}, the 6G architecture is systematically decomposed into four principal layers: Infrastructure, Networking, Orchestration, and Application.

\input{tables/tab1}

\textbf{Infrastructure}: The infrastructure layer originates with a distributed continuum of computing resources, spanning in‐network, edge, regional, and central nodes, architected to accommodate the exponential scale of connected devices and looming petabit‐per‐second bandwidth demands. These nodes are interlinked by high‐capacity wired and wireless technologies, integrating heterogeneous media: subterranean links that surmount underground propagation challenges; terrestrial networks optimized for ground‐level coverage; aerial relays using \acp{UAV} for on‐demand connectivity in remote or disaster‐affected areas; and satellite constellations that extend reach across diverse altitudes and terrains. Building on this foundation, the layer incorporates a spectrum of \acp{PoA} - from gNodeB to \ac{Wi-Fi} access points, \ac{UAV} relays, and satellite terminals - and computational entities such as near-edge, far-edge, and moving-edge nodes, all designed to meet stringent quality-of-service thresholds while adapting to the high dynamism of user mobility and resource availability. Operating as a vast distributed computing pool, this architecture delivers pervasive connectivity and real-time interaction with the physical world, enabling continuous data acquisition and processing with immense capacity, agility, and compliance with strict quality-of-service requirements.

\textbf{Networking}:
\blue{
6G redefines the traditional networking paradigm into a distributed computing continuum, where the edge functions as the mid-tier substrate bridging users and cloud resources. Within this continuum, 6G enables seamless migration, offloading, and cooperation among fog, edge, and cloud nodes through unified interfaces. The edge layer thus acts not merely as a communication intermediary but as a computational plane that hosts latency-sensitive service functions closer to users, leveraging 6G’s ultra-low latency and network programmability.
}

The 6G networking layer is founded on a distributed switching and forwarding framework designed to meet diverse service requests. At its basic level, decentralized routing algorithms replace monolithic control planes, enabling each network element to make independent path-selection decisions based on local topology and load. Building on this foundation, semantic-aware routing further refines packet delivery by incorporating content- or context-sensitive metadata - such as user intent, application type, or data priority - into forwarding decisions. Beyond packet steering, the networking layer introduces comprehensive network slicing mechanisms to partition the physical infrastructure into multiple coexisting virtual networks. End-to-end slicing spans access, transport, and core domains; radio network slicing isolates spectrum and air-interface resources; and application-based slicing tailors logical networks to specific service profiles \cite{6garchitecture_dogra}. Each slice operates as an autonomous, performance-guaranteed partition, ensuring isolation and efficient resource allocation. Crucially, 6G extends slicing with dynamic reconfiguration capabilities, allowing on-the-fly adjustment of routing policies, bandwidth allocations, and service functions in response to real-time user fluctuations and emerging stringent quality-of-service requirements.

\textbf{Orchestration}:
Decoupling the service logic from the underlying physical infrastructure, the orchestration layer concentrates on unifying heterogeneous domain resources\blue{, spanning operators, vertical industries, and sovereign network entities, }into a single network-aware, detectable, and programmable computing pool \cite{10816182}. \blue{It adopts a federated paradigm that enables secure and policy-aware interoperability across administrative boundaries while preserving domain autonomy. Such cross-domain operations introduce tailored requirements for federated authentication, mutual authorization, and policy reconciliation to ensure verifiable and privacy-preserving orchestration in multi-stakeholder environments \mbox{\cite{AIORA}}. Within this context, {\ac{SRIDS}} functions as a fundamental \red{component of the orchestration ecosystem} by maintaining an abstract knowledge of services and resources. Rather than relying on full visibility of \red{raw or fine-grained} telemetry across domains, {\ac{SRIDS}} operates on metadata that represents resource capabilities and service availability at an abstract level. This abstraction allows the orchestrator to coordinate resource allocation without compromising data sovereignty or domain confidentiality \mbox{\cite{crossDomain}}. This interoperability is facilitated by defining frameworks for unified APIs and service registries across domains \mbox{\cite{3gpp-TS29222}}, while blockchain-assisted audit trails and public key infrastructures reinforce accountability and trust.}

\blue{To manage the increasing security and complexity of multi-domain orchestration, emerging 6G systems embrace hybrid governance models that combine centralized policy enforcement for global coordination with decentralized execution for scalability and local autonomy \mbox{\cite{10078076}}. Local orchestration agents implement domain-specific compliance and trust policies, whereas a federated orchestrator ensures end-to-end service integrity, enabling secure {\ac{SRIDS}} across trust boundaries.}

This \blue{hierarchical yet federated orchestration architecture} effectively addresses the scale challenge posed by 6G’s exponential growth in connected devices and service diversity. Leveraging real-time telemetry and predictive analytics, autonomous representation engines continuously ingest and correlate user demand, computing and network resource availability, and end-to-end quality-of-service metrics to maintain an evolving, graph-structured model of the system state. Dynamic allocation engines then draw upon this model to provision computing resources and trigger reconfiguration procedures, thereby accommodating the inherent dynamism of 6G environments. The integration of digital twins (virtual replicas of physical domains) furnishes precise state estimation and what-if analysis, empowering the orchestrator to optimize resource utilization and fault mitigation at scale.

Building on these capabilities, the orchestration layer ensures efficient, end‐to‐end service provisioning through \ac{SRIDS} tasks \cite{6garchitecture_tarik}. \ac{SRIDS} maintains a catalog of available service instances and their performance profiles, supports dynamic indexing by context or intent, and automates discovery and binding to the optimal resource pool. In what follows, \ac{SRIDS} fundamentals are examined in detail.
\begin{itemize} 
    \item \textit{Service}: A service is formally defined as a function that maps a set of inputs to a corresponding set of outputs. Its behavior and deployment are captured in a \textit{service description}, which comprises:
    
    \textit{a) Function}: The service’s interfaces, accepted inputs, core processing logic, and expected outputs.
        
    \textit{b) Component}: A detailed inventory of executable components, their interdependencies, resource footprints, and the data‐exchange schemes they employ. 
        
    \textit{c) Requirement}: End-to-end quality-of-service targets - such as latency, throughput, context-specific thresholds - together with security and policy constraints.
    
    When a service consists of multiple components, it is inherently a composite entity. At a higher level of abstraction, multiple services can be orchestrated into workflows that realize capabilities beyond those of any single service. \textit{Service chaining} exemplifies this approach by linking discrete services to fulfill complex tasks.

    \item \textit{Registration \& Indexing}: This enables service providers to submit formal service descriptions to the orchestrator’s registry and receive unique identifiers that enable unambiguous reference and invocation. The registry enforces real-time consistency by immediately reflecting service introductions and removals, ensuring its catalog always represents the live set of offerings. Concurrently, advertisement mechanisms broadcast essential metadata - like functionalities and requirements - to potential entities (users or other registries in the case of utilizing non-centralized architectures), while an integrated index organizes active service instances and their network endpoints for rapid retrieval. This cohesive architecture underpins \ac{SRIDS} by maintaining an accurate, up-to-date inventory of active service instances and their locations, thereby enabling efficient discovery, binding, and timely access in the highly dynamic 6G landscape.

    \item \textit{Discovery}: This constitutes an automated mechanism that identifies the feasible services and instances for each user request, where a request specifies the required functionality and quality of experience, by evaluating the current system state (including location‐awareness metrics), the functional and quality-of-services targets defined in service descriptions and the real-time status of registered instances in the registry. By abstracting deployment details, this process enables users to locate and invoke services without prior knowledge of their endpoints or configurations. In highly dynamic 6G environments, achieving optimal performance requires a discovery mechanism that incurs minimal overhead while efficiently matching requests to available instances based on comprehensive metrics such as latency and throughput.
    
    \item \textit{Selection}: This is the algorithmic determination of the optimal service and instances from among those identified by the discovery process. It is driven by high-level policies that seek to optimize system-wide metrics - such as energy efficiency and total served requests - while satisfying each service’s functional and quality-of-service constraints. Because registry entries and runtime conditions may fluctuate, the selection mechanism continuously adapts by re-evaluating candidate instances against up-to-date monitoring data. Furthermore, by incorporating predictive analytics to forecast near-term variations in demand and resource status, the protocol can proactively adjust its choices, thereby enhancing decision robustness and overall performance.
\end{itemize}

\textbf{Application}: The application layer elevates the abstraction from individual service invocations to \textit{system-level composition and full-lifecycle management}. It recasts the services registered by the orchestration layer as modular, reusable building blocks and exposes intent-oriented interfaces through which users, such as software developers, municipal operators, or content creators, can assemble these blocks into persistent, higher-order, immersive use cases. A dedicated tool-chain supports the deployment, monitoring, upgrade, and monetization of such composites. At run time, the layer decomposes each use case into fine-grained service requests, delegates their execution to the orchestrator, ingests the resulting telemetry, and enforces closed-loop control to reconcile user quality-of-experience requirements and system-level policies with real-time state fluctuations.

\begin{figure}[t!]
    \centering
    \includegraphics[width=2.5in]{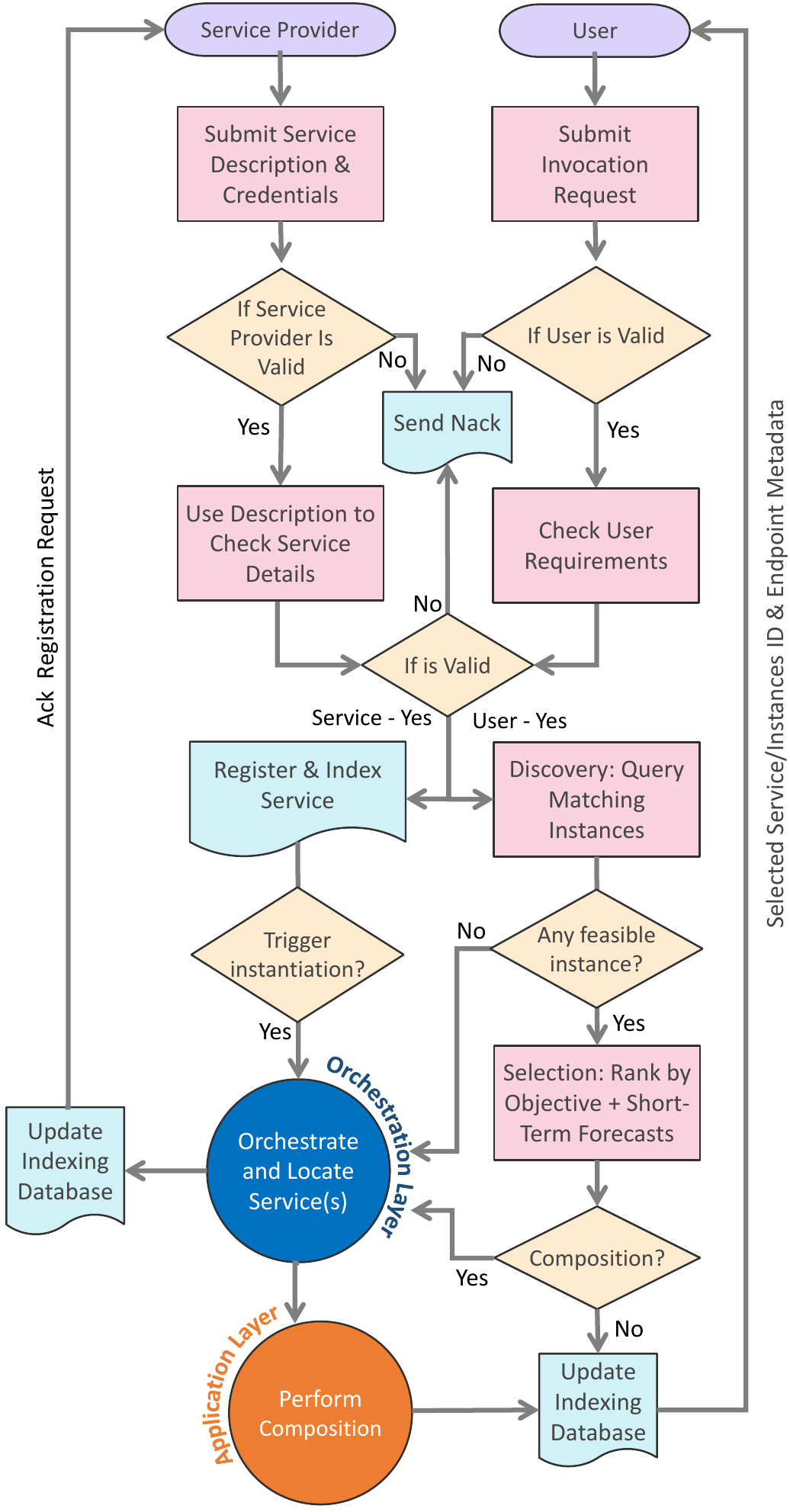}
    \vspace{-7pt}
    \caption{The detailed \blue{workflow illustrates the SRIDS process}, outlining the systematic stages for service and advertisement requests from users and service providers, respectively.}
    \label{fig:discovery-flowchart}
    \vspace{-6pt}
\end{figure}

\subsection{SRIDS: A Workflow}
Fig.~\ref{fig:discovery-flowchart} illustrates the \ac{SRIDS} workflow as two tightly coupled flows-provider registration plus indexing, and user discovery plus selection. On the provider side, a service provider submits a service description and credentials. The registry validates both the provider's identity and the description’s contents, returning a \ac{NACK} if either check fails. Once validated, the registration \& indexing module assigns a unique service identifier, persists the description in the real-time indexing database, tags it by context and intent, and ACKs the registration. The orchestration layer \blue{is then triggered to} locate the corresponding service instances \blue{if the registered service requires immediate instantiation; otherwise, the process terminates after indexing} with live endpoint metadata. On the consumer side, a user issues an invocation request specifying desired functionality and quality of experience targets. After credentials and requirement checks (again yielding a \ac{NACK} on failure), the discovery module queries the indexed catalog for matching instances. \blue{If no feasible instance is found, the orchestration layer dynamically instantiates the necessary service functions. Besides, if a composite service is required, the orchestration layer performs the composition by chaining all the necessary service functions and coordinating their deployment, while the application layer assists by providing high-level intent or composition logic}. \red{Afterwards, the orchestrator updates the registry to reflect the instantiated or composed workflow; otherwise, the index is directly updated.} The selection module ranks the available candidates based on policy constraints and short-term demand forecasts, and the final output (a selected service identifier including endpoint metadata) is returned for seamless instantiation by the networking layer.

\subsection{Futuristic SRIDS}\label{ssec_futuristic_srids}
The deployment of \ac{SRIDS} in next-generation systems entails several non-trivial challenges. To recognize, consider a continuous health-monitoring service, named Health Guardian, that (i) ingests real-time physiological and environmental streams, (ii) detects anomalous conditions, (iii) enriches alerts with patient-specific context, and (iv) leverages \acp{LLM} to generate natural-language health summaries and notifications for both patients and caregivers. This exemplary service is composed of the following components:
\begin{itemize}
    \item \textit{Ingestion}: Securely acquires and forwards raw vital-sign (e.g., electrocardiogram, oxygen saturation, motion) and ambient (e.g., air-quality, temperature) sensor data.
    
    \item \textit{Pre-processing}: Performs noise reduction and extracts features such as heart-rate variability and gait instability.
    
    \item \textit{Anomaly-Detection}: Applies threshold-based rules and lightweight machine-learning classifiers to identify early warning signs (e.g., arrhythmias, fall risk).
    
    \item \textit{Context-Enrichment}: Integrates detected anomalies with electronic-health-record data and environmental context to produce a comprehensive alert profile.
    
    \item \textit{\ac{LLM}-Inference}: Hosts a transformer-based model on computing nodes powered by graphics processing units to generate personalized, context-aware health recommendations in natural language.
    
    \item \textit{Notification}: Delivers summaries and alerts via push notifications, in-home displays, or messaging gateways.
\end{itemize}

\begin{figure}[t!]
    \centering
    \includegraphics[width=3.2in]{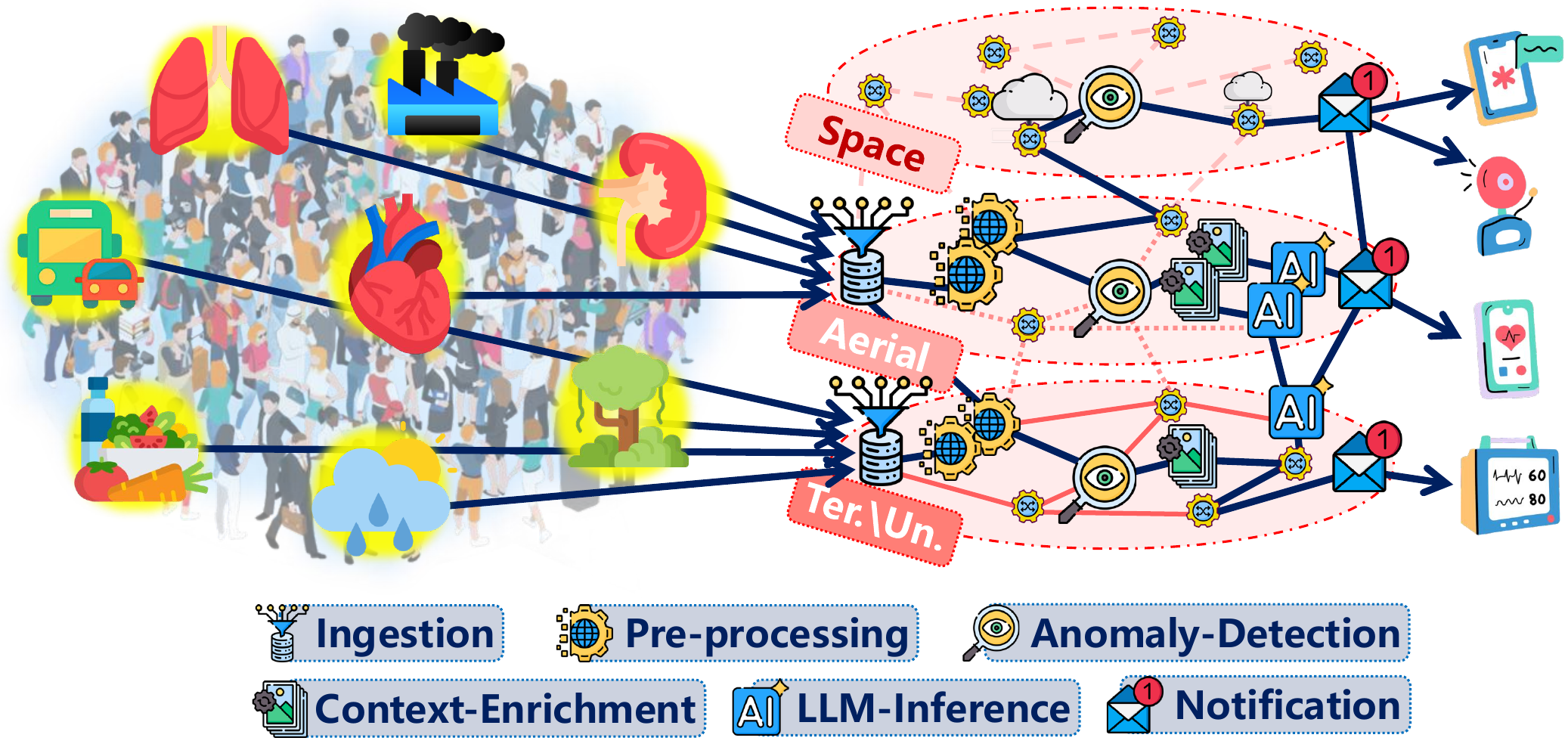}
    \vspace{-7pt}
    \caption{Health Guardian, a continuous health-monitoring service.}
    \label{fig:health_guardian}
    \vspace{-6pt}
\end{figure}

\begin{figure}[t!]
    \centering
    \includegraphics[width=2.5in]{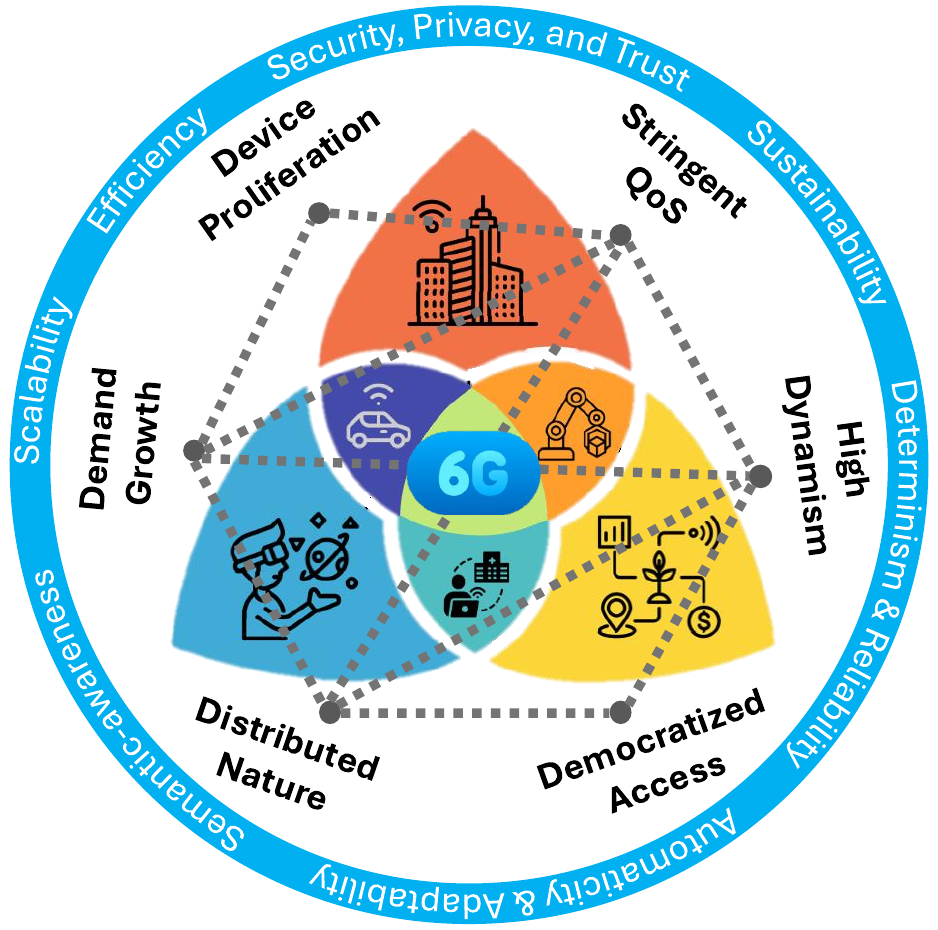}
    \vspace{-6pt}
    \caption{The challenges and primary design objectives associated with the implementation of SRIDS for future 6G use cases.}
    \label{fig:srids_challenges}
    \vspace{-6pt}
\end{figure}

In the simplest design, these components must be executed serially in the designated order, or they may be organized in a more complex graph, as illustrated in Fig.~\ref{fig:health_guardian}.
The challenges stem from four primary factors:
\begin{enumerate}
    % \item \textbf{Scale}: The number of connected devices per user in 6G is experiencing exponential growth, with projections indicating that the total device count will soon exceed the global population \cite{statistaNumber}. This substantial increase will significantly elevate bandwidth requirements, potentially reaching petabit-per-second levels in the near future \cite{BANAFAA2023}. In the Health Guardian exemplar, each patient may deploy 20–30 wearables plus multiple ambient monitors. Therefore, millions of concurrent data streams must be ingested and forwarded in a metropolitan area, sustaining hundreds of gigabits per second of raw input. 

    \item \textbf{Scale}: \blue{The proliferation of connected devices in 6G exhibits \red{a sustained and substantial} growth, driven by the convergence of human-centric, machine-type, and ambient intelligence devices. Recent projections indicate that the total number of connected devices will soon exceed the global population, reflecting dense device-per-area growth and the emergence of per-user micro-networks \mbox{\cite{statistaNumber, ITU2160}}. In practical terms, individuals may concurrently operate tens of interconnected wearables and sensors, typically 20-30 personal and ambient devices in healthcare scenarios.} This increase will significantly elevate bandwidth requirements, potentially reaching petabit-per-second levels in the near future \cite{BANAFAA2023}. Hence, millions of concurrent data streams must be ingested and forwarded in a metropolitan area, sustaining hundreds of gigabits per second of raw input.
    
    \item \textbf{Stringency}: 6G use cases and services demand stringent quality-of-service parameters, including ultra-low latency, ultra-high reliability and peak data rates, support for massive device densities, minimal jitter, and precise localization accuracy \cite{6GHealth}. In the Health Guardian exemplar, the Ingestion and pre-processing components must sustain end-to-end sensor-to-alert latency below 1 ms to detect arrhythmias or falls in real time, while the Anomaly-Detection and \ac{LLM}-Inference components guarantee $\geq 99.9999\%$ availability so that no critical event is missed. These components also handle terabit-scale aggregated throughput at jitter below 100 $\mu s$. Centimeter-level localization of patient wearables ensures accurate context enrichment (e.g., home vs. ambulance). Moreover, as the deployment rate of services like Health Guardian accelerates in dense urban centers, 6G systems are expected to be required to serve up to $10^8 \text{ devices/km}^2$ \cite{5Gto6G, ITU2160}.
    
    \item \textbf{Dynamism}: The 6G system state exhibits high dynamicity, originating from two main sources: users and infrastructure. Users in 6G environments are expected to be highly mobile and able to modify their service requirements frequently \cite{UserMobility}. Additionally, due to the distributed nature of computing and network resources in 6G, their availability, capacity, and utilization patterns are expected to fluctuate rapidly \cite{deeplearningbasedservice}. In the Health Guardian exemplar, patient mobility - transitioning from home to clinic to ambulance - triggers on-the-fly re-binding of different components to successive edge nodes. At the same time, competing 6G workloads may preempt edge computing or network resources, causing swift capacity shifts. 

    \item \textbf{Universality}: 6G aims to democratize access, enabling users without understanding the underlying technical architecture to request services with a high degree of heterogeneity. For instance, rather than being constrained by specific protocols, users should be able to describe their functional requirements using natural language \cite{6GNaturalLanguage, Autonomousnetworkorchestrationframework}. In the Health Guardian exemplar, caregivers and clinicians might state "\textit{I need a continuous electrocardiogram with 30-second storage of pre-event data when arrhythmia is detected}," automatically adjusting the pre-processing component to activate QRS complex detection algorithms. A nurse practitioner could request "\textit{add fall detection with higher sensitivity for this patient with osteoporosis},"  prompting the modification of the standard Anomaly-Detection component to incorporate new accelerometer thresholds and additional gyroscope inputs.    
\end{enumerate}
These challenges collectively define the design objectives for implementing \ac{SRIDS} in 6G systems, as shown in Fig.~\ref{fig:srids_challenges}.
Any proposed \ac{SRIDS} mechanism should be \textit{scalable}, since it should support millions of simultaneous connections with enormous bandwidth demands, while service implementation may span vast geographical areas, leveraging numerous computing and network resources. \textit{Efficiency} should be prioritized in \ac{SRIDS} algorithms, as excessive processing times would make maintaining stringent quality-of-service requirements (such as ultra-low latency) impossible in highly dynamic environments. \textit{Security} and \textit{privacy} are fundamental, given the distributed nature of 6G infrastructure, requiring robust protection against unauthorized access and safeguards against potential risks stemming from malicious attacks, misbehaving users or service providers, fraudulent service component instances, or compromised nodes. \blue{To ensure \textit{trust}, measurable properties, such as transparency, accountability, reliability, integrity, and privacy preservation, should be defined that collectively promote verifiable and explainable operations.} \textit{Sustainability} represents another critical design consideration, as without energy-efficient operation across the entire service lifecycle, the massive power consumption required to meet such immense demands could render many anticipated use cases economically or environmentally unfeasible. 

Any comprehensive \ac{SRIDS} mechanism should also emphasize \textit{determinism} and \textit{reliability}, ensuring services remain uninterrupted despite the highly dynamic nature of 6G, where user mobility and fluctuating resource availability create state changes. \textit{Automaticity} and \textit{adaptability} represent other critical design objectives, as \ac{SRIDS} solutions should autonomously reconfigure in real-time without human intervention to accommodate rapidly shifting states across distributed infrastructure. Finally, \textit{Semantic-awareness} should be embedded throughout the system, allowing it to interpret, understand, and fulfill user requests expressed in any unstructured form by users without technical expertise, translating abstract descriptions into precise technical implementations while adapting to previously unobserved functionalities.

\blue{Looking ahead, {\ac{SRIDS}} should evolve in synergy with the {\ac{GenAI}}-driven future. Generative models autonomously construct service blueprints, synthesize configurations, and anticipate demand fluctuations \mbox{\cite{GenAIFuture}}. Within such systems, {\ac{SRIDS}} provides the trusted substrate for registration, indexing, discovery, and selection, while {\ac{GenAI}} serves as the adaptive reasoning layer providing personalized and intent-aligned provisioning. This evolution complements the broader 6G AI ecosystem, where centralized AI manages global optimization, federated AI supports cross-domain learning, distributed multi-agent AI enables local autonomy, and {\ac{GenAI}} extends these by generating novel compositions and adaptive policies on demand. Integrating {\ac{GenAI}} into {\ac{SRIDS}} thus shifts orchestration from reactive to predictive and generative, enabling continuous self-optimization across the 6G continuum.}

\subsection{Standardization Activities}  
The \ac{SRIDS} ecosystem has received attention across both \blue{formal standards bodies and industry/open-source orchestration initiatives. Below, we summarize the most relevant activities and indicate how each contributes to SRIDS functionality.}

\blue{Early work on service registries and orchestrator workflows was rooted} in \ac{OASIS} specifications. \ac{UDDI} specification formalizes an \ac{XML}–based service registry that realizes the registration \& indexing module: service providers submit formal service descriptions and obtain unique identifiers, while \ac{SOAP}–based lookup operations are invoked to perform discovery. In parallel, \ac{OASIS}’s \ac{SPML} standardized the orchestrator’s lifecycle workflows (provisioning and credential management), thereby ensuring that the registry’s catalog remains synchronized with live service instances \cite{OASIS-SPML, webDiscovery}.

Within the \ac{3GPP}, the \ac{CAPIF} integrated a centralized service registry into the mobile-core orchestration layer. Governed by the \ac{SA5} technical specification group and its \ac{CFTF}, \ac{CAPIF} catalogs service instances and endpoints, each annotated with quality-of-service and security profiles, and exposes northbound \ac{REST} and \ac{RPC} interfaces for discovery and selection. This unified registry guarantees interoperable access to services across network slices and administrative domains, enabling orchestrators to bind user requests to optimal instances based on governing policies and real-time telemetry \cite{3gpp-tr}.

At the network edge, the \ac{ETSI}'s \ac{MEC} service framework defines a localized service registry for edge-hosted instances. Accompanying protocols - the MEC \ac{SAP} and the MEC service directory - propagate context-tagged metadata into downstream discovery caches. Standardized RESTful \acp{API} enable edge orchestrators to execute Selection, enforcing policies and bindings, in accordance with quality-of-service criteria, across distributed edge clusters \cite{ETSI_TS}. \blue{The ETSI} \ac{NFV-MANO} \blue{framework established the canonical lifecycle management and descriptor model for services, defining catalogs, descriptors, and orchestrator roles, thus providing a mature foundation for {\ac{SRIDS}} service descriptor expectations \mbox{\cite{etsi-nfvmano}}. ETSI Zero-touch network and Service Management (ZSM) extended this by introducing intent-driven, closed-loop automation for end-to-end management through integrated policy, analytics, and automation planes \mbox{\cite{etsi-zsm}}. Together, these standards enable {\ac{SRIDS}} engines to place, compose, and reconfigure services using lifecycle semantics abstractions}\red{, ensuring that they operate consistently across heterogeneous 6G domains.}

\blue{CAMARA, a Linux Foundation project \mbox{\cite{camara}}, and the Global System for Mobile Communications Association (GSMA) Open Gateway \mbox{\cite{gsma}} defined harmonized, developer-friendly operator northbound APIs for service management and reference implementations. By standardizing how operators expose capabilities such as network resources and quality-of-service controls, CAMARA reduces the heterogeneity {\ac{SRIDS}} should abstract from and enables consistent, automated discovery and invocation across multi-operator, cross-domain {\ac{SRIDS}} deployments.}

\blue{Open-Source platforms like Open Network Automation Platform (ONAP) operationalized {\ac{SRIDS}}-relevant primitives through production-grade orchestration and catalog engines, including service-design tools and run-time management APIs \mbox{\cite{onap}}. In these platforms, model-driven descriptors are demonstrated in practice, as well as northbound/southbound interfaces that can be integrated or emulated by {\ac{SRIDS}}. Their work informs {\ac{SRIDS}} implementation patterns for lifecycle synchronization as well as operation and management. Likewise, Telemanagement (TM) Forum’s service catalog models define standardized information schemas and APIs for service specifications and partner-oriented interactions \mbox{\cite{tm-forum}}. These information models are particularly useful for multi-stakeholder {\ac{SRIDS}} deployments, as they define canonical data schemas and operational APIs for catalog management, subscription, and monetization workflows that {\ac{SRIDS}} can adopt for cross-domain interoperability.}

These standards furnish complementary pieces of the {\ac{SRIDS}} puzzle: {\ac{UDDI}}/{\ac{SPML}} and TM Forum defined registry and catalog semantics; {\ac{CAPIF}} \red{and CAMARA standardized northbound interfaces and discovery primitives for edge services}; and NFV-MANO, ZSM, and ONAP contribute lifecycle, intent, and management primitives for dynamic instantiation, composition, and runtime coordination. Aligning {\ac{SRIDS}} with these efforts enables practical deployments to exploit existing API, descriptor models, \red{and automation frameworks that bind across edge-cloud and multi-domains.}

%% file: tables/tab1.tex
\begin{table}[t!]
\caption{List of abbreviations and acronyms.}
\vspace{-10pt}
\label{table:acronym-list}
\centering
\begin{tabular}{ll}
\toprule
\textbf{Acronym} & \textbf{Full form/expansion} \\
\midrule
6G      & Sixth-Generation \\
AI      & Artificial Intelligence \\
AMC     & Ad-hoc Mobile Cloud \\
API     & Application Programming Interface \\
AR      & Augmented Reality \\
BFT     & Byzantine-Fault-Tolerant \\
CA      & Certificate Authority \\
CAN     & Content Addressable Network \\
% CFN     & Compute First Networking \\
CoAP    & Constrained Application Protocol \\
DAG     & Directed Acyclic Graph \\
DHT     & Distributed Hash Table \\
DNS     & Domain Name System \\
GenAI   & Generative AI \\
HTTP    & HyperText Transfer Protocol \\ 
ID      & Identifier \\
IoT     & Internet of Thigs \\
IIoT    & Industrial Internet of Things \\
IOTA    & Internet of Things Application \\
\red{IP}& \red{Internet Protocol} \\
LLM     & Large Language Model \\
MANET   & Mobile Ad Hoc Network \\
MEC     & Multi-access Edge Computing \\
ML      & Machine Learning \\
MSN     & Mobile Social Network \\
\red{NACK}& \red{Negative ACknowledgment} \\
NFV     & Network Function Virtualization \\
\red{NoSQL}   & \red{No Structured Query Language} \\
ONS     & Object Name Service \\
OWL     & Web Ontology Language \\
PKI     & Public Key Infrastructure \\
PoA     & Point of Attachment \\
RDF     & Resource Description Framework \\
REST    & Representational State Transfer \\
RPC     & Remote Procedure Call \\
SAP     & Service Advertisement Protocol \\
SLP     & Service Location Protocol \\
SOAP    & Simple Object Access Protocol \\
SPML    & Service Provisioning Markup Language \\
SRIDS   & Service Registration, Indexing, Discovery, and Selection \\
\red{TCP}     & \red{Transmission Control Protocol} \\
TLS     & Transport Layer Security \\
UAV     & Unmanned Aerial Vehicle \\
\red{UDP}     & \red{User Datagram Protocol} \\
UPnP    & Universal Plug and Play \\
% Wi-Fi   & Wireless Fidelity \\
\bottomrule
\end{tabular}
\vspace{-6pt}
\end{table}

%% file: sections/sec3.tex
\section{Literature Review} \label{sec:literatureReview}

\input{tables/tab2}

Extensive literature surveys conducted over the past decade have significantly advanced the understanding of state-of-the-art service provisioning methodologies. These comprehensive analyses have systematically categorized the available approaches and critical challenges pertaining to the \ac{SRIDS} workflow. The research has meticulously examined the specialized mechanisms for service registration, indexing (including advertisement and request handling), discovery, and selection (including service composition). These functionalities have been evaluated across diverse infrastructures and from multiple theoretical perspectives, establishing a robust taxonomic foundation for further investigation, particularly from an architectural standpoint. Table \ref{table:surveys_comparison} presents a comparative analysis of these recent surveys, illustrating the evolution of conceptual frameworks in this domain. \blue{While we compare existing surveys to expose their thematic limitations, subsequent sections include recent studies that advance {\ac{SRIDS}}.}

Pervasive computing serves as the foundation for the infrastructures anticipated in future 6G systems, where a multitude of resources with varied capabilities are generally accessible. Zhu \textit{et al.} \cite{serviceDiscoveryInPervasiveEnvironment} established a taxonomic framework that systematically classifies \ac{SRIDS} functions across interconnected architectural dimensions. Their framework delineates 1) registration functionality through advertisement protocols (\textit{unicast}, \textit{multicast}, \textit{broadcast}); 2) indexing mechanisms via service naming conventions, state storage paradigms, and registry architectures; 3) discovery processes characterized by query formalization, network-topology domain boundaries, role-based authentication, and spatial parameters; and 4) selection methodologies incorporating \textit{deterministic} and \textit{nondeterministic} algorithms enhanced by contextual parameters that establish quantifiable metrics for service evaluation. Heidari \textit{et al.} \cite{HeidariCloudDiscovery2022} extended the review by employing cloud computing as the foundational substrate, and conducted a two-phase systematic literature review to evaluate cloud service discovery mechanisms, identifying critical challenges and future research directions. Their investigation methodically classified \textit{centralized}, \textit{decentralized}, and \textit{hybrid} architectural paradigms, quantitatively assessing performance across multiple operational metrics.

The advancement of wireless communication technologies has necessitated fundamental adaptations to \ac{SRIDS} methodologies. Ververidis \textit{et al.} \cite{manetServiceDiscovery2008} conducted a comprehensive analysis of service advertisement, discovery, and selection protocols within \ac{MANET}. The researchers established a taxonomic classification of discovery architectures applicable to \ac{MANET} environments, systematically evaluating their comparative advantages and limitations. Additionally, the authors examined service description formalisms, registry maintenance mechanisms, protocol interoperability requirements, and service selection algorithms. In a parallel investigation, Gavrilovska \textit{et al.} \cite{adHocNetwork-discovery} analyzed the inherent challenges of \ac{SRIDS} implementation within dynamic Ad-Hoc network topologies. Their research methodology incorporated multiple analytical dimensions, including architectural frameworks, context-awareness mechanisms (with particular emphasis on spatial localization and semantic interpretation), and security protocol implementations.

Service-oriented paradigms deployed across wireless networks depend critically on robust \ac{SRIDS} methodologies. For example, \acp{MSN}, which synthesize mobile communication protocols with service-oriented architectural frameworks, facilitate opportunistic resource and service allocation among proximate devices. Resultantly, \ac{SRIDS} protocols should incorporate mechanisms for identifying and accessing services provisioned by neighboring devices. Through systematic analysis of extant literature, Girolami \textit{et al.} \cite{GIROLAMI-MSN-discovery} developed a comprehensive framework of \ac{SRIDS} methodologies within \ac{MSN} environments. Their analytical model outlined the sequential discovery process, encompassing advertisement protocols, query formulation, selection algorithms, and service delivery mechanisms. The concept of inter-device service provisioning can be extended to inter-network paradigms, wherein Meshkova \textit{et al.} \cite{p2pSurvey2008} examined service discovery solutions spanning multiple network domains, including point-to-point overlay networks and non-\ac{IP} architectures. Their investigation concluded in the formulation of a taxonomic classification for discovery architectures. 

\begin{figure*}[!t]
\vspace{-4pt}
\centerline{\includegraphics[width=6.9in]{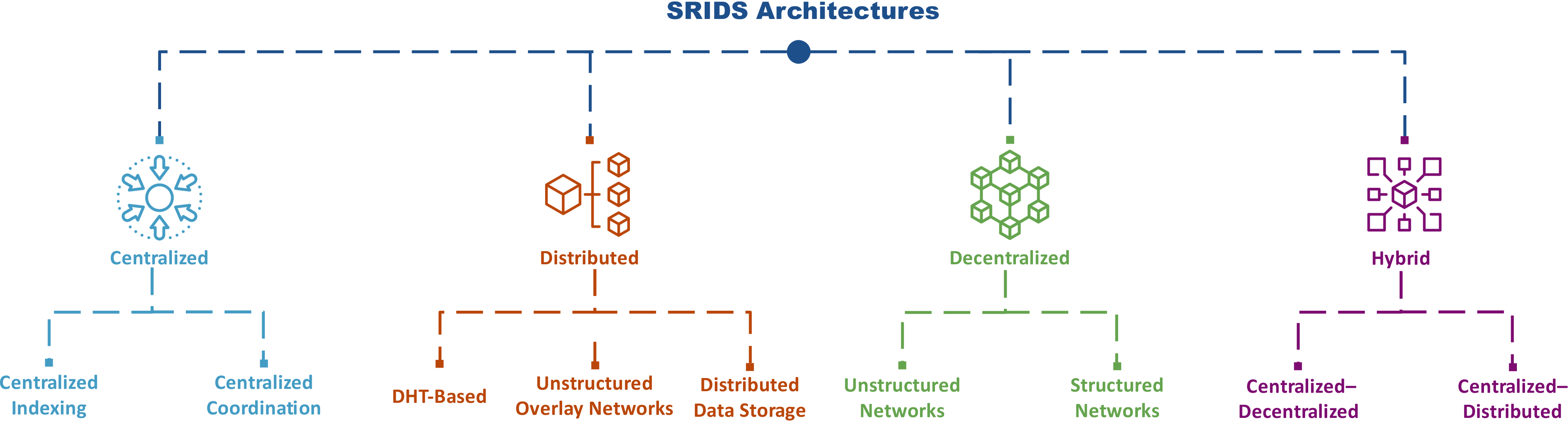}}
\vspace{-14pt}
\caption{The proposed categorization of SRIDS architectures based on their processes.}
\label{fig:architecture_categories}
\vspace{-5pt}
\end{figure*}

Contemporary application ecosystems rely fundamentally on sophisticated \ac{SRIDS} mechanisms. Zorgati \textit{et al.} \cite{serviceDisoceryIoT2019} conducted a comprehensive analytical evaluation of service discovery challenges within \ac{IoT} infrastructures. Their investigative framework established a classification taxonomy: \textit{protocol-based} and \textit{semantic-aware}. In a parallel investigation, Pourghebleh \textit{et al.} \cite{Pourghebleh2020} implemented a systematic literature review protocol to evaluate discovery methods within \ac{IoT} ecosystems. Their analytical framework formulated a classification system: \textit{context-aware}, \textit{energy-aware}, \textit{quality-of-service-aware}, and \textit{semantic-aware} methodologies. The functional attributes of each taxonomic category in these two studies were methodically assessed according to multiple performance criteria, including adaptability, dynamism, sustainability, efficiency, reliability, and scalability. Additionally, Achir \textit{et al.} \cite{ACHIR-Iot-Discovery2022} formulated a hierarchical taxonomic framework for the systematic classification of service discovery and selection methodologies in \ac{IoT} environments. Their analytical model established a tripartite categorization system: 1) description methodologies, comprising \textit{syntactic-based}, \textit{semantic-based}, \textit{resource-based}, and \textit{hybrid} approaches; 2) discovery and selection algorithms, encompassing \textit{semantic-based}, \textit{nature-inspired}, \textit{protocol-based}, \textit{context-based}, and \textit{quality-of-service-based} methodologies; and 3) architectural paradigms.

A subset of research literature has systematically evaluated \ac{SRIDS} mechanisms focused on specific technologies. Nazarabadi \textit{et al.} \cite{dhtArchitectureSurvey2021} conducted a comprehensive analysis investigating the implementation viability of \ac{DHT}-assisted architectures as foundational \ac{SRIDS} frameworks. \ac{DHT} structures function as distributed key-value repositories, providing essential service capabilities including data storage and replication, query resolution, as well as load balancing. The investigators examined \ac{DHT} implementation methodologies that enable service providers to register provisioned services and identified critical parameters for developing robust \ac{DHT}-assisted service registries. This investigation established a systematic taxonomic classification of contemporary \ac{DHT}-based methodologies and implementations, encompassing multiple dimensions of system architecture.

While existing surveys predominantly emphasize the architectural modules of \ac{SRIDS}, they fall short of addressing the forward-looking design objectives and challenges outlined in Section~\ref{ssec_futuristic_srids} from an architectural standpoint, an aspect identified as central to \ac{SRIDS} development in Section~\ref{sec:introduction}. These studies offer limited insights into the functionality of \ac{SRIDS} within multi-user, multi-service scenarios, where achieving \textit{scalability} to support millions of concurrent connections is paramount. Achir \textit{et al.} \cite{ACHIR-Iot-Discovery2022} omits a comprehensive analysis of \textit{determinism} and \textit{reliability} in service delivery and mobility management. Current research predominantly focuses on static infrastructures, thereby overlooking the critical requirements of \textit{automaticity} and \textit{adaptability} necessary for dynamic edge-cloud environments. Pourghebleh \textit{et al.} \cite{Pourghebleh2020} presented a narrow definition of mobility, failing to account for the broader system dynamicity required to ensure \textit{reliability} under fluctuating network conditions. Implementation challenges affecting \textit{efficiency} remain insufficiently investigated despite Zhu \textit{et al.}'s \cite{serviceDiscoveryInPervasiveEnvironment} preliminary work. Most of the studies inadequately address \textit{security}, \textit{privacy}, and \textit{trust} essential in distributed architectures, while neglecting \textit{semantic-awareness} capabilities necessary for interpreting unstructured user requests.

Notably, \blue{while prior literature provided coverage of the enablers and adjacent mechanisms, to date, no comprehensive survey has examined the entire {\ac{SRIDS}} pipeline tailored to 6G-enabled edge-cloud continuum infrastructures. Therefore, we systematically analyze related works across other domains to reveal their transferable principles, highlight their architectural limitations, and outline how SRIDS must evolve to meet the demands of future {\ac{GenAI}}-driven and 6G networks.} This paper also examines the efficacy of current \ac{SRIDS} mechanisms in addressing future design objectives from an architectural perspective, and where deficiencies exist. \red{Accordingly, the survey complements prior studies by bridging survey-level insights with recent novel architectural paradigms designed for emerging 6G systems, enabling a unified perspective that spans both legacy SRIDS paradigms and emerging service-management frameworks relevant to 6G deployments.}

% This paper examines the efficacy of current \ac{SRIDS} mechanisms in addressing the future design objectives from an architectural perspective, and where deficiencies exist, explores novel architectural paradigms specifically designed for the emerging 6G systems.

%% file: tables/tab2.tex
\begin{table*}[t!]
\centering
\caption{Summary of SRIDS surveys with emphasis on architectural aspects.}
\vspace{-5pt}
\label{table:surveys_comparison}
\def\arraystretch{1.7}
\begin{tabularx}{\textwidth}{lX X}
\toprule
\textbf{Survey} & \textbf{Focus areas} & \textbf{Research methods} \\
\midrule
Zhu \textit{et al.} \cite{serviceDiscoveryInPervasiveEnvironment} 
& Taxonomic framework for SRIDS functions in \textit{pervasive computing} environments 
& Categorization of registration (advertisement protocols), indexing (naming, persistence, registry architecture), discovery mechanisms, and selection algorithms. \\

Gavrilovska \textit{et al.} \cite{adHocNetwork-discovery} 
& SRIDS implementation challenges in dynamic \textit{Ad-Hoc networks} 
& Analysis of architectural frameworks, context-awareness mechanisms, and security protocols. \\

Meshkova \textit{et al.} \cite{p2pSurvey2008} 
& Service discovery across multiple network domains, including \textit{peer-to-peer overlays} 
& Taxonomic classification of discovery architectures for resource-constrained networks. \\

Ververidis \textit{et al.} \cite{manetServiceDiscovery2008} 
& Service advertising, discovery, and selection in \textit{MANETs} 
& Classification of discovery architectures, description formalisms, registry maintenance, and selection algorithms. \\

Girolami \textit{et al.} \cite{GIROLAMI-MSN-discovery} 
& SRIDS methodologies in \textit{mobile social networks} 
& Analysis of advertisement protocols, query formulation, selection algorithms, and service delivery mechanisms. \\

Zorgati \textit{et al.} \cite{serviceDisoceryIoT2019} 
& Service discovery challenges in \textit{IoT} infrastructures 
& Classification into protocol-based and semantic-aware methodologies. \\

Pourghebleh \textit{et al.} \cite{Pourghebleh2020} 
& Systematic review of service discovery in \textit{IoT} 
& Categorization into context-aware, energy-aware, QoS-aware, and semantic-aware approaches. \\

Achir \textit{et al.} \cite{ACHIR-Iot-Discovery2022} 
& Taxonomic framework for service discovery in \textit{IoT} 
& Categorization by description methods, discovery/selection algorithms, and architectural aspects. \\

Nazarabadi \cite{dhtArchitectureSurvey2021} 
& \textit{DHT-assisted architectures} as SRIDS frameworks 
& Analysis of DHT implementations for service registration and registry development. \\

Heidari \textit{et al.} \cite{HeidariCloudDiscovery2022} 
& Service discovery mechanisms in \textit{cloud computing} 
& Systematic review classifying architectures (centralized, decentralized, hybrid) with performance assessment. \\

This work 
& SRIDS architectures in the \textit{dynamic 6G edge-cloud continuum} 
& Classification into centralized, distributed, decentralized, and hybrid architectures with subcategories. \\
\bottomrule
\end{tabularx}
\end{table*}

%% file: sections/sec4.tex
\section{SRIDS Architectures}\label{sec:architecture}

\ac{SRIDS} architectures have evolved through various approaches, traditionally classified into centralized and non-centralized categories. Centralized architectures employ unified service information management, thereby simplifying \ac{SRIDS} implementation. Conversely, non-centralized architectures eliminate the central manager, enabling autonomous operation of individual modules. To establish a comprehensive taxonomic foundation for futuristic \ac{SRIDS}, we extend this classification framework and propose four distinct architectural classes: \textit{centralized}, \textit{decentralized}, \textit{distributed}, and \textit{hybrid}, as illustrated in Fig.~\ref{fig:architecture_categories}. Non-centralized approaches are characterized by two paradigms distinguished by their cooperation and decision-making mechanisms. Distributed \ac{SRIDS} maintains coordinated operation wherein \ac{SRIDS} processes are distributed across participating modules with systematic information dissemination. Decentralized approaches achieve complete autonomous operation, where each module executes \ac{SRIDS} functions independently. Hybrid architectures integrate elements from both centralized and decentralized/distributed methodologies.

\blue{While often used interchangeably, distributed and decentralized differ in coordination and state awareness. In a distributed {\ac{SRIDS}}, functions are partitioned across nodes that collaborate via shared protocols or synchronization layers to maintain global consistency, as in federated registries with replicated indices. A decentralized {\ac{SRIDS}}, by contrast, lacks coordinating logic; each node independently discovers, registers, and selects services based on local views, as seen in blockchain-based or autonomous edge domains.} \red{While it is imperative to distinguish between these two paradigms, recognizing both is necessary to ensure that SRIDS architectures can align effectively with heterogeneous 6G environments with differing requirements.}

\renewcommand{\arraystretch}{1.38}
\input{tables/tab3}

\begin{figure}[!t]
    \vspace{-4pt}
    \centerline{\includegraphics[width=3.3in]{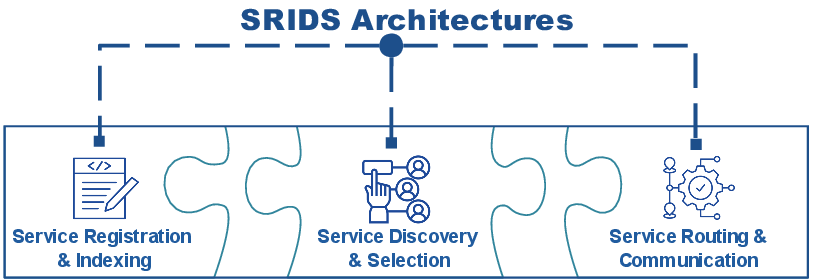}}
    \vspace{-4pt}
    \caption{The complementary categorization of existing SRIDS architectures based on their functionalities.}
    \label{fig:architecture_functional_categories}
    \vspace{-5pt}
\end{figure}

To complement the architectural categorization and gain deeper insights into the operational characteristics of \ac{SRIDS} approaches, we further classify the existing literature according to their functional roles, as illustrated in Fig.~\ref{fig:architecture_functional_categories}. While the architectural taxonomy offers a horizontal perspective on system structure, this functional categorization provides a vertical view into the internal mechanisms that drive service-related operations.
% Given the scarcity of studies encompassing all \ac{SRIDS} modules, we further classify the existing literature according to their functional roles, as illustrated in Fig.~\ref{fig:architecture_functional_categories}.

Each functional category addresses a distinct \blue{yet interdependent} aspect of \ac{SRIDS}, and their integration is necessary to achieve comprehensive system functionality. The first category, \textit{Service Registration \& Indexing}, pertains to the structuring, storage, and maintenance of service data \blue{across potentially heterogeneous domains}. The second category, \textit{Service Discovery \& Selection}, investigates the mechanisms for identifying available services and matching service requests to the most appropriate service instances. The third category, \textit{Service Routing \& Communication}, \blue{extends beyond conventional data forwarding to encompass mechanisms that maintain reachability, coordination, and interoperability} within the \ac{SRIDS} framework. \blue{In futuristic 6G environments, where services can be composed together, \textit{Service Routing \& Communication} becomes an enabler, which refers to the logical and control-plane} interactions among nodes as well as the dissemination of information \blue{to ensure consistent service delivery.} Within the architectural classifications, one or more of these functional categories may be present. Utilizing both architectural and functional classification schemes, we analyze the current state-of-the-art in \ac{SRIDS} research. \blue{As summarized in Table~{\ref{table:discovery-architectures_summary}}, existing {\ac{SRIDS}} architectures exhibit complementary strengths across individual functions; however, none integrate all functionalities concurrently under the dynamic and user-centric conditions envisioned for 6G networks.}

\input{sections/sec4/ssec4_1}
\input{sections/sec4/ssec4_2}
\input{sections/sec4/ssec4_3}
\input{sections/sec4/ssec4_4}

%% file: tables/tab3.tex
\begin{table*}[t!]
\centering
\caption{Comparative summary of well-known SRIDS architectures, highlighting their architectural and functional categories, as well as key features. \blue{The comparison reflects the extent to which each architecture jointly supports SRIDS functional components in different environments}.}
\vspace{-5pt}
\label{table:discovery-architectures_summary}
\scriptsize
\begin{tabularx}{\textwidth}{lcccccccX}
\toprule
\textbf{Architecture} & \textbf{Cen.} & \textbf{Dis.} & \textbf{Dec.} & \textbf{Hyb.} & \textbf{Re./In.} & \textbf{Di./Se.} & \textbf{Ro./Co.} & \textbf{Key features} \\
\midrule
\textit{Napster}              & \checkmark &   &   &   & \checkmark &   &   & Client–server search platform with TCP/IP \\
\textit{SLP-based}            & \checkmark &   &   &   & \checkmark & \checkmark &   & Client–server service discovery and advertisement with TCP/IP \\
\textit{Discovery Broker}     & \checkmark &   &   &   & \checkmark & \checkmark &   & Broker-based indexing mechanism with representative discovery \\
\textit{DNS-based}            & \checkmark &   &   &   & \checkmark &   &   & Client–server domain name lookup \\
\textit{Kubernetes Discovery} & \checkmark &   &   &   &   &   & \checkmark & Clustered DNS-based container orchestration with built-in service discovery \\
\textit{Detection Mesh}       & \checkmark &   &   &   & \checkmark &   & $\sim$  & Hierarchical ontology structure with global registry \\
\textit{Chord}                &   & \checkmark &   &   & \checkmark &   &   & Peer-to-peer DHT-based routing and lookup \\
\textit{Kademlia}             &   & \checkmark &   &   & \checkmark &   & \checkmark & DHT-based peer-to-peer XOR routing \\
\textit{Pastry}               &   & \checkmark &   &   & \checkmark &   & \checkmark & Structured fault-tolerant DHT-based peer-to-peer prefix routing \\
\textit{CAN}                  &   & \checkmark &   &   &   & \checkmark &   & DHT-based peer-to-peer coordinate-space routing \\
\textit{Skip Graph}           &   & \checkmark &   &   &   &   & \checkmark & Peer-to-peer multi-overlay routing with TCP/IP \\
\textit{Cycloid}              &   & \checkmark &   &   &   &   & \checkmark & Peer-to-peer fault-tolerant routing \\
\textit{Kinaara}              &   & \checkmark &   &   & \checkmark & $\sim$ &   & Multi-tier overlay with dual-layer ring for indexing \\
\textit{HandFan}              &   & \checkmark &   &   &   &   & \checkmark & Peer-to-peer geometric resource identification for load balancing \\
\textit{IOTA}                 &   & \checkmark &   &   & \checkmark & $\sim$ &   & Peer-to-peer publish–subscribe ledger and transaction settlement with DAG \\
\textit{Gnutella}             &   &   & \checkmark &   & \checkmark & \checkmark &   & Peer-to-peer service discovery via controlled flooding and hop limits \\
\textit{Freenet}              &   &   & \checkmark &   & \checkmark &   & $\sim$  & Peer-to-peer anonymous registration with encrypted descriptions \\
\textit{SD-AMC}               &   &   & \checkmark &   &   & \checkmark &   & Peer-to-peer publish–subscribe device-centric resource-aware discovery \\
\textit{Tree-based}           &   &   & \checkmark &   & $\sim$ &   & \checkmark & Peer-to-peer scalable and fault-tolerant routing \\
\textit{Social-based}         &   &   & \checkmark &   & \checkmark & $\sim$ &   & Logical graph overlays using trusted neighbors for discovery \\
\textit{Jini}                 &   &   &   & \checkmark & $\sim$ & \checkmark &   & Service-oriented discovery using lookup servers with TCP and UDP \\
\textit{BitTorrent}           &   &   &   & \checkmark & \checkmark &   &   & Peer-to-peer dynamic registry with one or multiple trackers \\
\textit{CoAP-based}           &   &   &   & \checkmark & \checkmark & $\sim$ &   & Centralized cloud composition with decentralized fog indexing via CoAP \\
\textit{eDonkey}              &   &   &   & \checkmark & \checkmark &   &   & Directory servers combined with peer-to-peer substrates fragmenting content \\
\textit{Istio}                &   &   &   & \checkmark &   & \checkmark &   & Proxy-based control plane with distributed data plane using TLS for discovery \\
\bottomrule
\end{tabularx}
\begin{tablenotes}
\footnotesize
\item Cen.: centralized, Dis.: distributed, Dec.: decentralized, Hyb.: hybrid, Re.: registration, In.: indexing, Di.: discovery, Se.: selection, Ro.: routing, Co.: communication. Symbol $\sim$ indicates partial support.
\end{tablenotes}
\vspace{-7pt}
\end{table*}

%% file: sections/sec4/ssec4_1.tex
\subsection{Centralized}\label{ss_centralized}
In a centralized architecture, \ac{SRIDS} processes are governed by central modules or directories that serve as the entry point for users' or service providers' queries. Directories can be hierarchical, resembling trees with levels of directories branching off from a central point. Each level may represent a different category, domain, or subdomain, depending on the architecture and may have different functionalities. 

A centralized \ac{SRIDS} architecture presents distinct advantages and limitations. The advantages include enhanced efficiency in execution and deployment, streamlined implementation processes, improved sustainability in maintenance operations, and accelerated development cycles. Additionally, centralized architectures demonstrate consistent service availability through unified control and management mechanisms. However, dependence on a singular directory for coordination constitutes a primary limitation of these architectures. This dependency compromises reliability by introducing a single point of failure, wherein directory malfunction results in complete \ac{SRIDS} system failure. Centralized architectures exhibit reduced reliability compared to distributed or decentralized approaches, as a single entity governs the entire operational environment. Furthermore, the scalability of centralized architectures is constrained due to potential bottlenecks as the volume of registered services and requests increases. Finally, concentrating service information within a central repository raises significant privacy concerns and compromises security - a breached directory may expose sensitive information and violate privacy protocols.

\subsubsection{Centralized Indexing} \hspace{\parindent} 
As a subcategory of centralized architectures, centralized indexing leverages a central index registry or directory server to store and manage available services and their metadata information. For users and service providers, the central index provides a point of reference for registering or requesting services. Considering the vast amount of service-related information, a centralized repository must be capable of consistently curating substantial volumes of data while managing multiple access challenges. To address the challenges, researchers tried a variety of approaches, such as Li \textit{et al.} \cite{centralIndexingStorage} proposed a distributed \ac{NoSQL} database for central-indexing storage that supports parallel processing for handling large volumes of data. By using the object identifier as the row key, event timestamp as the column identifier, and event index content as the cell value, the updated storage schema optimizes the efficiency of accessing data records, resulting in commendable performance concerning concurrent access. Following that, centralized indexing architectures are described.

\begin{figure}[t!]
    \centering
    \includegraphics[scale=0.34]{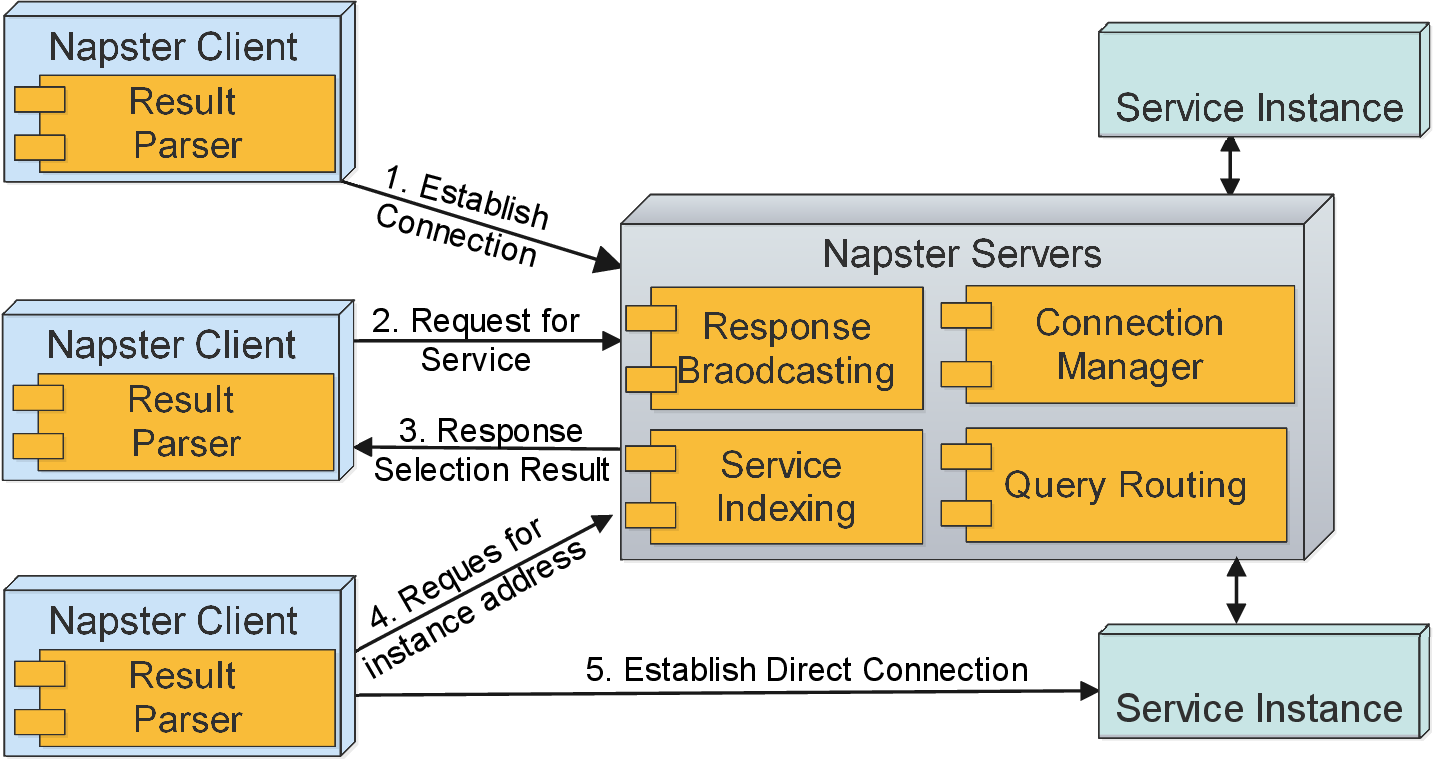}
    \vspace{-2pt}
    \caption{Napster SRIDS architecture and selection process \cite{howe2000napster}.}
    \label{fig:napster}
    % \vspace{-5pt}
\end{figure}

\paragraph{Napster} \hspace{\parindent} 
The Napster \ac{SRIDS} architecture provides centralized service registration and indexing within a peer-to-peer environment by employing a central server to collect, index, and manage service-related metadata, encompassing service names, descriptions, and network locations, to enable efficient discovery and selection \cite{howe2000napster}. Building upon this capability, the workflow unfolds as follows: the Napster client transmits a query specifying the desired service name or keywords to the central server (Fig.~\ref{fig:napster}), which executes a service selection operation against its indexed repository and returns a list of potential instance matches with corresponding metadata; the client then parses and displays these matches for user evaluation; once the user selects an instance, the server provides its \ac{IP} address, and the client establishes a direct peer-to-peer connection to that service instance, thereafter sending all subsequent requests directly and bypassing the central server.

\paragraph{\ac{SLP}-based} \hspace{\parindent} 
A centralized‐indexing approach based on the \ac{SLP} provides the capability to aggregate, index, and manage service advertisements within a unified directory, thereby enabling efficient resolution of service queries and reducing network traffic. Building on this capability, Chaudhry \textit{et al.} \cite{slpArchitecture2006} employed a directory proxy agent that collects service metadata from heterogeneous network domains and forwards it to a central \ac{SLP} directory; upon receiving a client request, the directory proxy agent queries the directory index and returns a concise list of nearby service instances, which reduces discovery overhead and accelerates network‐selection decisions. Extending this workflow to fog computing, Davoli \textit{et al.} \cite{DavoliArchitecture2021} integrated an \ac{SLP} service agent on each fog node that publishes service descriptors via the fog hypervisor service manager to an \ac{SLP} directory agent; the directory agent then centralizes these descriptors into its index, supporting critical orchestration functions and enhancing scalability by maintaining an up‐to‐date catalog of all fog‐hosted services for subsequent discovery and binding.

\paragraph{Discovery Broker} \hspace{\parindent} 
Broker-based \ac{SRIDS} architectures furnish a centralized indexing capability by aggregating service advertisements and metadata within a single discovery broker, complemented by a semantic matching engine to enable flexible, semantically rich query matching over named data networking interfaces \cite{brkerArchitecture}. Building on this capability, the workflow proceeds in two phases: during the \emph{indexing phase}, users issue service‐lookup interests to the discovery broker, which consults the semantic matching engine to perform semantic matching against its indexed registry of available services and then returns matched service descriptions; during the \emph{content retrieval phase}, users issue interests directly to the selected service instance to fetch content. Concurrently, service providers invoke \emph{registration} and \emph{deregistration} operations by sending update messages to the broker, which maintains and updates its local service table accordingly.

Chirila \textit{et al.} \cite{device_discovey_2016} enhanced broker-based centralized indexing by introducing an offline semantic clustering engine and similarity metrics to organize registered services into clusters, thereby enabling representative service discovery. Building on this centralized index, the registration of new or updated services triggers dynamic reclustering: the broker first performs offline clustering on the initial service set, then, upon each registration event, updates the cluster membership and recomputes a cluster representative via defined similarity metrics. At query time, the broker delegates the user’s lookup to its semantic matching engine, which identifies the best‐matching cluster representative and returns the associated service description to the user.

Nguyen \textit{et al.} \cite{hierarchicalIndexing} proposed a hierarchical, broker-based centralized indexing framework across edge, regional, and central computing nodes to distribute indexing load while retaining a unified, central registry at each tier. Leveraging this capability, the process comprises two stages: in the \emph{service initialization stage}, incoming service deployments are propagated from edge nodes to regional nodes and optionally to central resources, and metadata are stored in four centralized tables - registered service store, forwarding information base, pending interest table, and online service store; in the \emph{service discovery stage}, a user first queries its local edge broker, which, if no local match is found or quality-of-service requirements are unmet, escalates the query to regional and then to central nodes. The aggregator module forwards requests, the service discovery component retrieves matching entries from the centralized tables, and the broker returns service endpoints. Although this hierarchical brokering optimizes computation at the central controller, it incurs additional latency in inter-tier communication.
 
Amadeo \textit{et al.} \cite{togetherICN} introduced a social-aware broker that centralizes indexing by maintaining user profiles, service metadata, and social context to perform socially informed semantic matching in social-information-centric networking. Exploiting this capability, users submit interest packets to the broker, which uses combined semantic and social-relevance metrics to match requests to providers; upon a successful match, the social-aware broker returns service descriptors drawn from its central index, enabling users to retrieve content directly from the selected providers. This centralized, socially aware matching process enhances discovery precision and reduces redundant network traffic.

\subsubsection{Centralized Coordination} \hspace{\parindent} 
Centralized coordination architectures instantiate a single authoritative control plane that orchestrates all interactions among users, service providers, and intermediary modules. Distinct from centralized indexing, whose primary function is to store service metadata in a central registry, centralized coordination focuses on managing invocation sequences, enforcing synchronization constraints, and applying governance policies from a central mediator. This mediator validates incoming requests, arbitrates access control, consolidates coordination logic, and monitors invocation flows to simplify per-service implementations and reduce distributed agreement overhead. To counteract its inherent vulnerability as a performance bottleneck and single point of failure, the central entity integrates fault‐tolerant mechanisms (like active-standby replication, consensus‐based leader election, and periodic state checkpointing) to guarantee high availability.

The centralized coordination paradigm further enables advanced \ac{SRIDS} functionalities: by maintaining a global view of available services, their capabilities, and dependency structures, the coordinator can compose atomic services into complex workflows, define precise sequences, and orchestrate multi‐step processes to fulfill composite objectives. Moreover, it supports transactional workflows and distributed commit protocols, ensuring atomicity, consistency, isolation, and durability across multiple service endpoints, by sequencing commit or rollback operations centrally, thereby preserving system integrity in multi‐service transactions. The following details architectures that leverage centralized coordination.

\paragraph{\ac{DNS}-based} \hspace{\parindent} 
\ac{DNS}‐based coordination offers a globally consistent name‐resolution service by mapping human‐readable domain names to network endpoints. It leverages a hierarchical namespace, delegated zones, and distributed caching under a central authority to provide unified registration and indexing of service identifiers across a geographically distributed infrastructure. Building on these capabilities, the \ac{DNS} resolution workflow proceeds as follows: a \ac{DNS} user issues a lookup request to its configured recursive resolver, which first inspects its local cache for the requested record. If absent, the resolver iteratively queries the root servers to locate the appropriate top‐level domain servers, then forwards the request to the authoritative name server, as the central coordinator, for the target zone. Upon receiving the definitive IP address record, the resolver caches it according to its time‐to‐live and returns it to the user.

\begin{figure}[t!]
    \centering
    \vspace{-10pt}
    \includegraphics[scale=0.42]{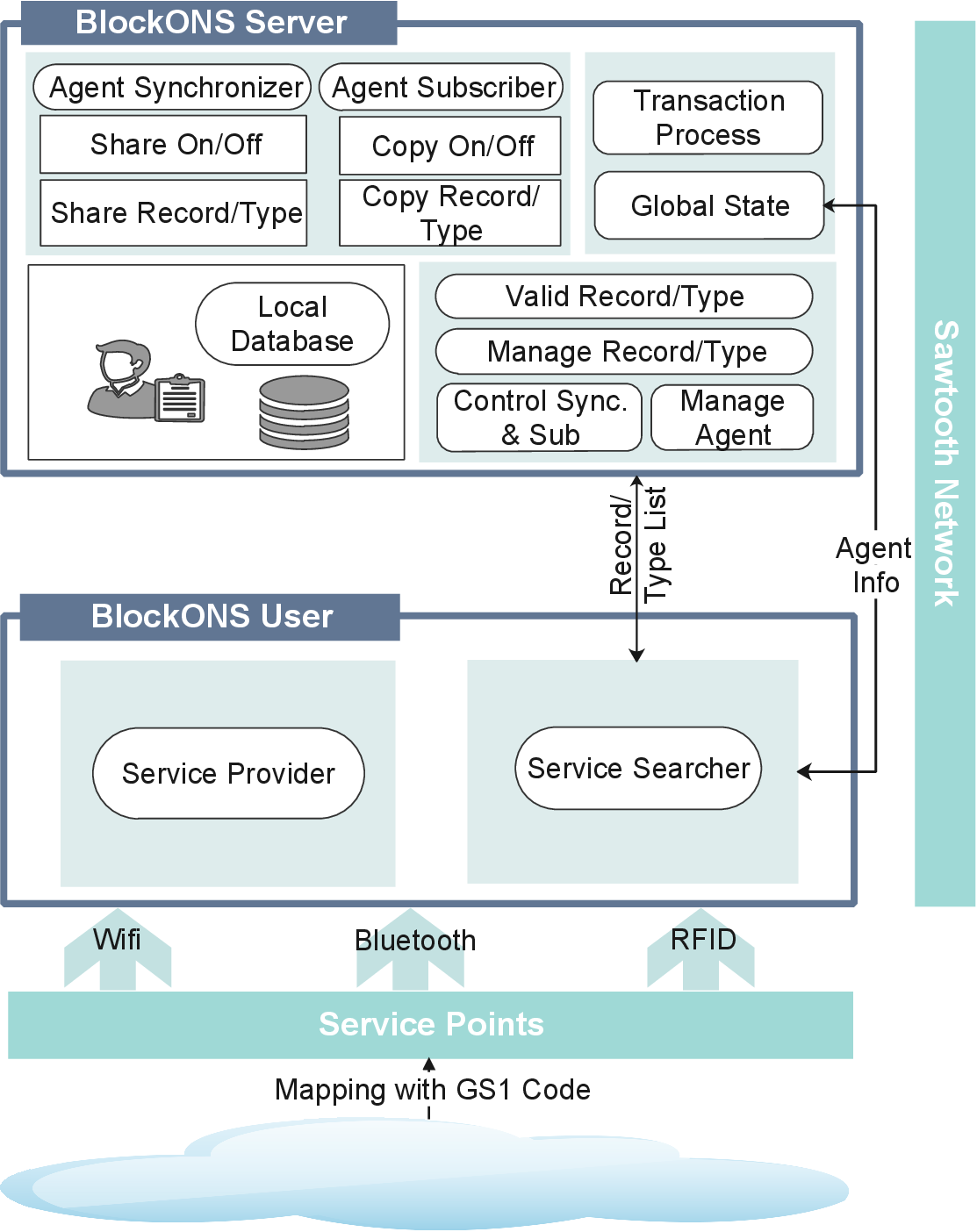}
    \vspace{-30pt}
    \caption{Blockchain-based Object Name Service (BlockONS) SRIDS architecture \cite{blockONS}.}
    \label{fig:blockOns}
    % \vspace{-5pt}
\end{figure}

% use of ONS in \ac{DNS} (https://www.gs1.org/sites/default/files/docs/epc/ons_2_0_1-standard-20130131.pdf)
The \ac{ONS} adapts \ac{DNS}‐based coordination to the \ac{SRIDS} domain by employing the \ac{DDDS} algorithm to resolve \ac{GS1} identification keys into service metadata and network locations \cite{gs1Ons}. \ac{ONS} furnishes a centralized registry that associates logical object names with service endpoints, enabling dynamic discovery across heterogeneous and dispersed environments. In the \ac{ONS} workflow, service providers publish metadata (e.g., service name, network address) into designated \ac{DNS} zones. A user issues a \ac{DDDS}‐formatted query for a \ac{GS1} key, which the recursive resolver routes through the \ac{ONS} namespace until it reaches the authoritative server. The server, following its central coordination, responds with the corresponding service record, allowing the user to connect directly to the specified endpoint. Although \ac{ONS} yields a unified view of available services, it inherits \ac{DNS} vulnerabilities, such as cache poisoning, spoofing, and local \ac{DNS} cracking, and exhibits limited fault tolerance, making it susceptible to service disruption and data tampering.

Building on \ac{ONS} capabilities, several architectures enhance \ac{DNS}‐based \ac{SRIDS}: Mitsugi \textit{et al.} \cite{migusi2014} integrated \ac{ONS} with \ac{UPnP} in an electronic product code framework, where a central coordination server fuses \ac{GS1} lookups and \ac{UPnP} advertisements into a comprehensive resource‐constrained device inventory. Jara \textit{et al.} \cite{digcovery} proposed Digcovery, a centralized \ac{IoT} discovery architecture that couples \ac{DNS} queries with Elasticsearch indices to register and resolve heterogeneous devices (e.g., ZigBee sensors) via standardized \acp{API}, thereby achieving scalable universal resource discovery. Korea \textit{et al.} \cite{blockONS} introduced BlockONS (Fig.~\ref{fig:blockOns}), which augments \ac{ONS}’s centralized authority with blockchain‐based consensus and off‐chain replication. In BlockONS, a central coordination agent manages service data models and scaling rules, while nodes participate in distributed consensus to verify record updates. The agent enforces role‐based replication policies, refines service metadata for efficient indexing, and orchestrates fault‐tolerant recovery, thereby mitigating \ac{DNS} cache poisoning and single‐point‐failure risks. Horvath \textit{et al.} \cite{horvath2025} embedded geospatial metadata into \ac{ONS} records within a 6G edge‐cloud continuum. An authoritative server, acting as the central coordinator for geographic delegation, applies proximity and load‐balancing policies to direct users to the nearest edge service instance. This centralized coordination of location‐aware \ac{DNS} responses reduces end‐to‐end latency and enhances quality‐of‐service in geographically distributed deployments.

\paragraph{Kubernetes Service Discovery} \hspace{\parindent} 
Kubernetes furnishes a unified coordination layer that automatically registers, indexes, and resolves containerized microservices within a cluster \cite{burns2018kubernetes}. Its core capabilities include a built-in name‐resolution mechanism (CoreDNS or kube‐dns), a consistent \ac{REST} \ac{API} for resource registration and querying, and programmatic clients (kubectl, client‐libraries) that enable both user agents and internal controllers to discover services and pods without hard-coded addresses. In operation, each microservice deployment is annotated with a \ac{DNS}‐compatible service name; when a user issues a lookup via the Kubernetes \ac{DNS} service, the name server consults its registry to translate the logical service name into one or more ClusterIP addresses. Concurrently, the Kubernetes \ac{API} server and controller manager maintain an in-memory view of active services, exposing endpoints through standard \ac{API} calls. Applications and other services invoke the Kubernetes \ac{API} or perform \ac{DNS} queries to resolve service names, thereby obtaining the current network locations of their peers and ensuring seamless inter-service communication.

\begin{figure}[t!]
    \centering
    \includegraphics[scale=0.33]{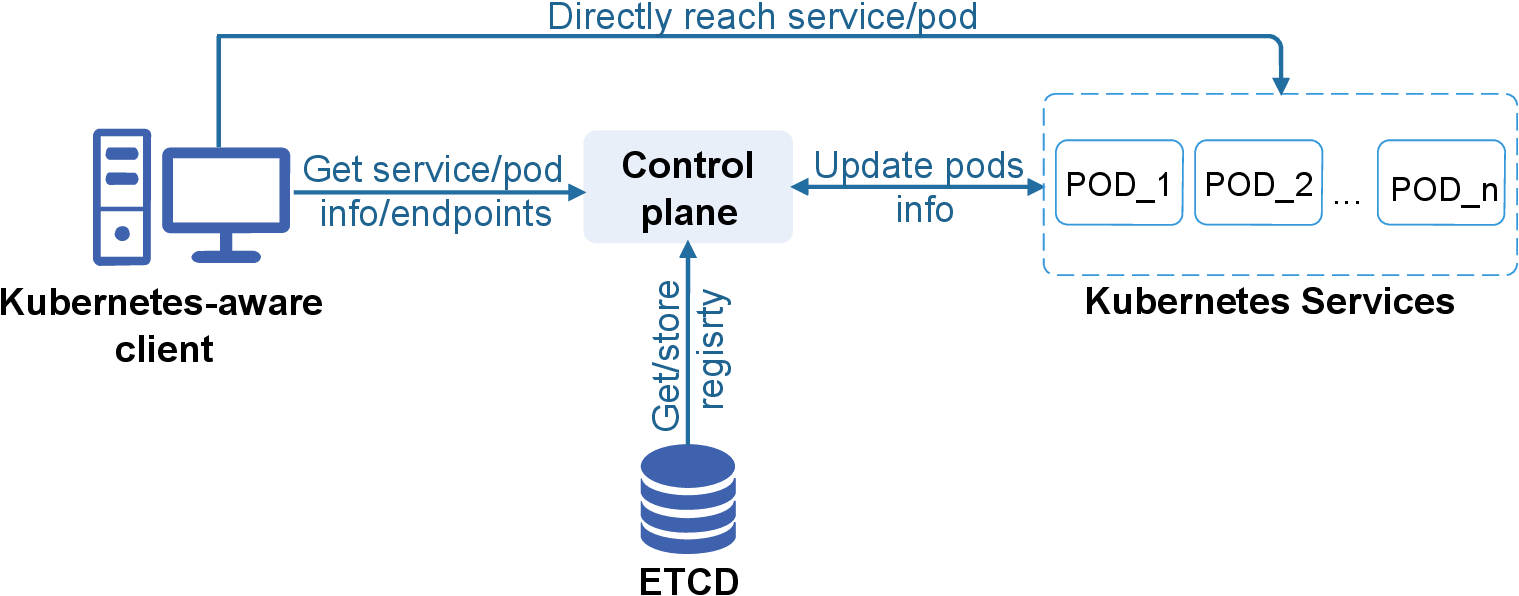}
    \vspace{-8pt}
    \caption{Kubernetes SRIDS architecture and its process \cite{burns2018kubernetes}.}
    \label{fig:kubernetes_discovery_arch}
    % \vspace{-5pt}
\end{figure}

The heart of Kubernetes lies in its central control plane, as depicted in Fig.~\ref{fig:kubernetes_discovery_arch}, which orchestrates registration, indexing, and consistency of cluster state through a distributed key–value store, known as ETCD. The control plane’s primary capabilities comprise atomic transactions for resource creation and update, watch semantics for change notifications, and strong consistency guarantees across multiple control-plane replicas. When a service object is created or modified, the \ac{API} server commits the record to ETCD; the core \ac{DNS} module (CoreDNS) and kube‐proxy each subscribe to watch streams to receive coordinated updates. CoreDNS regenerates \ac{DNS} records in real time, while kube‐proxy configures packet‐level rules on each node to route traffic to the appropriate pod \acp{IP}. This tightly coupled workflow ensures that any change in service definition (whether addition, update, or deletion) is propagated atomically to all clusters, maintaining a single source of truth and preventing split‐brain scenarios in service routing.

To accommodate the ephemeral nature of pods, Kubernetes introduces service objects that present a stable virtual \ac{IP} and act as coordinated proxies for a dynamically changing set of endpoints. This mechanism’s capabilities include label-selector-based membership, session-affinity configuration, and built-in load balancing across backend pods. Under this model, the control plane continually reconciles the desired set of pods (determined by label selectors) with the actual endpoints list, updating the endpoints \ac{API} object in ETCD and triggering synchronized updates in kube-proxy. Traffic to the virtual \ac{IP} is then distributed evenly across healthy pods, guaranteeing resiliency despite pod churn. Complementing this native \ac{SRIDS} solution, HashiCorp Consul offers a multi-datacenter coordination service with analogous capabilities \cite{consulWebsite}: secured service registration with mutual \ac{TLS}, a gossip protocol for distributing catalog updates, and health‐check driven membership flags \cite{Consul2021}. Upon startup, Consul agents register their local services to a central catalog where periodic health probes and gossip dissemination ensure that all agents maintain an up-to-date view of service availability. Users query the Consul \ac{DNS} or \ac{API} to resolve service names into \ac{IP} addresses, enabling dynamic routing, load balancing, and failover in geographically dispersed Kubernetes clusters.

\begin{figure}[t!]
\centering
\includegraphics[scale=0.35]{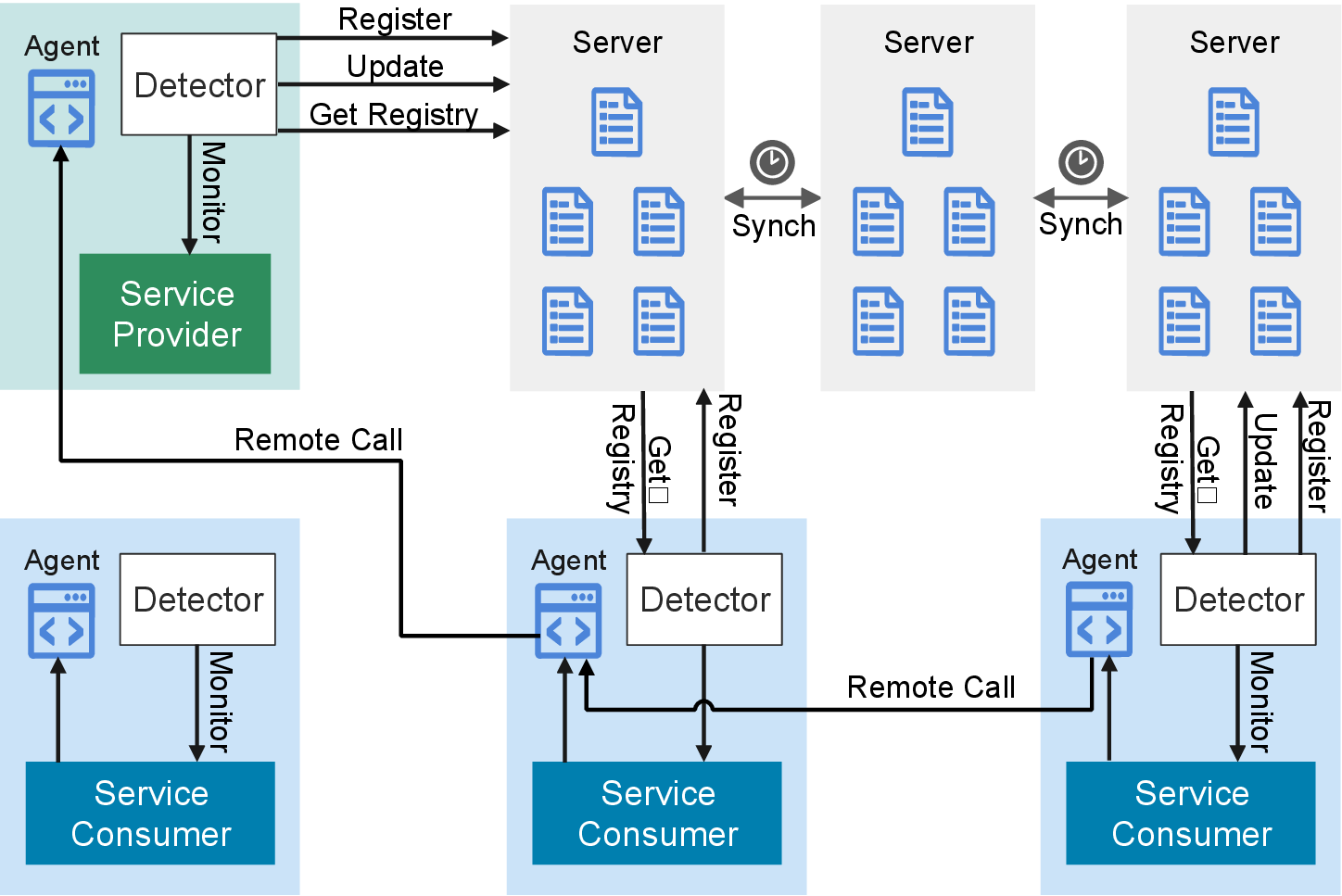}
% \vspace{-3pt}
\caption{Detection Mesh SRIDS architecture that serves the service registration and indexing functionalities \cite{detectionMesh}.}
\label{fig:detection_mesh}
% \vspace{-5pt}
\end{figure}

\paragraph{Detection Mesh} \hspace{\parindent} 
The detection mesh architecture of Zhu \textit{et al.} \cite{detectionMesh} establishes a centrally coordinated, hierarchical framework for service registration and indexing in \ac{SRIDS}. As depicted in Fig.~\ref{fig:detection_mesh}, its chief capabilities comprise a three‐tier module structure (agent, detector, and server) that cleanly separates request forwarding, health monitoring, and global registry maintenance. This maintains a hierarchical ontology structure with abstract and instance layers for semantic categorization and load balancing, and an extensible coordination plane that enforces consistency of service metadata across distributed environments while supporting high‐level policy orchestration. Building on these capabilities, microservice instances first register with their local detector, which authenticates the service and relays its description to the central server. The server, acting as the coordination nexus, updates the global service registry and propagates incremental index updates back to all detectors. Agents then poll their associated detectors to discover available services: each lookup is scoped by the hierarchical ontology structure taxonomy so that only the relevant abstract or instance layer is queried, thereby reducing coordination overhead. All registration, indexing, and cache‐invalidation operations are centrally orchestrated, ensuring that every agent, detector, and server maintains a consistent, up‐to‐date view of the entire service landscape.  

%% file: sections/sec4/ssec4_2.tex
\subsection{Distributed}\label{ss_distributed}
A distributed, or directory‐less, \ac{SRIDS} architecture dispenses with any single coordinating entity by empowering individual participants to share responsibility for registration, indexing, discovery, and selection. Rather than forwarding all requests and updates to a central registry, each node contributes to the maintenance of service metadata, propagating advertisements, processing queries, and relaying information according to cooperative protocols. In doing so, the system realigns the traditional client–server paradigm into a peer‐centric interaction model, wherein providers and users collectively uphold the integrity and availability of the service catalog.  

Building on this cooperative foundation, distributed \ac{SRIDS} exhibits several key strengths. First, the elimination of a single point of failure inherently enhances fault tolerance and resilience: individual participants may fail or become compromised without precipitating a total system outage. Second, the security risks associated with targeting a lone directory server are mitigated, since adversaries must compromise multiple, geographically and logically dispersed nodes to disrupt SRIDS. Third, by distributing decision‐making and encouraging self‐organization, the architecture scales organically as nodes join or depart; dynamic registration mechanisms ensure that updates propagate swiftly, thus preserving operational continuity even under high churn. Despite these benefits, directory‐less \ac{SRIDS} introduces notable challenges. The lack of a trusted central authority necessitates that each node perform additional cryptographic verification and trust management. This imposes computational overhead that may degrade performance and energy efficiency - an acute concern in resource‐limited edge–cloud environments. Furthermore, ensuring consistency of service views across peers requires sophisticated synchronization schemes; absent a unified directory, nodes can develop divergent catalogs, leading to stale or conflicting discovery results. Finally, the decentralized topology amplifies the susceptibility to distributed denial‐of‐service attacks and routing perturbations unless robust protection mechanisms, such as secure gossip dissemination or consensus‐based agreement protocols, are employed to preserve coherence and prevent malicious interference.

\subsubsection{DHT-Based} 
\begin{figure}[t!]
    \centering
    \includegraphics[scale=0.35]{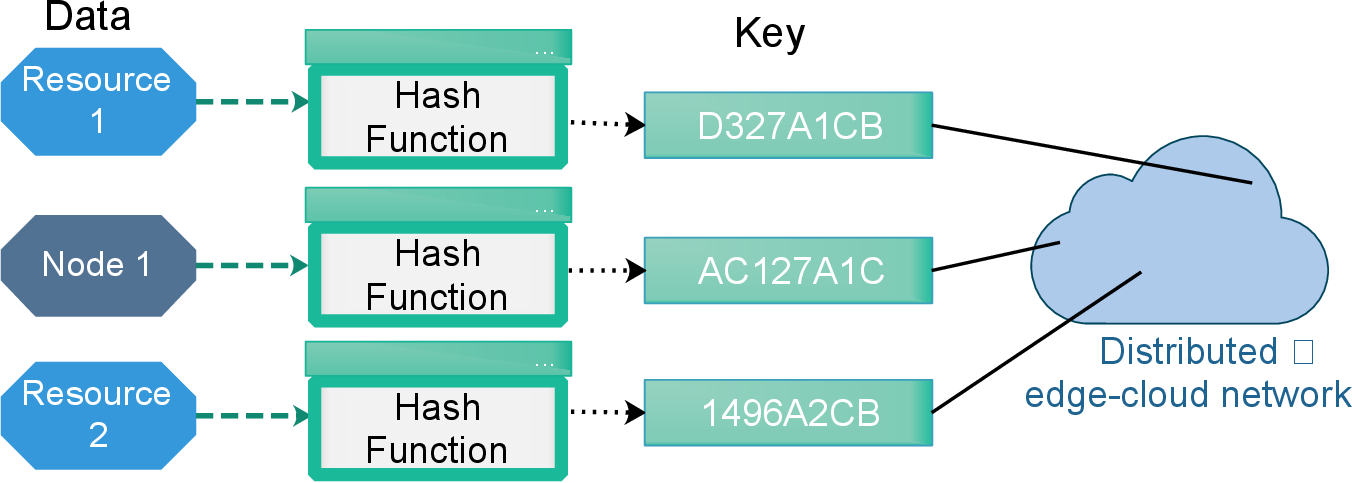}
    \vspace{-4pt}
    \caption{Using a hash function to generate service and node ID in DHT.}
    \label{fig:dht_format}
    \vspace{-5pt}
\end{figure}

\hspace{\parindent} 
A \ac{DHT}‐based distributed \ac{SRIDS} architecture realizes service provision without any centralized directory by organizing edge–cloud nodes into a structured overlay network whose routing and storage responsibilities are governed by collision‐resistant hash functions. Each node in the overlay network acquires a unique identifier via a secure hash of its network address, and every service or resource description is likewise assigned a hashed key. The key–value mapping, where the key denotes the resource or service \ac{ID} and the value indicates the node \ac{ID} responsible for that entry, is then stored at the node whose identifier is closest to the key in the identifier space (Fig.~\ref{fig:dht_format}). 

Building on this foundation, service registration proceeds by hashing the service’s canonical identifier, routing the registration message through the overlay’s logarithmic‐time lookup protocol (for example, using finger tables or routing buckets), and depositing the service metadata at the designated node. Service indexing follows the same procedure for each service instance, storing its network endpoint in the local registry of the responsible node. When a user initiates discovery, the user computes the hash of the desired service \ac{ID} and invokes the overlay's lookup routine, via \ac{TCP}, \ac{UDP}, or \ac{RPC}, to locate the node holding the corresponding key–value entry. Upon resolution, the node returns the service \ac{ID} and instance address, at which point the user may establish a direct session with the selected service instance.  

By distributing both metadata storage and query processing evenly across all participants, \ac{DHT}‐based \ac{SRIDS} achieves inherent fault tolerance (no single node failure incapacitates the lookup mechanism) and provable correctness, since the hash‐based assignment and routing algorithms guarantee that each existing key maps to exactly one responsible node. The overlay scales gracefully: each node maintains only a logarithmic number of routing pointers and processes lookups in logarithmic time, while the uniform key distribution yields natural load balancing among nodes. Nonetheless, the decentralized \ac{DHT} substrate incurs certain costs. Storage efficiency is diminished because nodes must maintain not only their own data but also routing tables and, in many implementations, replicated entries for fault tolerance. The overlay’s stabilization routines, required to reassign keys upon node joins or departures, generate substantial overhead in highly dynamic environments, potentially degrading lookup performance. Moreover, the absence of a trusted central authority complicates security: \ac{DHT} overlays are vulnerable to man‐in‐the‐middle, Sybil, and routing‐manipulation attacks unless augmented with robust cryptographic or distributed trust mechanisms. These trade‐offs should be carefully managed to harness the scalability and resilience advantages of \ac{DHT}‐based distributed \ac{SRIDS}. A detailed description of architectures based on \ac{DHT} follows.

\begin{figure}[t!]
    \centering
    \includegraphics[scale=0.49]{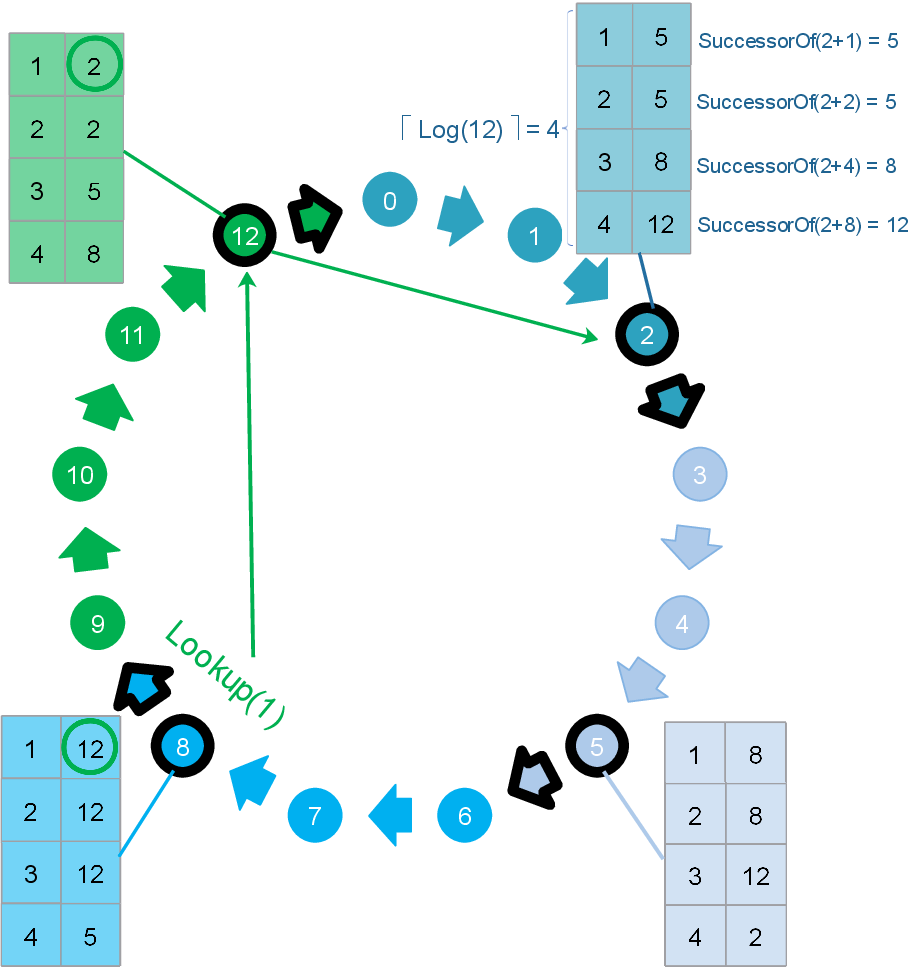}
    \vspace{-4pt}
    \caption{An example of a chord DHT overlay network with 4 nodes and 9 service instances. Nodes in bold are peers in the network, while others are resources or service instances.}
    \label{fig:chord}
    \vspace{-5pt}
\end{figure}

\paragraph{Chord} \hspace{\parindent} 
A Chord‐based distributed \ac{SRIDS} architecture organizes participating nodes into a self‐organizing overlay ring underpinned by a \ac{DHT}. Each node is assigned a unique identifier via a collision‐resistant hash, and each service is similarly mapped to a key in the identifier space. The ring structure, combined with consistent hashing, ensures that responsibility for storing each $(\mathrm{key},\mathrm{value})$ pair is distributed evenly among the nodes \cite{chordLoadBalancing} (Fig.~\ref{fig:chord}). Upon joining the network, a node hashes its \ac{IP} to locate its position in the ring and acquires its successor and predecessor pointers. It then takes over responsibility for the key‐range between its predecessor and itself, migrating the corresponding service entries from the former successor. Every node maintains a routing (finger) table of size $m=\lceil\log_2 N\rceil$, where the $i^\text{th}$ entry point is the successor of $(p + 2^{i-1})\bmod 2^m$ for node $p$. Service registration follows the same hashing procedure: the service's canonical identifier is hashed, and the registration message traverses the overlay, guided by finger‐table entries, until it reaches the node responsible for that key, which then stores the service metadata. Service discovery invokes an analogous lookup workflow. A user computes the hash of the target service \ac{ID} and issues a request to its local Chord node. If the responsible node is not itself, the request is forwarded to the closest preceding node in the finger table, which repeats this process in a decentralized fashion until the lookup converges on the correct node. For instance, in Fig.~\ref{fig:chord}, a lookup for service identifier 1 originating at node 8 is successively routed through nodes 12 and 2 (before arriving at node 5), to the designated successor that holds the desired service entry.  

The Chord overlay confers several benefits for \ac{SRIDS}. Lookups complete in $\mathcal{O}(\log N)$ hops, enabling scalable resolution even with large numbers of nodes and services. The fully decentralized design eliminates single points of failure, providing resilience to node departures or failures without interrupting discovery or indexing operations. Moreover, Chord's stabilization protocol accommodates dynamic membership: when nodes join or leave, routing tables and successor lists are updated to preserve the ring's consistency and service availability. Nevertheless, maintaining Chord's overlay incurs overhead. Periodic stabilization requires $\mathcal{O}((\log N)^2)$ messages to update routing tables and successor lists \cite{lua2005survey}, which is costly in environments with high node churn or mobility. To mitigate this, Chord2 introduces a small set of more stable “bootstrap” peers that serve as persistent index servers \cite{joung2007chord2}. Hybrid‐Chord further enhances performance by layering multiple Chord rings and employing constant‐size successor lists to reduce lookup latency \cite{Flocchini}.

Recent Chord extensions introduce context-aware functional partitioning to enhance the discovery and registry of services in smart city environments \cite{Chord2024}. Services are clustered into functional context networks based on dynamic consumption patterns, and each context is managed via a dedicated, trimmed Chord overlay constructed from only the most central nodes (as determined by closeness centrality) within the context network. This targeted, multi-Chord design enables reductions in both search and communication latency, while preserving traditional Chord rings' logarithmic lookup efficiency. The proposed architecture maintains adaptability by periodically recomputing context structures and routing tables based on temporal service usage, ensuring robustness to node mobility and changing network topologies.

\begin{figure}[t!]
    \centering
    \includegraphics[scale=0.205]{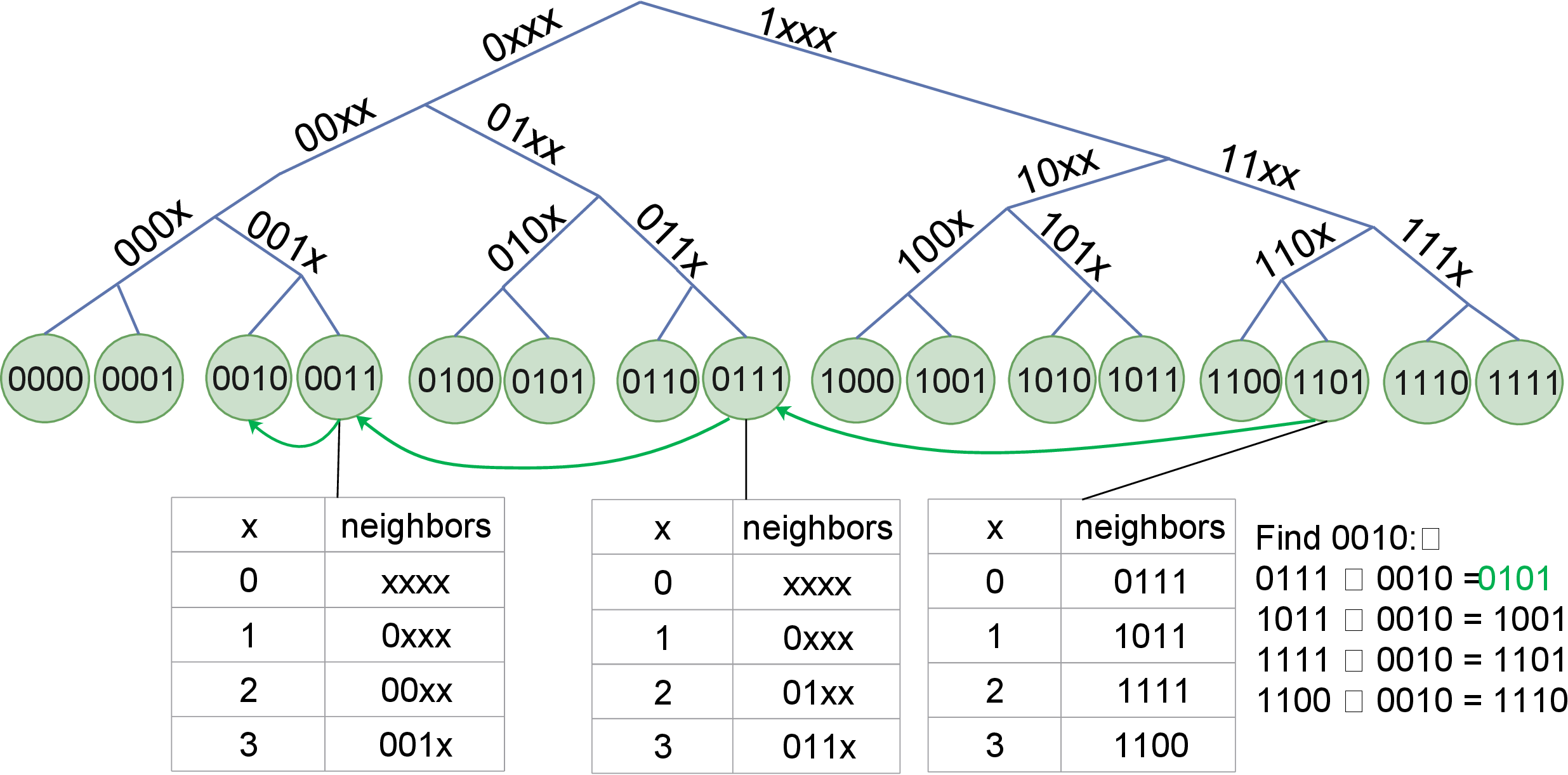}
    \vspace{-13pt}
    \caption{The identifier space and exemplified routing tables in a Kademlia DHT overlay network with $N$ = 16 nodes.}
    \label{fig:kademlia}
    \vspace{-5pt}
\end{figure}

\paragraph{Kademlia} \hspace{\parindent} 
Kademlia is a \ac{DHT}‐based overlay network designed to support distributed key–value storage and lookup in peer-to-peer environments. Widely adopted in large‐scale peer‐to‐peer systems \cite{crosby2007analysis}, Kademlia assigns each peer a random $\lceil\log N\rceil$‐bit Node \ac{ID}, where $N$ denotes the total number of participants (Fig.~\ref{fig:kademlia}). Service or resource identifiers are hashed into the same identifier space, enabling a uniform distribution of $(\mathrm{key}, \mathrm{value})$ pairs across the network. Peers maintain routing tables segmented into $X$‐buckets ($0\le X<\lceil\log N\rceil$), each bucket storing contacts whose Node \acp{ID} share exactly $X$ leading bits with the local Node \ac{ID}. The distance between any two identifiers $a$ and $b$ is computed by the bitwise XOR metric, $d(a,b)=a\oplus b$, which underpins both routing decisions and load balancing. 

Building on this foundation, Kademlia’s lookup workflow proceeds iteratively: to locate a key, a requester computes its hash and queries the peer in its routing table closest (in XOR distance) to that key. If the respondent is not responsible for the key, it returns $k$ of the closest known contacts, and the requester repeats the process with these candidates. This converge‐to‐closest scheme resolves lookups in $\mathcal{O}(\log N)$ message exchanges. Join and leave operations similarly invoke $\mathcal{O}(\log N)$ stabilization messages, during which peers update bucket entries upon receiving \acp{RPC}, thereby maintaining an up‐to‐date routing state. To mitigate lookup hotspots, Kademlia incorporates pair caching: peers temporarily store observed $(\mathrm{key}, \mathrm{value})$ pairs for requests they forward, broadening the set of potential storage nodes and smoothing query load. Persistent bucket refreshes triggered by incoming messages ensure resilience to churn and preserve the routing table's binary‐tree–like structure.  

% \begin{figure}[t!]
%     \centering
%     \includegraphics[scale=0.34]{figures/psychrosm.eps}
%     \vspace{-16pt}
%     \caption{PhysarumSM architecture that uses Kademlia to provide service registration functionality \cite{PhysarumSM}.}
%     \label{fig:physarumSM}
%     \vspace{-5pt}
% \end{figure}

Several \ac{SRIDS} platforms have extended Kademlia to address multi‐attribute queries, edge coverage, and dynamic cloud environments. Khethavath \textit{et al.} \cite{khethavath2013introducing} overlayed a game‐theoretic auction model atop Kademlia to enable simultaneous, attribute‐rich resource requests, assigning resources to the highest bidders in exclusive allocations. Murturi \textit{et al.} \cite{murturi2019edge} deployed additional edge nodes and a local search engine, interconnected via a Kademlia‐based edge‐to‐edge communication module, to expand \ac{SRIDS} coverage for \ac{IoT} resources. PhysarumSM \cite{PhysarumSM} adopted Kademlia within a three‐tier peer-to-peer architecture - comprising resource metrics, telemetry and registry, and control/management layers - to furnish a distributed service registry, optimize proxy caching, and support independent \ac{SRIDS} functionality in dynamic multi‐tier cloud environments.
% (Fig.~\ref{fig:physarumSM}).
To address the limitations of standard Kademlia in adversarial settings, a secure \ac{SRIDS} architecture was proposed wherein Kademlia's operations are fortified with cryptographic primitives to ensure confidentiality, integrity, and authenticity of service metadata during registration and discovery \cite{SKademliaBased2024}. This design preserves Kademlia’s logarithmic lookup efficiency while establishing a secure, fully distributed discovery process suitable for heterogeneous, trustless IoT environments.

\begin{figure}[t!]
    \centering
    \includegraphics[scale=0.217]{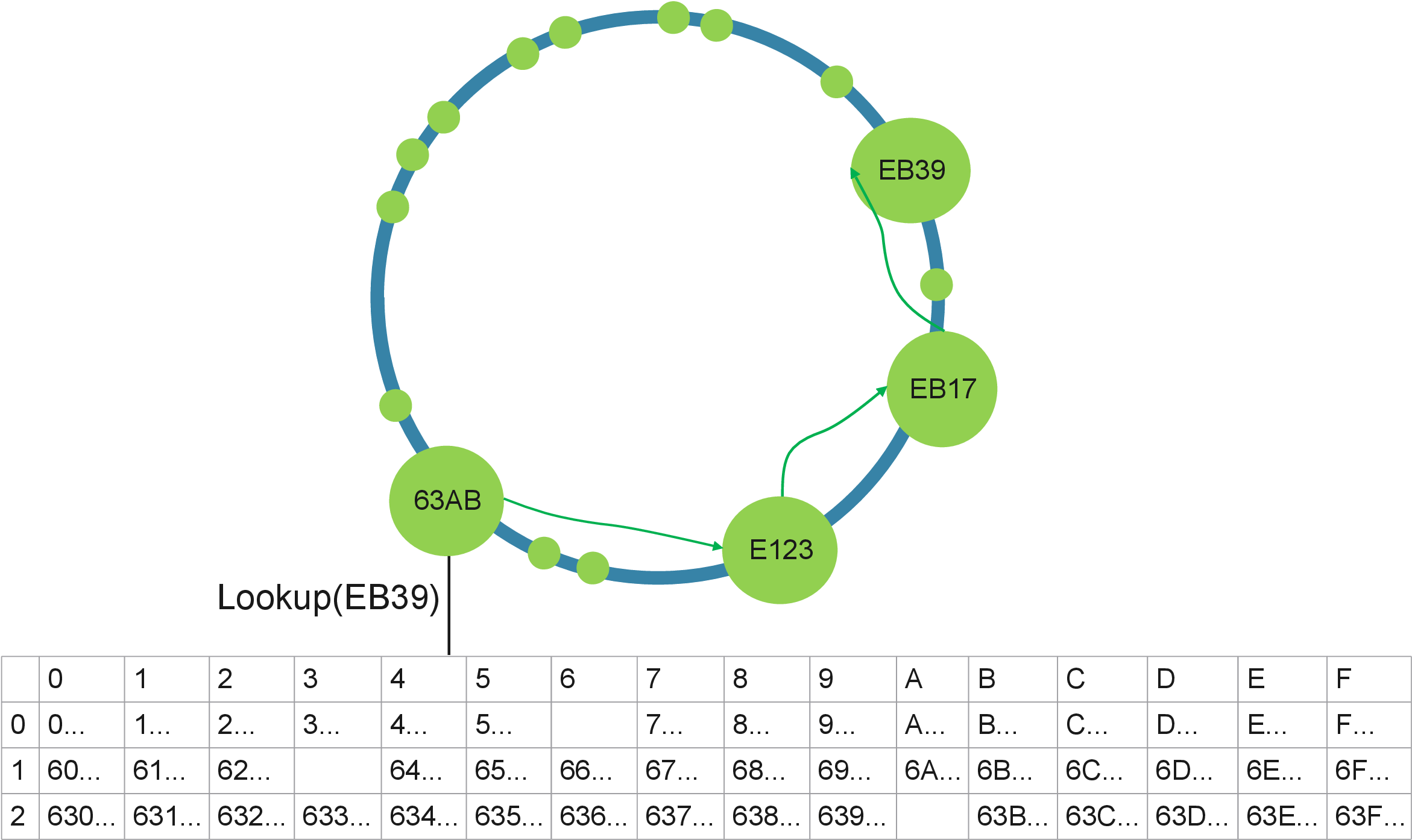}
    \vspace{-10pt}
    \caption{Pastry scheme with $N$ = 10000 and $B$ = 16 \cite{dhtArchitectureSurvey2021}. Green arrows indicate a lookup example from node 63AB for the target identifier EB3E.}
    \label{fig:pastry}
    \vspace{-2pt}
\end{figure}

\paragraph{Pastry} \hspace{\parindent}
Pastry is a \ac{DHT}‐based overlay network designed to support efficient routing, as well as service registration and indexing in large‐scale distributed systems \cite{pastry2001}. Each peer (of $N$ participants) and each resource or service is assigned a randomly generated identifier of $\lceil\log_b N\rceil$ digits in base $b$ (commonly $b=16$), mapping the identifier space onto a self‐organizing ring (Fig.~\ref{fig:pastry}). A Pastry node maintains a routing table of size $(\lceil\log_b N\rceil-1)\times b$, where the entry in row $i$ and column $j$ points to a node whose identifier shares exactly $i$ most‐significant digits with the local node and whose $(i+1)^{\mathrm{th}}$ digit equals $j$. Rows for which no such node exists remain empty. In addition to the routing table, each node keeps a leaf set of the numerically closest peers, which facilitates final‐hop delivery and resilience to node churn.  

Building on this structure, Pastry’s service registration and lookup processes employ prefix‐based routing. To register a service, its canonical identifier is hashed into the Pastry namespace, and the registration message is forwarded at each hop to the neighbor whose identifier shares the longest common prefix with the target key. Once the message reaches a node within the appropriate leaf set, that node stores the $(\mathrm{service ID},\,\mathrm{metadata})$ pair. Service discovery proceeds via an identical prefix‐routing workflow: a requester hashes the desired service name, iteratively routes through the overlay toward the closest prefix match, and upon reaching the responsible node, retrieves the stored metadata (including location, availability, and attributes), which it then uses to establish direct communication with the service instance.  

Pastry’s consistent‐hashing scheme yields a uniform distribution of service entries and query load, while its combination of routing tables and leaf sets ensures $\mathcal{O}(\log_b N)$ routing hops under stable conditions. Fault tolerance is achieved by replicating state across the leaf set: when a node fails, its neighbors detect the absence via periodic probes and redirect requests to alternate nodes responsible for the same key range. The absence of any centralized coordinator endows Pastry with robustness to dynamic joins and departures, as nodes locally update routing and leaf‐set entries to preserve overlay connectivity and routing correctness. 

Pastry has been applied to \ac{SRIDS} architectures to exploit its prefix‐locality and self‐organizing properties. Liew \textit{et al.} proposed a Pastry‐based structured peer-to-peer model wherein service identifiers are hashed and then registered under a fixed‐length prefix of the full key to emphasize proximity in discovery. This prefix‐routing scheme preserves Pastry’s locality guarantees, reducing lookup latency in practice. Service lookup incurs $\mathcal{O}(\log_{B}M)$ hops, where $B$ is the routing‐table base and $M$ is the space of possible keys, and the non-centralized, cooperative architecture supports dynamic, efficient discovery of heterogeneous cloud resources.

\subsubsection{Unstructured Overlay Networks} \hspace{\parindent}   
Unstructured overlay networks form a subclass of distributed \ac{SRIDS} architectures in which peers establish direct, peer‐to‐peer links without enforcing a global, key‐based topology. Each node maintains a local neighbor set, selected according to proximity, resource similarity, or at random, and constructs a virtual overlay atop the underlying physical network. In contrast to structured, \ac{DHT}‐based overlays, unstructured networks do not bind service advertisements or queries to deterministic identifier spaces; instead, they employ mechanisms such as flooding, gossip, or random walks to propagate and locate service metadata.  

In an unstructured overlay network, a service provider registers by broadcasting its advertisement to its immediate neighbors. Those neighbors forward the advertisement according to the chosen diffusion protocol - flooding every neighbor, probabilistically gossiping to a subset, or forwarding along random walks - until a time‐to‐live bound is reached. A user seeking a service issues a query that likewise diffuses through successive neighbor sets until a matching advertisement is discovered or the query expires. This workflow requires no global reorganization when nodes join or depart, thereby simplifying deployment and supporting highly dynamic topologies. Despite these advantages, the lack of deterministic routing in unstructured overlays incurs communication overhead: queries may traverse redundant paths, and periodic neighbor‐set refresh messages are necessary to maintain connectivity and consistency under churn. Under unreliable or rapidly changing network conditions, these overheads can degrade \ac{SRIDS} performance and limit scalability. The following paragraphs examine representative unstructured-overlay-based \ac{SRIDS} architectures, such as epidemic and random‐walk protocols. analyzing their trade‐offs between flexibility, message complexity, and discovery effectiveness. 
% Architectures based on unstructured overlay networks are examined in detail below.

\begin{figure}[t!]
    \centering
    % \vspace{-3pt}
    \includegraphics[scale=0.13]{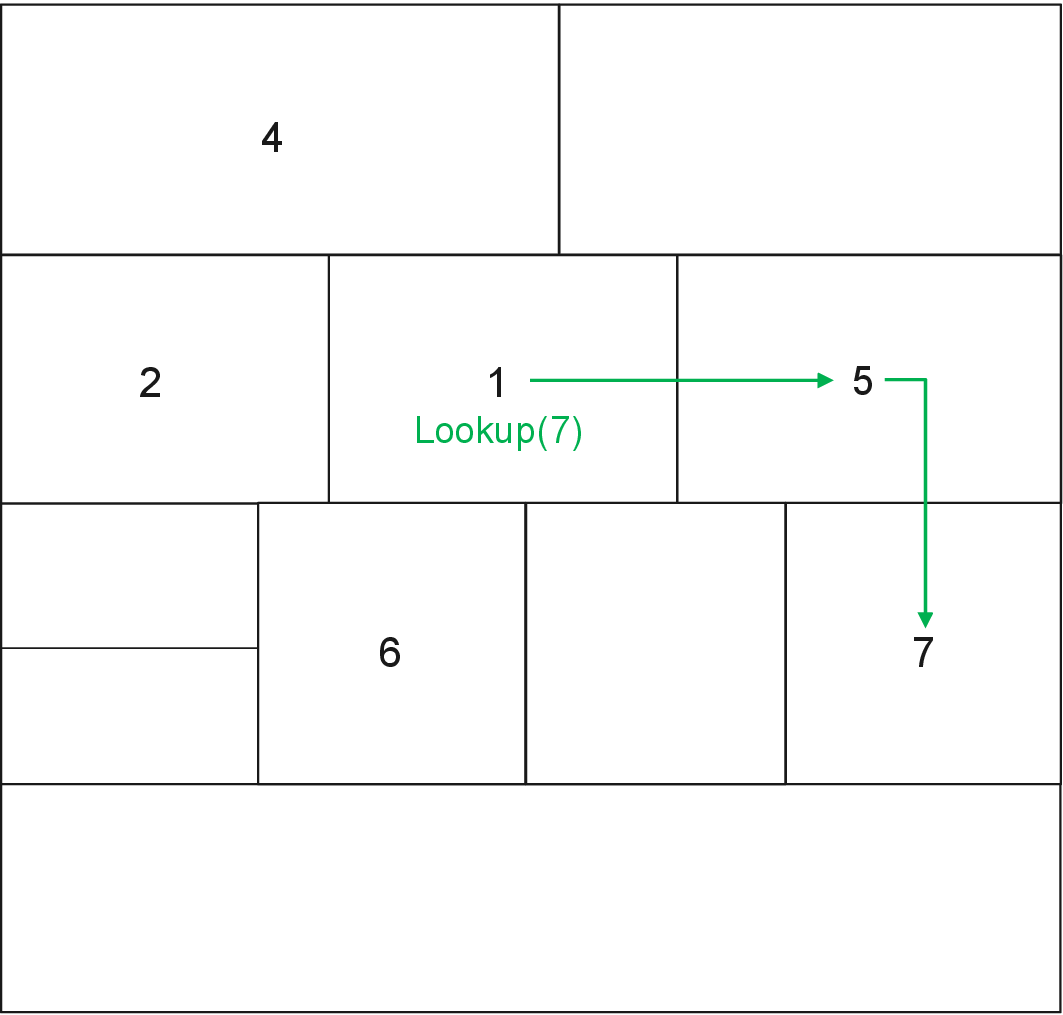}
    % \vspace{-3pt}
    \caption{An example of a 2-dimensional space CAN overlay network \cite{ratnasamy2001scalable}. The coordinate neighbor set of the node with an identifier equal to 1 is \{2,4,5,6\}.}
    \label{fig:can}
    \vspace{-5pt}
\end{figure}

\paragraph{\ac{CAN}} \hspace{\parindent} 
\ac{CAN} is an unstructured overlay network that nonetheless shares certain principles with \ac{DHT}‐based systems \cite{ratnasamy2001scalable}. Rather than organizing peers in a logical ring, \ac{CAN} partitions a $d$‐dimensional Cartesian coordinate space into hyper-rectangular zones, each managed by a super‐node responsible for all keys whose hash values fall within its coordinate range (Fig.~\ref{fig:can}). Each super‐node maintains pointers to its neighboring zones, and keys are mapped uniformly into the space via a collision‐resistant hash function. Routing employs greedy forwarding: upon receiving a lookup request, a node forwards the message to the adjacent supervisor whose zone centroid is closest to the target coordinates, thereby progressively reducing the Euclidean distance to the destination \cite{p2pSurvey2008}. To accommodate churn, zones split when new nodes join and merge when nodes depart, preserving full coverage without overlap. The average lookup cost in \ac{CAN} scales as $O\bigl((d \cdot N)^{1/d}\bigr)$ messages, where $N$ is the total number of nodes, and each join or leave event triggers $2d$ update messages to inform immediate neighbors of zone reassignments. An important extension is the notion of parallel realities: by applying multiple independent hash functions to the same key, \ac{CAN} instantiates several coordinate replicas, thereby enhancing data replication, reducing query latency, and improving fault tolerance at the expense of additional storage and maintenance overhead. \ac{CAN}’s self‐organizing nature ensures that zone assignments and routing tables adapt automatically to membership changes without centralized coordination.  

Osamy \textit{et al.} \cite{ADSDA2019} devised an adaptive distributed \ac{SRIDS} architecture for \ac{IoT} networks built upon \ac{CAN}, characterized by highly dynamic sensor data. Their method first divides the geographical area into hexagonal virtual cells - sized according to node transmission range - and elects the node with the highest residual energy as the cell coordinator. Service registration follows a push model: individual nodes advertise their services to neighbors, and coordinators propagate these advertisements to adjacent cells, transforming the overlay into a distributed directory. Service discovery exploits this directory: a requester issues a query to any neighbor advertising the desired service, thereby obviating the need for targeted searches. When a node moves between cells, it employs a pull mechanism to contact local neighbors, obtain the incumbent coordinator’s identity, and seamlessly integrate into the new cell’s service registry. This three‐layer architecture - perception, discovery, and application - ensures environment‐aware, user‐centric service registration, discovery, and selection without relying on centralized entities.

\begin{figure}[t!]
    \centering
    \includegraphics[scale=0.341]{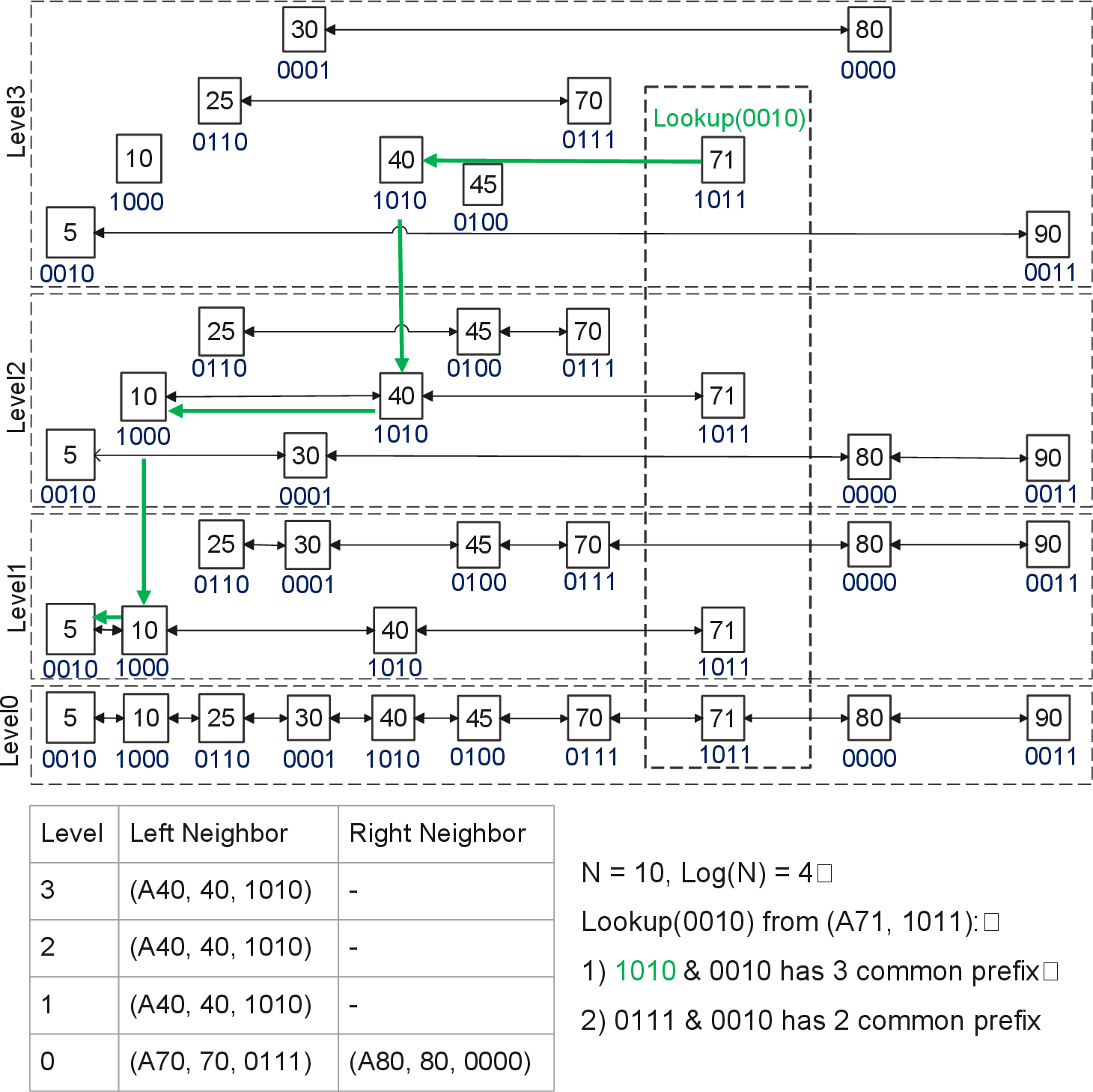}
    \vspace{-1pt}
    \caption{An example Skip Graph overlay network with 10 nodes and 4 levels with exemplified routing table \cite{dhtArchitectureSurvey2021}.}
    \label{fig:skipGraph}
    \vspace{-6pt}
\end{figure}

\paragraph{Skip Graph} \hspace{\parindent}  
Skip Graph is an overlay network architecture devised to support efficient routing for \ac{SRIDS} \cite{skipGraph}. The overlay comprises $N$ nodes organized into $\mathcal{O}(\log N)$ hierarchical levels. Each node possesses a unique non‐negative integer \emph{numerical \ac{ID}} and a binary \emph{name \ac{ID}} of length $\mathcal{O}(\log N)$ bits. At each level, nodes link to exactly one peer sharing the same prefix of the name \ac{ID} and to their immediate numerical predecessor and successor. Accordingly, every node maintains a lookup table with one row per level and two columns indicating its left and right neighbors in numerical ordering. This multi‐level structure bounds overlay diameter and underpins logarithmic‐scale service lookup.  

Routing in the Skip Graph proceeds via two complementary workflows. In numerical‐ID–based navigation, a query bearing the target numerical \ac{ID} begins at the highest level: if the target \ac{ID} is smaller than the local node’s, the packet is forwarded to the left neighbor; if larger, to the right neighbor. When no close neighbor exists at the current level, the algorithm descends one level and repeats, ensuring convergence to the target in $\mathcal{O}(\log N)$ hops (Fig.~\ref{fig:skipGraph}, horizontal arrows). For name‐ID–based lookups, each hop selects the neighbor whose name \ac{ID} maximizes the length of the common prefix with the target, thereby pruning the search space and accelerating the \ac{SRIDS} process. Vertical arrows in Fig.~\ref{fig:skipGraph} illustrate internal prefix‐matching computations during traversal.  

To enhance the Skip Graph’s resilience under dynamic membership, recent works introduce predictive stabilization and churn‐modeling techniques. Interlaced \cite{Interlaced2021} is a fully distributed algorithm that forecasts node departures and proactively repairs neighbor pointers to preserve connectivity and bound query latency, all without increasing message complexity. Complementarily, the sliding window De Bruijn graph is employed as an adaptive learning mechanism to estimate per‐node availability probabilities, tuning stabilization parameters to observed churn behaviors, and thereby improving overall stability for robust \ac{SRIDS} performance.

\begin{figure}[t!]
    \centering
    \includegraphics[scale=0.38]{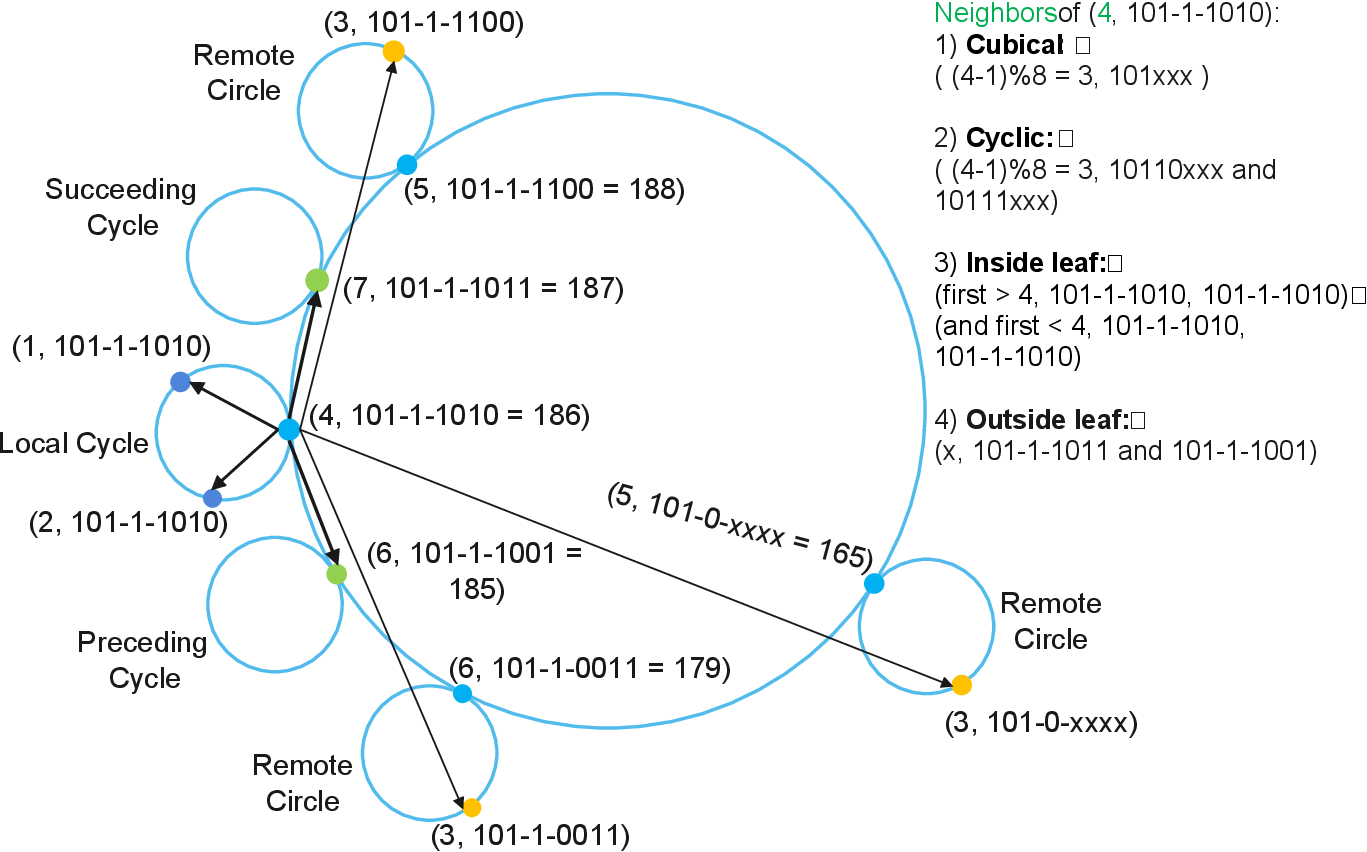}
    \vspace{-14pt}
    \caption{An example of a 8-dimensional Cycloid overlay network \cite{dhtArchitectureSurvey2021}.}
    \label{fig:cycloid}
    \vspace{-5pt}
\end{figure}

\paragraph{Cycloid} \hspace{\parindent}  
Cycloid is an overlay architecture for \ac{SRIDS} based on the constant‐degree \ac{DHT} known as $d$‐dimensional cube‐connected cycles \cite{cycloid}. It embeds two cyclic rings into a hypercube structure: each vertex of a $d$‐dimensional cube is replaced by a cycle of $d$ nodes, yielding up to $N \!\!=\!\! d \! \cdot \! 2^d$ participants (Fig.~\ref{fig:cycloid}). This design achieves routing in $\mathcal{O}(d)$ hops while each node retains only $\mathcal{O}(1)$ neighbors. Each Cycloid node is identified by a pair $(k,a_{d-1}a_{d-2}\dots a_0)$, where the \emph{cyclic index} $k \! \in \! \{0,\dots,d-1\}$ locates it within its local cycle and the \emph{cubical index} $\!(a_{d-1}\dots a_0)\!$ denotes its hypercube vertex. The node with maximum $k$ in each cycle serves as the \emph{primary node}, representing that cycle in the main ring. Key assignment parallels Pastry’s approach: given $\mathrm{hash}(key)$, the cyclic index is $\mathrm{hash}(key)\bmod d$ and the cubical index is $\lfloor\mathrm{hash}(key)/d\rfloor$. Data are stored in the cycle whose cubical index shares the longest common prefix with the key’s index, ensuring placement in the closest peer in the hypercube metric. Dynamic joins and departures, however, complicate the maintenance of balanced key distribution and fully connected overlays.

Cycloid's neighbor table comprises seven entries: two \emph{cubical neighbors}, two \emph{cyclic neighbors}, two \emph{inside leaf neighbors}, and one \emph{outside leaf neighbor}, with proposals to augment leaf sets to four nodes each (two predecessors and successors), raising the total to eleven neighbors for enhanced resilience. For a node $(k,a_{d-1}\dots a_0)$, a cubical neighbor shares cyclic index $(k-1)\bmod d$ and maximizes common bits in the cubical index; cyclic neighbors lie adjacent in numerical order of cubical indices with matching bit‐prefix length and cyclic index $(k-1)\bmod d$; inside leaf neighbors are the immediate predecessors and successors on the local cycle; outside leaf neighbors connect to the adjacent cycles in the hypercube embedding.  

Routing is governed by the Most Significant Different Bit (MSDB) between source and destination cubical indices. Upon receiving a lookup, a node compares its $k$ to the destination’s MSDB. If $k<\mathrm{MSDB}$, the request is forwarded to an outside leaf neighbor; if $k=\mathrm{MSDB}$ and the destination resides in the outside leaf set, it is delivered directly; otherwise, the packet is routed to the cubical neighbor. If cubical forwarding is not feasible, the node selects the nearest of its cyclic neighbors or an inside leaf neighbor. This multi‐dimensional greedy forwarding ensures steady convergence toward the target with a constant-degree state and logarithmic‐scale traversal, supporting efficient distributed lookup and data placement in \ac{SRIDS} systems.

\paragraph{Kinaara} \hspace{\parindent}  
Kinaara is a ring‐like, distributed multi‐tier overlay architecture tailored to service indexing in unstructured \ac{SRIDS} environments \cite{Kinaara2017}. Conceptually akin to Cycloid, Kinaara interconnects edge and cloud resources via a dual‐layer ring of mediators, while preserving flexibility in resource discovery and allocation across heterogeneous clusters. The architecture comprises three logical entities: the cloud server, a set of mediator nodes, and end users. Mediators organize themselves into proximal clusters, each covering resources in physical or logical proximity, and dynamically elect a primary mediator per cluster. Primary mediators maintain bidirectional channels both to the central cloud service and to adjacent mediators in the ring. Within each cluster, individual resources are assigned local keys for intra‐cluster indexing, whereas mediators themselves each hold a unique global key that positions them in the overarching mediator ring. 

Kinaara’s indexing workflow unfolds as follows: When a new resource joins, the cluster’s primary mediator consults a similarity table - constructed via a chosen metric such as Hamming distance - to identify the predecessor and successor nodes between which the newcomer should be inserted. These relationships form an unstructured neighbor set that adapts dynamically to resource attributes rather than to a rigid identifier space. Resource departure is handled by active failure detection: neighboring nodes detect timeouts or heartbeat lapses, promptly notify the mediator, and trigger local reconfiguration to preserve connectivity. Service lookup in Kinaara is mediated by the primary mediator. Upon receiving a request, the mediator evaluates service attributes, such as priority and diversity criteria, and initiates a breadth‐first search across both its local cluster and across other clusters via the mediator ring. At each step, candidate nodes are probed for instance availability; once a suitable service is located, the mediator reserves the instance by setting an on‐hold timeout before returning its endpoint to the requester. This breadth-first-based search eschews global hashing in favor of attribute‐guided expansion, enabling discovery across loosely structured overlays. To support mobility and churn, Kinaara employs heartbeat protocols and network‐level timeouts that inform applications of potential disconnections. While these mechanisms detect lost links and trigger mediator‐led reorganization, they do not provide immediate localization of moving resources, which remains an area for further enhancement in highly dynamic edge environments.

\paragraph{HandFan} \hspace{\parindent}  
HandFan is an unstructured overlay \ac{SRIDS} architecture proposed as a flexible, scalable alternative to \ac{DHT}‐based systems in resource‐constrained \ac{IoT} environments \cite{Handfan2022}. It replaces ring or hypercube topologies with a geometric embedding: each participating node or resource key is mapped to a vertex of a convex regular $N$‐gon inscribed in a common circumcircle. This $N$‐gon representation yields a uniform angular distribution of identifiers, thereby facilitating load balancing among heterogeneous nodes regardless of their capacity or degree. The mapping scheme first hashes each node or service identifier to an integer in $[0,N-1]$, then assigns it to the corresponding $N$‐gon vertex. Each node computes its polar angle $\theta$ on the circumcircle and populates a finger table of size $m+1$, where $m=\lceil\log N\rceil$. Table entries are ordered by increasing angle and contain pairs $(v_i,\theta_i)$ denoting the vertex identifier and its angular coordinate. 

Routing in HandFan employs a deterministic, angle‐based greedy forwarding mechanism. To locate a target node~$\beta$, the source node~$\alpha$ compares $\beta$’s polar angle $\theta_\beta$ against its own interval $[\theta_\alpha,\theta_{\mathrm{succ}(\alpha)}]$. If $\theta_\beta$ falls within this range, $\alpha$ returns its successor’s address. Otherwise, $\alpha$ selects from its finger table the entry whose angle $\theta_i$ is the largest value less than or equal to $\theta_\beta$ and forwards the lookup to that node. Repeating this process converges to $\beta$ in $\mathcal{O}(\log N)$ hops, achieving efficient service discovery without centralized coordination.

\subsubsection{Distributed Data Storage} \hspace{\parindent} 
Distributed data storage architectures for \ac{SRIDS} distribute service metadata management by dispersing information across multiple storage nodes to enhance fault tolerance and reliability. In these architectures, the global set of service descriptions and index entries is partitioned into smaller units - commonly referred to as shards, partitions, or zones - each of which is assigned to one or more storage nodes. Partitioning may employ consistent‐hashing schemes, which map service identifiers uniformly onto node responsibilities, or range‐based partitioning, which allocates contiguous identifier intervals to ensure balanced load. By directing registration and lookup operations to the node(s) holding the relevant partition, this approach achieves scalable distribution of service data and supports parallel query processing.

Building on partitioning, distributed storage relies on replication to guarantee availability under node failures or network partitions. A straightforward replication strategy maintains a fixed number of replicas per partition, whereas adaptive protocols modulate the replication factor in response to observed node churn or workload intensity. When a primary node becomes unreachable, secondary replicas assume responsibility for serving registration and discovery requests, thereby preventing data loss and service interruption. Nonetheless, replication entails the challenge of preserving consistency and propagating updates across dispersed replicas. \red{Various consistency models are available, ranging from consensus-based protocols (that require replicas' agreement before acknowledging writes) to eventual consistency schemes (that provide for temporary divergence but eventually converge asynchronously)}. Coordination among replicas introduces additional messaging overhead and can increase operation latency, necessitating careful selection of replication and consistency mechanisms to balance system performance, availability, and complexity. Subsequent paragraphs examine representative \ac{SRIDS} implementations that instantiate these partitioning, replication, and consistency principles. 
% The subsequent subsection presents a representative architecture that employs distributed data storage.

\paragraph{\ac{IOTA}} \hspace{\parindent} 
\ac{IOTA} is a distributed‐ledger–based \ac{SRIDS} architecture that employs its native cryptocurrency and a \ac{DAG} data structure (known as the Tangle) to decentralize service registration, indexing, and discovery across all participating nodes \cite{iota}. By replicating the Tangle on every \ac{IOTA} node, the architecture inherently disperses service metadata, ensuring fault tolerance, redundancy, and tamper‐evident storage without relying on a central directory. Early blockchain‐inspired approaches - such as the hello‐message scan of Daza \textit{et al.} \cite{Daza2017}, the gossip‐based semantic multicast of Ruta \textit{et al.} \cite{Ruta2017}, and the API‐driven configuration discovery of Manevich \textit{et al.} \cite{Manevich208} - demonstrate the value of distributed ledgers for secure \ac{SRIDS}. \ac{IOTA} advances these concepts by replacing linear chains with the Tangle \ac{DAG}, in which each new transaction endorses two previous transactions, thereby concurrently validating and attaching service registrations to the global ledger.  

In \ac{IOTA}’s workflow, a service provider first creates a Tangle transaction embedding its unique identifier, location, service description, performance metrics, and access credentials. Before appending this registration to the \ac{DAG}, the node issuing the transaction performs a lightweight proof‐of‐work to validate two unconfirmed “tip” transactions, propagating consensus without mining pools. Once confirmed by subsequent arrivals, the registration becomes an immutable ledger entry accessible to all peers. When an \ac{IoT} device seeks service instances, it composes a query transaction specifying functional or contextual requirements and similarly validates two tips via proof‐of‐work. This query transaction disseminates through the network, triggering participating nodes to respond with transactions that reference matching service registrations. The requesting device collects these responses and selects the optimum instance based on proximity or quality‐of‐service parameters before establishing a direct connection. Upon service termination, the device may issue a finalization transaction or rely on the natural expiry of masked authenticated messaging channels to retire the association.  

\begin{figure}[t!]
    \centering
    \includegraphics[scale=0.63]{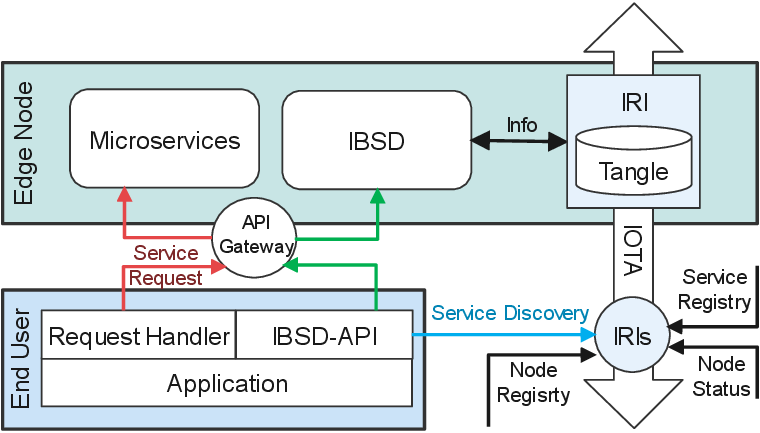}
    \vspace{-5pt}
    \caption{The architecture of IOTA-Based services discovery \cite{IOTAelectronics2021}.}
    \label{fig:iotaIBS}
    \vspace{-3pt}
\end{figure}

Tang \textit{et al.} \cite{IOTAelectronics2021} extended \ac{IOTA}’s basic \ac{SRIDS} capabilities with the \ac{IBSD} framework (Fig.~\ref{fig:iotaIBS}). The proposed architecture integrates the \ac{IRI} daemon, an \ac{IBSD} indexing module, and a microservice layer on edge nodes. \ac{IBSD} distributes and indexes service metadata via \ac{IRI}, exposes RESTful \acp{API} for query execution, and leverages masked messaging transaction chains to publish dynamic updates (e.g., identification codes). By maintaining an up‐to‐date list of edge‐hosted service instances, \ac{IBSD} reduces search ranges within the Tangle, accelerates discovery, and enhances resilience against network attacks. This distributed indexing mechanism thus preserves \ac{IOTA}’s decentralization while achieving efficient, secure service discovery in large‐scale \ac{IoT} environments.

%% file: sections/sec4/ssec4_3.tex
\subsection{Decentralized}\label{ss_decentralized}
Non‐centralized \ac{SRIDS} architectures encompass both distributed and decentralized paradigms. In distributed approaches, essential functions are allocated among subsets of nodes according to predefined overlay or partitioning schemes, yet overall coordination persists via structured protocols or partial hierarchies. Decentralized architectures extend this concept by conferring full \ac{SRIDS} capability on every participant, thereby dispensing with any specialized or coordinating entity and enabling true peer‐to‐peer interactions. In a decentralized \ac{SRIDS}, each node independently maintains its own service repository, advertises its offered services to neighboring peers, and processes incoming discovery queries without recourse to a central or specialist directory. Service announcements propagate through gossip or flooding mechanisms, while lookup requests traverse multiple peer links until matching entries are found. At each hop, nodes consult local indices and may respond directly or forward the query further, allowing the requester to aggregate responses and select the optimal service instance based on proximity, performance metrics, or cost criteria.

The foremost advantage of decentralization is its elevated fault tolerance and reliability. By replicating both state and control across all participants, the architecture avoids single points of failure: individual node outages or compromises cannot incapacitate the global discovery process. Autonomy at each peer also simplifies scaling, since newcomers require no centralized registration; they immediately contribute to and benefit from the collaborative maintenance of the service catalog. This egalitarian model balances workloads and enhances system robustness and availability. Nevertheless, decentralization introduces significant challenges. Maintaining consistency and synchronization across independently evolving local indices demands sophisticated update and convergence protocols to prevent stale or conflicting service views. Moreover, reliance on multi‐hop, peer‐based communication for both advertisement and lookup generates increased traffic and may incur higher latency compared to structured or partially centralized distributed architectures. These trade‐offs between autonomy and communication overhead must be carefully managed when deploying decentralized \ac{SRIDS} in resource‐constrained or latency‐sensitive environments.

\subsubsection{Unstructured Networks} \hspace{\parindent}
Unstructured networks constitute a specialized class of decentralized \ac{SRIDS} architectures in which each node assumes both service provider and user roles, thereby eliminating any central coordinator. In a typical unstructured network workflow, a peer registers its available services by updating a local index and propagates this metadata to neighboring nodes through gossip or controlled flooding. Discovery requests are issued as lookup messages that traverse successive peer links until one or more nodes possessing matching entries respond, after which the originator aggregates replies and selects an appropriate service instance.

The foremost advantage of unstructured peer-to-peer \ac{SRIDS} lies in its self-organizing, egalitarian design: new participants immediately expand the \ac{SRIDS}’s storage and forwarding capacity, yielding near-linear scalability and improved load balancing. By replicating metadata across multiple peers, the architecture enhances fault tolerance and maximizes resource utilization without reliance on centralized directories. Nevertheless, the absence of a trusted authority introduces significant security and privacy concerns. Peers should establish mutual trust, often via cryptographic authentication or reputation systems, to prevent malicious registrations, Sybil attacks, and data tampering. Furthermore, unstructured or loosely structured overlays can incur high network overhead if naive flooding is used; therefore, efficient routing algorithms are essential to constrain lookup latency and control-plane traffic. In what follows, further discussion about architectures utilizing this approach can be found.

\paragraph{Gnutella} \hspace{\parindent}  
Gnutella \cite{kan2002gnutella} exemplifies an unstructured, decentralized \ac{SRIDS} overlay in which each peer, identified by a unique key, maintains knowledge of its immediate neighbors without recourse to any central directory. Upon joining the overlay, a new node locates an active Gnutella participant (the bootstrap peer) to obtain initial neighbor information and integrates into the peer graph. Service discovery relies on controlled flooding: a peer originates a query message that is broadcast to all adjacent nodes; each recipient inspects its local service registry and, if no matching entry is found, forwards the query to its own neighbors. This propagation continues until either a provider is reached or the query’s hop limit expires. While flooding ensures broad coverage in highly dynamic peer‐to‐peer environments, it incurs substantial bandwidth overhead, redundant message transmissions, and an elevated risk of peer churn due to network saturation. To mitigate these inefficiencies, the Gia framework \cite{chawathe2003making} replaces blind flooding with a stochastic search mechanism: peers perform random walks of limited length, directing queries toward subsets of neighbors to locate services in proximity rather than across the entire overlay. By bounding the number of parallel walks, Gia reduces control‐plane traffic and accelerates discovery in densely connected regions.  

Building upon Gia’s principles, Gnutella2 \cite{stokes2003gnutella2} introduces a two‐tier architecture of leaves and hubs. Leaves function as ordinary peers, each maintaining a persistent connection to a designated hub. Hubs aggregate and redistribute service metadata by exchanging hashed keyword summaries (known as query routing tables) among themselves. Upon issuing a lookup, a leaf submits its query to its hub; if the hub’s index contains relevant entries, it returns the address of a suitable provider; otherwise, it forwards the query to other hubs in an iterative fashion. This super‐peer model concentrates routing state within hubs, thereby curtailing overall message complexity and alleviating leaf‐level workloads. Nevertheless, the reliance on hub availability introduces new vulnerability vectors: excessive query loads on hubs can degrade performance, and the compromise or failure of a hub may disrupt service access for all attached leaves, potentially facilitating denial‐of‐service attacks.

Manolito \cite{manolotio}, another Gnutella‐derived design, substitutes \ac{UDP} for \ac{TCP} to streamline query and response exchanges and integrates end‐to‐end encryption to bolster peer anonymity \cite{p2pSurvey2008}. New participants contact an \ac{HTTP}‐based network gateway to retrieve a candidate list of peers, from which they select connections based on measured bandwidth capacities. This approach preserves the fully decentralized ethos of Gnutella while enhancing privacy and reducing transport‐layer overhead, albeit at the cost of more complex peer selection logic and the absence of reliable transport guarantees inherent in \ac{TCP}.

\paragraph{Freenet} \hspace{\parindent}  
Freenet \cite{clarke2001freenet} is an unstructured overlay designed to realize secure and anonymous \ac{SRIDS} by distributing encrypted service descriptors across a network of equal‐capability peers. Each node contributes local storage and bandwidth, and service metadata is addressed via content‐based identifiers computed as the \ac{SHA}-1 hash of the descriptor. This cryptographic naming both guarantees uniqueness and conceals semantic content, thereby preserving confidentiality during discovery. The Freenet lookup workflow proceeds in two phases. Initially, a requesting node computes the target service’s \ac{SHA}‐1 identifier and issues a query that each intermediate peer forwards to the neighbor whose identifier lies closest to the target in the overlay’s identifier space, a steep-ascent hill–climbing strategy guided by each node’s routing table. When a query encounters a dead-end or revisits a prior node, Freenet invokes a backtracking procedure to explore alternate neighbor paths, preventing infinite loops without centralized control. Once the node storing the requested descriptor is reached, the service information propagates back along the reverse path to the originator, and each hop caches the descriptor locally. Under storage pressure, nodes employ probabilistic eviction of the oldest or least‐accessed entries to manage limited cache space.  

By replicating service descriptions along successful query routes and fragmenting knowledge of descriptor locations, Freenet obviates single points of failure, enhances data availability, and delivers strong anonymity and censorship resistance. However, its unstructured nature entails logarithmically increasing lookup latencies and substantial message overhead as the overlay scales. This limits performance and responsiveness relative to structured or hybrid decentralized \ac{SRIDS} architectures.

\begin{figure}[t!]
    \centering
    \includegraphics[scale=0.282]{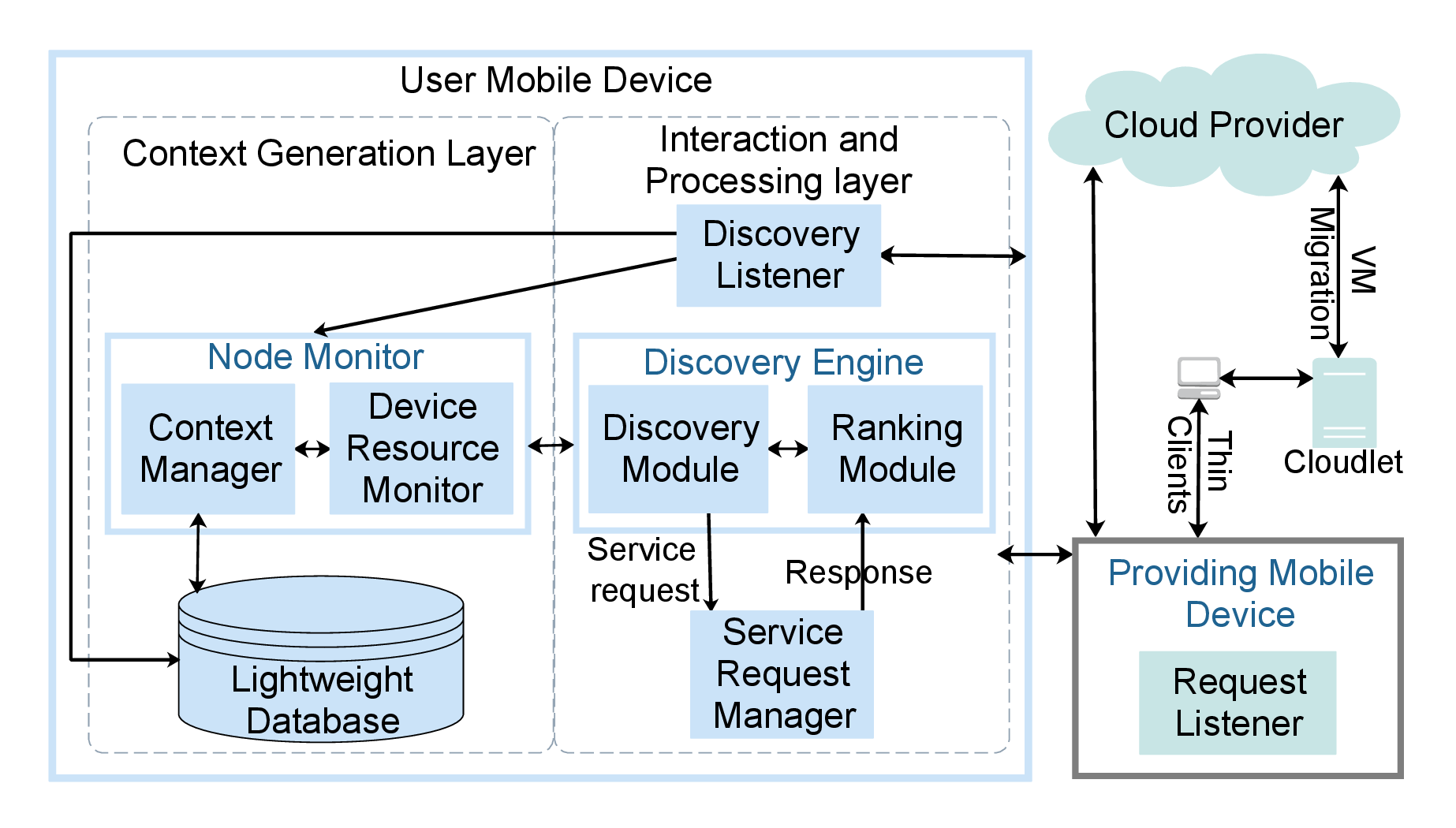}
    \vspace{-18pt}
    \caption{Resource-aware SD-AMC's SRIDS architecture \cite{sdamc}.}
    \label{fig:sdamc}
    % \vspace{-6pt}
\end{figure}

\paragraph{\ac{SD‐AMC}} \hspace{\parindent}  
\ac{SD‐AMC} \cite{sdamc} is a fully decentralized, resource‐aware \ac{SRIDS} architecture tailored to dynamic ad‐hoc mobile cloud environments. It leverages peer‐to‐peer interactions among mobile nodes, each acting simultaneously as a service requester and host, to adaptively discover and provision services under stringent resource constraints. The architecture is organized into two functional strata: a \emph{context‐generation} layer and an \emph{interaction‐and‐processing} layer (Fig.~\ref{fig:sdamc}). Within the former, the node monitor continuously samples non‐static device metrics (battery level, available memory, storage capacity) and exposes this information via a lightweight database. A subordinate context manager within the node monitor organizes device‐specific preferences to guide discovery decisions. The interaction layer employs two mediating modules: the service request manager mediates all user‐to‐device query exchanges, while the request listener orchestrates inter‐peer request forwarding and response retrieval by consulting the local database of hosted services. Core discovery logic resides in the discovery engine, which applies a lightweight, keyword‐based matching and ranking algorithm to identify and prioritize candidate service providers based on both semantic relevance and current resource availability.  

Although \ac{SD‐AMC} defines a comprehensive set of modules and peer‐to‐peer workflows for adaptive service matching, implementation details and empirical validation remain sparse. A concrete usage scenario or prototype evaluation, demonstrating end‐to‐end discovery latency, ranking accuracy, and resilience under node mobility, would substantiate the architecture’s efficacy in real‐world deployments.

\subsubsection{Structured Networks} \hspace{\parindent}  
Structured networks constitute virtual communication fabrics that sit atop the physical infrastructure to enable efficient \ac{SRIDS} in dynamic, collaborative, and non‐centralized environments. These networks adopt structured topologies (like a tree) to optimize routing performance and information dissemination. In distributed \ac{SRIDS} architectures, an overlay network forges predetermined logical links among participating nodes, forming a virtual layer above the underlying network. A decentralized structured network overlay is a virtual layer while keeping all peers on an equal footing. It underpins the cooperative propagation of service metadata, the forwarding of discovery queries, and the shared maintenance of the service catalog without reliance on any single directory.

Extending this concept, fully decentralized \ac{SRIDS} architectures exploit structured networks to achieve autonomous operation: each node independently preserves information about its immediate neighbors and executes local protocols to publish service advertisements, relay lookup requests, and collect responses. By harnessing locally maintained neighbor relationships and decision logic at each node, the system achieves resilient, adaptive routing even under high churn. Through the combination of flexible topology design and per‐node autonomy, structured overlay‐based approaches strike a balance between scalability, robustness, and adaptability. The subsequent paragraphs examine representative structured networks and delineate their specific workflows within this decentralized \ac{SRIDS} framework.

\paragraph{Tree‐based} \hspace{\parindent}  
Tree‐based overlays implement decentralized \ac{SRIDS} by embedding service and node identifiers within a prefix‐organized routing tree. In Tapestry \cite{Tapestry}, each node is assigned a unique node \ac{ID} in a base‐$B$ identifier space, and each service descriptor is hashed to a corresponding key. The common‐prefix ordering of node \acp{ID} and keys defines the logical tree structure that underlies all routing and indexing operations. Service registration in Tapestry begins when a provider computes the \ac{SHA}‐based key of its service description and issues a registration message to the node whose node \ac{ID} shares the longest common prefix with that key. At each hop, intermediate peers update the routing‐table entry corresponding to the current prefix length, thereby propagating the service index along the appropriate tree branch. Service discovery employs an analogous prefix‐routing workflow: a user hashes its query to derive the service key and forwards a lookup request through successive hops, each selecting the next peer whose node \ac{ID} extends the matched prefix by one digit. Upon reaching the responsible node, the lookup returns the service endpoint, and intermediate nodes may optionally cache the descriptor to accelerate future queries \cite{zhao2004tapestry}.  

The prefix‐tree topology guarantees lookup and maintenance costs of $\mathcal{O}(\log_{B}N)$ messages, where $N$ denotes the number of nodes and $B$ is the identifier base. Redundant routing‐table entries and proximity‐aware neighbor selection confer resilience to churn and single‐node failures. However, this structured approach introduces challenges in balancing the key space across branches, handling simultaneous failures that can sever multiple tree paths, and accommodating the route‐population delays experienced by newly joining nodes.

\paragraph{Social-based} \hspace{\parindent}
Social-based networks support decentralized \ac{SRIDS} by leveraging social interaction histories, or social proximity among nodes, to guide service discovery \cite{Social-based}. Instead of relying on strict structural placement like tree- or ring-based overlays, these approaches construct a logical graph of overlay nodes by relying on prior service exchanges, social affinity, or collaboration frequency. In this paradigm, each node autonomously registers its services, forwards lookup queries through its social links, and optionally propagates service advertisements to its trusted neighbors. Service discovery follows a two-tier process: initially attempting social-based resolution through friends and extended friends, and falling back on the network's name-based routing if the local overlay cannot satisfy the query. This hybrid resolution flow enhances both efficiency and adaptability by fusing locality with broader name-based reachability.

Social-based networks excel in dynamic and trust-sensitive environments, where rigid topologies are difficult to maintain or insufficiently expressive. Their decentralized nature confers resilience and fault tolerance, while localized decision-making reduces coordination overhead. However, service reachability is influenced by social graph density and node willingness to relay queries, which may lead to inconsistent discovery performance if trust links are sparse or unevenly distributed. Social-based networks introduce a semantically rich, interaction-driven alternative to structure-heavy overlay protocols, striking a pragmatic balance between scalability and contextual service relevance.

%% file: sections/sec4/ssec4_4.tex
\subsection{Hybrid}\label{ss_hybrid}
Hybrid \ac{SRIDS} architectures build upon the principles of non-centralized service provisioning by retaining the peer‐centric workload distribution and resilience inherent to directory-less systems while selectively reintroducing centralized modules where strict control or global consistency is indispensable. \red{Such designs use non-centralized overlays or peer-equivalent nodes to perform the SRIDS tasks: service descriptors propagate via gossip or controlled flooding, lookup requests traverse multi-hop overlays, and each participant maintains local indexes independently and participates equally in query routing}. Concurrently, coordination functions - such as global policy enforcement, metadata reconciliation, or schema governance - are delegated to a central module or hierarchy of controllers. Having this duality enhances the scalability and fault tolerance gains of fully distributed \ac{SRIDS}, yet benefits from a single locus of control for cross‐domain consistency, conflict resolution, and administrative oversight. 

The principal advantage of this hybrid approach lies in its ability to reconcile two previously opposing objectives: scalability and consistency. By decentralizing high-volume, latency-sensitive operations (e.g, service advertisement diffusion and query propagation), the system sustains throughput under heavy churn and avoids single-point bottlenecks. At the same time, centralized coordination modules enforce end-to-end policies, manage global registration conflicts, and reconcile versioning of service schemas, thereby preventing divergence from distributed indices. This fusion yields an adaptable design that accommodates dynamic resource fluctuations and heterogeneous trust domains, while guaranteeing coherent, system-wide governance and rapid convergence of states following topology changes. Nevertheless, hybrid \ac{SRIDS} architectures introduce distinct complexities. The coexistence of non-centralized overlays and centralized controllers gives rise to coordination overhead, as metadata must cross domain boundaries and be transformed between non-centralized formats and a unified central schema. Ensuring low‐latency synchronization without compromising fault isolation demands sophisticated consistency protocols, which can themselves become performance bottlenecks under extreme dynamics. Moreover, the partitioning of responsibilities requires careful trust management: centralized modules must authenticate and authorize updates from numerous non-centralized peers, and any compromise within this control plane may undermine the integrity of the entire \ac{SRIDS} fabric.

\subsubsection{Centralized–Decentralized Combination} \hspace{\parindent}  
The hybrid centralized–decentralized \ac{SRIDS} paradigm synergistically combines a central authority, responsible for service registry, global indexing, and policy coordination, with a peer‐to‐peer substrate that disseminates and caches service metadata. In this architecture, service providers first submit their descriptions and credentials to a central registry, which validates entries, enforces access controls, and assigns unique identifiers. Subsequently, the registry publishes indexed metadata to participating peers or super‐peer nodes, where decentralized protocols take over to propagate updates, resolve local queries, and maintain per‐node indexes without constantly consulting the central node.

In operation, service registration and indexing proceed via the following workflow. Upon receiving a new descriptor, the central module verifies its authenticity, records it in the master index, and pushes incremental updates (tagged with context or intent) to a designated set of overlay nodes. These nodes employ controlled flooding or gossip to distribute metadata across the overlay, ensuring multiple replicas and reducing lookup latency. When a user issues a discovery request, it first consults its nearest peer cache; if the local index cannot respond or lacks the required entry, the query escalates to the central coordinator, which returns the authoritative result. This dual‐path resolution guarantees both rapid local responses under normal conditions and consistent global answers when strict freshness or policy compliance is required. Subsequent paragraphs examine the concrete instantiations of this architecture and analyze their respective details.

\begin{figure}[t!]
    \centering
    \includegraphics[scale=0.235]{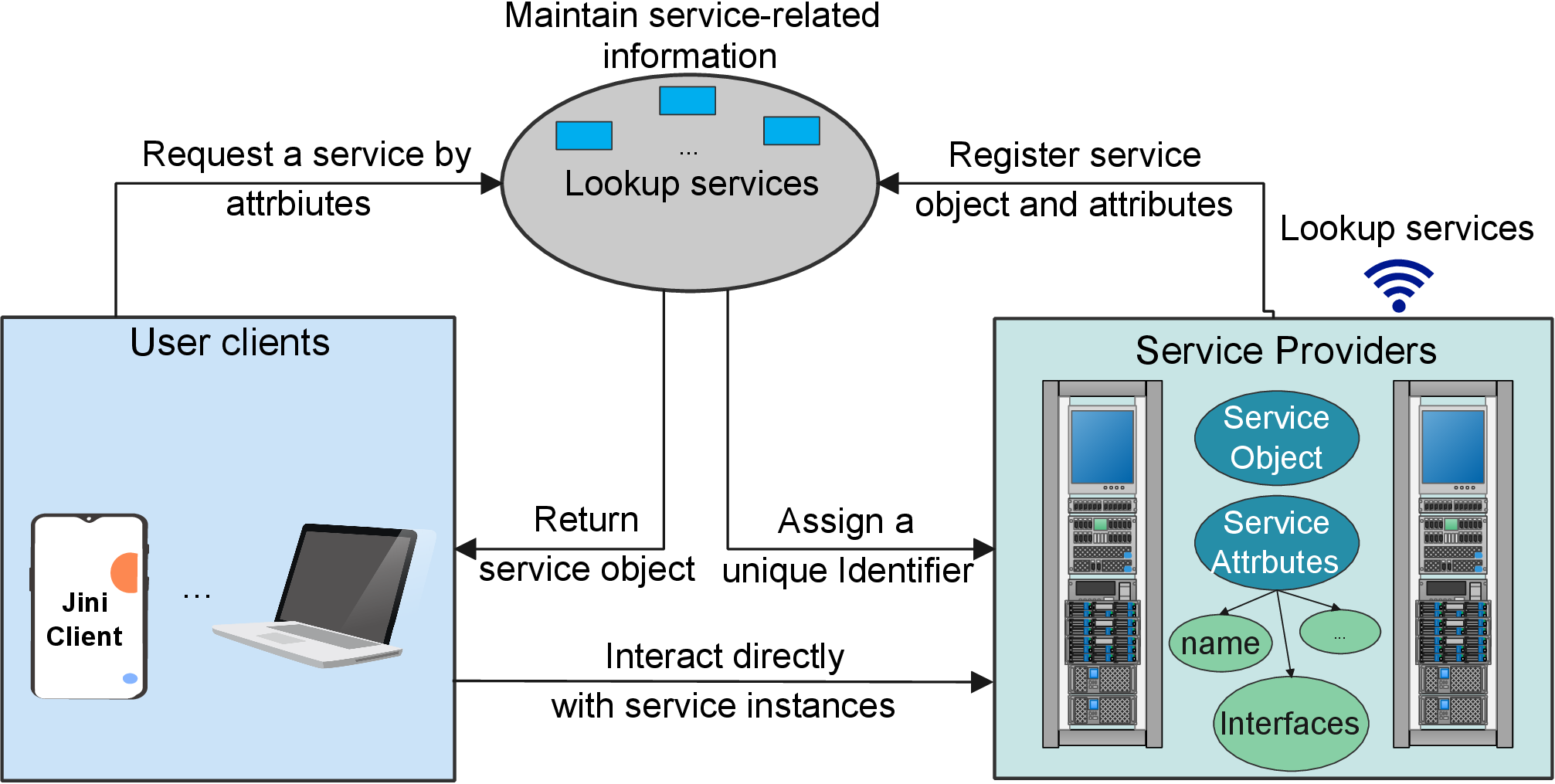}
    \vspace{-8pt}
    \caption{Jini SRIDS architecture and its process \cite{allard2003jini}.}
    \label{fig:jini}
    \vspace{-8pt}
\end{figure}

\paragraph{Jini} \hspace{\parindent}  
Jini \cite{allard2003jini} implements a hybrid centralized–decentralized \ac{SRIDS} architecture by combining centralized lookup directories with a federated overlay of peer lookup services (Fig.~\ref{fig:jini}). In this design, lookup services act as authoritative registries, maintaining dynamic indices of service proxies, interfaces, network endpoints, and descriptive attributes. To avoid a single point of failure and improve scalability, multiple lookup services autonomously discover one another and establish peer-to-peer federation, thereby replicating the directory state and coordinating query handling across the network.

Service registration in Jini begins when a provider submits its service object and metadata to one or more lookup services. Each lookup service assigns a unique identifier, updates its local index with the service’s interface signature and attribute set, and propagates the registration update to its federated peers. This synchronization ensures that all lookup services maintain a consistent, up-to-date view of available services. Service discovery proceeds as users issue queries, specifying interface types or attribute filters, to any lookup service. The contacted directory consults its local index and, if necessary, forwards the query to federated peers to aggregate additional matches. Upon resolution, the lookup service returns a list of matching service proxies, which the user then utilizes to establish direct communication with the selected instance. 

\paragraph{BitTorrent} \hspace{\parindent}  
BitTorrent \cite{cohen2003incentives} implements a hybrid \ac{SRIDS} architecture by combining directory‐based trackers for service registration and indexing with a fully decentralized peer‐to‐peer mechanism for metadata propagation and content retrieval. In this design, torrent files encapsulate the service description - file length, piece hashes, and piece size - and act as the registration artifact. Trackers maintain a dynamic registry of peers (seeders and leechers) for each torrent, thereby providing centralized indexing and peer‐discovery functionality. To mitigate single‐point‐of‐failure risks and improve scalability, multiple trackers or trackerless modes (via \acp{DHT}) can be employed, and super‐peer clustering strategies, such as those in Anatomic peer-to-peer, draw inspiration from BitTorrent’s incentives to distribute indexing load while preserving efficient lookup performance \cite{p2pSurvey2008}.  

The BitTorrent workflow begins with service advertisement: a provider generates a torrent descriptor and registers it with one or more trackers, which validate the descriptor, assign a unique torrent identifier, and record the initial peer set. A user first acquires the torrent file - via a web‐based directory or \ac{DHT} lookup - and then contacts the tracker to obtain a list of active peers. Thereafter, peers establish connections and exchange pieces according to the Tit‐for‐Tat incentive mechanism, which prioritizes uploads to those who have reciprocated in the past, and the rarest‐first piece‐selection policy \cite{crosby2007analysis}, which ensures rapid dissemination of the least‐replicated segments. Through the interplay of centralized peer discovery and decentralized, cooperative indexing and retrieval, BitTorrent attains efficient service registration, discovery, and distribution without reliance on a single authoritative server.

\begin{figure}[t!]
    \centering
    \includegraphics[scale=0.315]{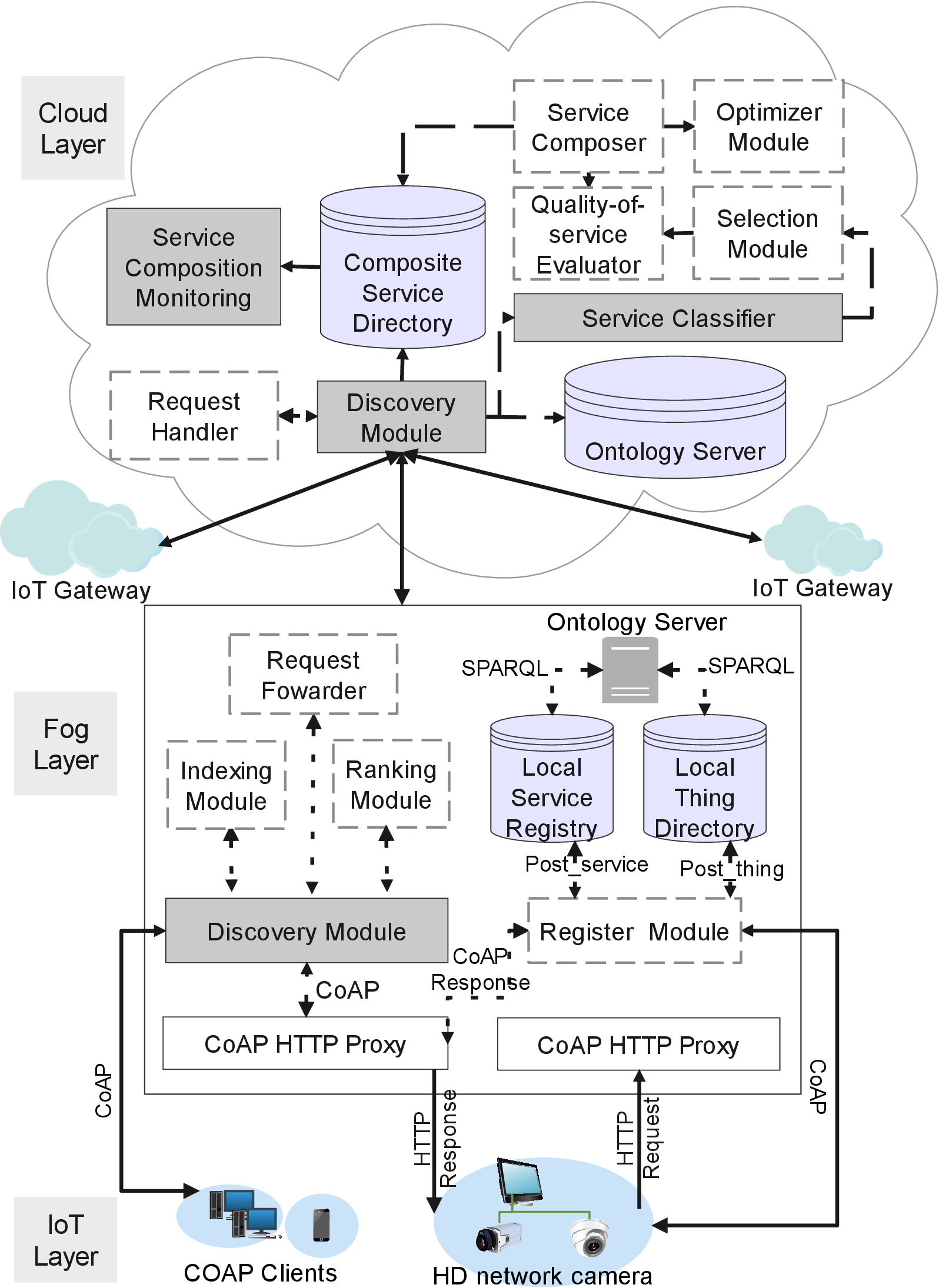}
    \vspace{1pt}
    \caption{A CoAP-based SRIDS architecture designed for IoT environments \cite{achir2022distributed}.}
    \label{fig:coap-based_architecture}
    \vspace{1pt}
\end{figure}

\paragraph{\ac{CoAP}-based} \hspace{\parindent}  
Hybrid \ac{CoAP}-based \ac{SRIDS} architectures exploit \ac{CoAP}’s lightweight messaging protocol to bridge resource‐constrained \ac{IoT} endpoints, fog‐tier peers, and cloud‐level registries, thereby combining decentralized indexing with centralized composition. Achir \textit{et al.} \cite{achir2022distributed} illustrated this paradigm by instantiating \ac{IoT} objects alongside network infrastructure and organizing fog nodes into a structured peer-to-peer overlay (Fig.~\ref{fig:coap-based_architecture}). Each gateway (relevant to \ac{PoA} defined in the Infrastructure) hosts both \ac{HTTP}–\ac{CoAP} and \ac{CoAP}–\ac{HTTP} proxies to translate between web and constrained environments, while a shared semantic model (grounded in basic ontologies) ensures interoperability among heterogeneous devices and services.

In their workflow, physical sensors and actuators at the \ac{IoT} layer register service descriptions via \ac{CoAP} POST requests to neighboring fog gateways. Upon receipt, gateways consult a local thing directory and service registry, annotate descriptors through an ontology server, and propagate registrations across the peer-to-peer overlay to achieve decentralized publication. Discovery and indexing modules at each fog peer resolve lookup queries by consulting locally cached indices or forwarding requests to overlay neighbors. Composition of higher‐order services, however, is deferred to the cloud layer, where abundant computing and storage resources host a global service directory. Composed sequences are cached for reuse: if a subsequent request matches an existing composition, the cloud registry returns the precomputed orchestration plan; otherwise, it invokes the composition engine to generate a new workflow that satisfies the requester's requirements.  

Ferdousi \textit{et al.} \cite{coap-based} extended this hybrid model by introducing context‐based analysis and virtual cloud‐based registries to mitigate scalability, interoperability, and bottleneck concerns. In their approach, refinery nodes extract service contexts from incoming \ac{CoAP} registrations and dynamically populate decentralized instances in the cloud. User request handlers then classify each discovery query, selecting the most efficient resolution mechanism - local fog lookup, peer overlay search, or cloud-based registry interrogation - while border routers route queries to the appropriate domain. This two‐tiered scheme leverages decentralized fog‐level resource sharing and fault tolerance alongside centralized, context‐aware registries that guarantee flexible, pay‐as‐you‐go composition and ensure global consistency.

\subsubsection{Centralized–Distributed Combination} \hspace{\parindent}  
The hybrid centralized–distributed \ac{SRIDS} architecture integrates the rigorous control of a central registry with the resilience and scale of distributed service publication. In this paradigm, the central authority is entrusted with global indexing, policy enforcement, and coordination tasks. This ensures efficient management of service metadata and the consistent application of system‐wide rules. Simultaneously, a network of peer nodes partakes in the dissemination and localized caching of service information, thereby distributing discovery queries' load and reducing reliance on a single directory.

In operation, service providers submit their descriptions and access credentials to the central registry, which validates the entries, assigns unique identifiers, and updates its master index. Incremental index updates - comprising service identifiers, interface signatures, and attribute filters - are then propagated to a selected set of peer nodes via a structured overlay or gossip protocol. These peers maintain local caches that accelerate lookup operations and participate in fault‐tolerant metadata replication. When a user issues a discovery request, it first consults the nearest peer cache; if the required entry is absent or if strict freshness is mandated by policy, the peer forwards the query to the central registry. The registry then returns the authoritative result, which the peer both relays to the user and incorporates into its local cache. This dual‐path workflow harmonizes low‐latency resolution under typical conditions with guaranteed consistency for critical operations, yielding a scalable yet controllable \ac{SRIDS} suitable for dynamic, large‐scale environments. The following is a detailed description of the hybrid architectures used.

\paragraph{eDonkey} \hspace{\parindent}  
eDonkey realizes a hybrid centralized–distributed \ac{SRIDS} architecture by coupling dedicated directory servers, which act as central dictionaries and global indices, with a peer‐to‐peer substrate that fragments and distributes service content \cite{heckmann2004edonkey}. In this design, directory servers maintain mappings of service descriptors to peer sources and assign users either a high‐\ac{ID} (unrestricted access) or low‐\ac{ID} (limited access) status. Service contents are partitioned into independent chunks and advertised across the peer-to-peer network, enabling concurrent retrieval and seamless composition. eDonkey's service discovery begins with a user's search request either via a \ac{TCP} connection to one or more directory servers or by issuing \ac{UDP} queries. The contacted server consults its index, performs full‐phrase or substring matching on the requested service name, and returns a list of known source peers. The user then establishes direct connections to selected peers and downloads the requisite chunks in parallel. Upon completion of chunk retrieval, the user reassembles them to reconstruct the complete service. This workflow leverages centralized lookup for rapid discovery and distributed chunk dissemination for scalable, fault‐tolerant access and supports service composition by aggregating segments from multiple sources.  

Qin \textit{et al.} \cite{revision_EDonkey} extended this architecture with a cross‐transmission layer based on a centralized multi‐media search engine that federates identifiers across multiple eDonkey‐style overlays. In their scheme, users first query the search engine to obtain composite identifiers spanning several peer-to-peer networks. Armed with these identifiers, they contact each overlay’s directory server to discover additional peers, and then orchestrate parallel chunk retrieval across the federated overlays. This cross‐overlay workflow preserves compatibility with the original peer-to-peer system while reducing user response time, minimizing access latency, ensuring quality-of-service, and providing enriched service metadata without disruptive changes to existing network modules.

\paragraph{Istio} \hspace{\parindent}  
Istio realizes a hybrid \ac{SRIDS} architecture by combining a centralized control plane, anchored in the Kubernetes \ac{API} server, with a distributed data plane of Envoy sidecar proxies \cite{istio_website}. The control plane comprises Pilot, Citadel, and Galley. Pilot monitors service and instance records in the Kubernetes registry, converts them into dynamic proxy configurations, and updates routing rules as deployments evolve. Citadel functions as a certificate authority: it exposes a certificate signing request endpoint, verifies submitted signing requests issued by node agents, signs them, and issues \ac{mTLS} certificates via the secret discovery service \ac{API}. Galley validates and distributes configuration artifacts to the control‐plane modules, ensuring consistency.  

In operation, service registration occurs when providers publish metadata and network endpoints to the Kubernetes \ac{API} server. Pilot retrieves these entries, generates per‐sidecar Envoy configurations, including local service registry snapshots and load‐balancing policies, and pushes them to each proxy. Envoy intercepts all service‐to‐service traffic, consults its local registry to select target instances, and applies its configured load‐balancing mode (round‐robin, random, or weighted). Envoy also performs periodic health checks: instances exceeding failure‐rate thresholds are evicted from the pool, while those with sustained successful checks are reinstated, yielding resilient, low‐latency routing. Istio’s security workflow is orchestrated by Citadel and the discovery service \ac{API}. Each node agent generates a private key and certificate signing request, submits the request to Citadel, and upon signature retrieval provisions Envoy with the certificate and key via the Envoy discovery service \ac{API}. This enables \ac{mTLS} between proxies and is repeated periodically to refresh credentials and maintain trust throughout the mesh \cite{istio2020}.  

\begin{figure}[t!]
    \centering
    \vspace{-4pt}
    \includegraphics[scale=0.31]{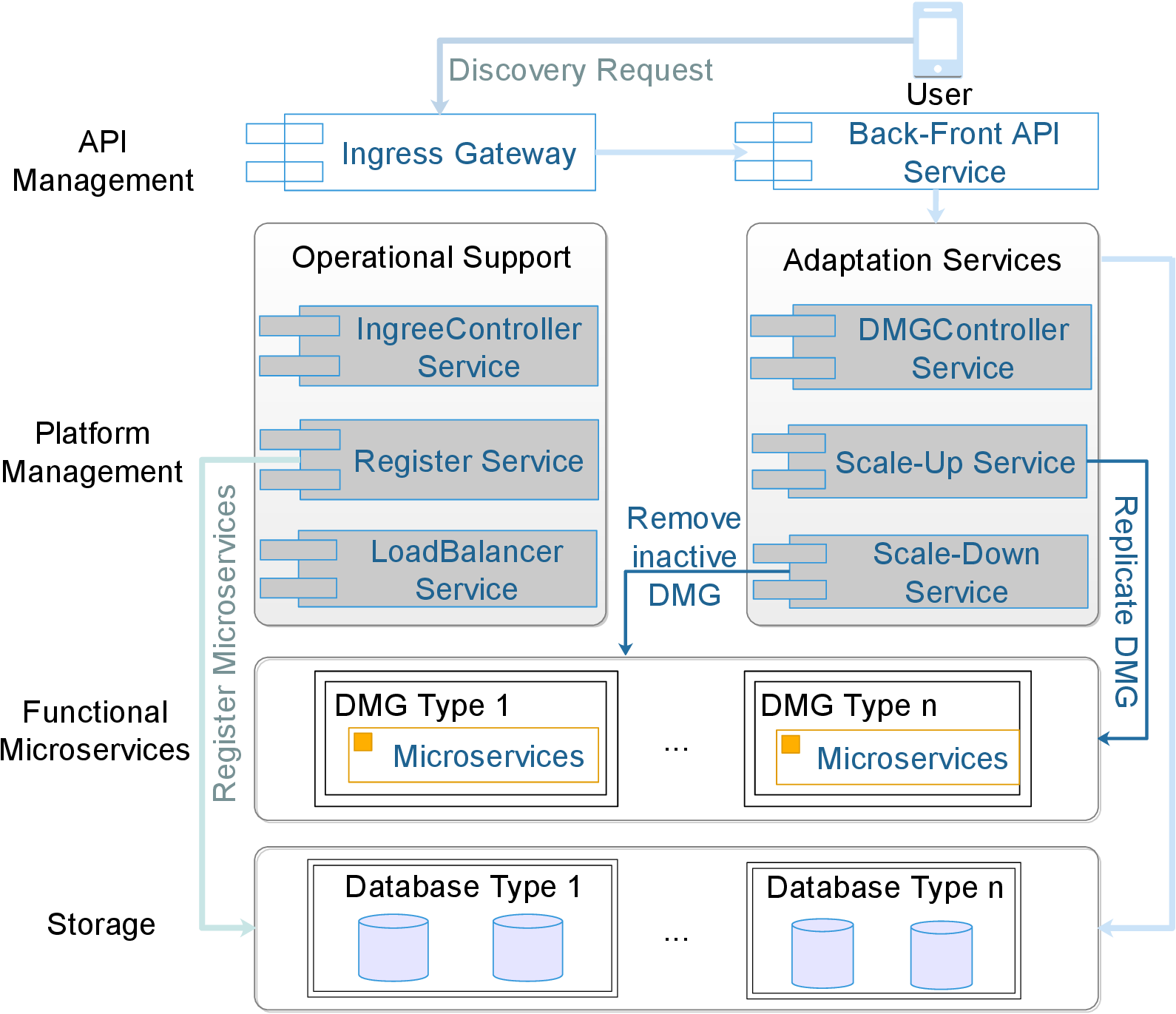}
    \vspace{-24pt}
    \caption{Overview of the Istio's data-driven SRIDS architecture \cite{Houmani2020}.}
    \label{fig:istio_architecture}
    \vspace{-8pt}
\end{figure}

Houmani \textit{et al.} \cite{Houmani2020} extended Istio’s mesh into a data‐driven \ac{SRIDS} framework by introducing profile‐matching and elastic scaling of data‐product replicas. Their architecture layers a register service - which logs microservice lifecycle events in a central database and notifies management services - beneath an IngressController that dynamically updates Istio routing rules. A LoadBalancer module distributes requests among replicas of each data type, while a backend‐for‐frontend \ac{API} gateway authenticates user requests against a Data Management Group (DMG) controller. The DMG controller verifies resource availability before discovery proceeds. When demand surges, a scale-up service instantiates additional DMG replicas; conversely, a scale-down decommissions idle replicas to conserve resources. By integrating these modules with Istio’s native control‐plane workflows (Fig.~\ref{fig:istio_architecture}), their system adapts to variable request rates while upholding response‐time and acceptance‐rate objectives.

\renewcommand{\arraystretch}{1.4}
\input{tables/tab5}

%% file: tables/tab5.tex
\begin{table*}[t!]
\centering
\caption{Comparison of SRIDS architectures based on the futuristic 6G system requirements.}
\vspace{-1pt}
\label{table:discovery-architectures_comparison}
\footnotesize
\begin{tabularx}{\textwidth}{lcccccccX}
 \toprule
        \textbf{Architecture} &
        \textbf{Reliability} &
        \textbf{Scalability} &
        \textbf{Adaptability} &
        \textbf{Determinism} &
        \textbf{Efficiency} &
        \textbf{Sustainability} &
        \textbf{Semantic-Awareness} &
        \textbf{Security} \\

\midrule
\textit{Napster} & Low & Low & Low & Medium & Low & Low & No & Low \\ 
        \textit{SLP-based} & Medium & Low & Low & Low & Medium & Low & No & Low \\ 
        \textit{Discovery Broker} & Low & Low & Medium & Medium & Medium & Medium & No & Low \\ 
        \textit{DNS-based} & Medium & Medium & Medium & Medium & Medium & Low & No & Low \\ 
        \textit{Kubernetes} & Medium & Medium & Medium & High & Medium & Medium & No & High \\ 
        \textit{Detection Mesh} & Medium & Low & Medium & Low & High & Medium & No & Medium \\ 
        \textit{Chord} & Medium & High & Low & Low & Medium & Low & No & Medium  \\ 
        \textit{Kademlia} & Medium & High & Low & Low & Medium & Low & No & Low \\ 
        \textit{Pastry} & High & Medium & Medium & Low & High & Low & No & Low \\ 
        \textit{CAN} & Medium & Medium & Medium & Medium & High & Low & No & Low \\ 
        \textit{Skip Graph} & Medium & Medium & Medium & Low & High & Medium & No & Medium \\  
        \textit{Cycloid} & Medium & Medium & Low & Low & Medium & Medium & No & Medium \\ 
        \textit{Kinaara} & Medium & Medium & High & Low & Medium & Medium & No & Medium \\ 
        \textit{HandFan} & Medium & Medium & Low & Medium & High & High & Yes & Low \\ 
        \textit{IOTA} & Medium & Medium & Medium & Medium & High & Low & Yes & High \\ 
        \textit{Gnutella} & Medium & Low & Medium & Low & Low & Medium & No & Low \\ 
        \textit{Freenet} & Medium & Low & Medium & Low & Low & Low & No & Medium \\ 
        \textit{SD-AMC} & Medium & Medium & High & Medium & High & Medium & Yes & Low \\ 
        \textit{Tree-based} & High & Medium & Medium & Low & High & Low & No & Medium \\ 
        \textit{Social-based} & High & Medium & High & Medium & Low & Low & Semantic prefix-match & Medium \\ 
        \textit{Jini} & Medium & Low & Low & Low & Medium & Medium & No & Medium \\ 
        \textit{BitTorrent} & Low & Low & Low & Low & Medium & Medium & No & Medium \\ 
        \textit{CoAP-Based} & Medium & Medium & Medium & Low & Medium & High & Content-based analysis & Medium \\ 
        \textit{eDonkey} & Medium & Medium & Medium & Low & Medium & Medium & Search on content & Medium \\ 
        \textit{Istio} & Medium & High & Low & Medium & High & High & No & High \\ 
\bottomrule
\end{tabularx}
% \vspace{-5pt}
\end{table*}

%% file: sections/sec5.tex
\section{Discussion}\label{sec:discussion}
Based on the projected 6G design objectives illustrated in Fig.~\ref{fig:srids_challenges}, this section systematically defines each objective and critically examines the extent to which these objectives are addressed within the architectures proposed in the \ac{SRIDS} literature. A comparative analysis of these proposed architectures is presented in Table \ref{table:discovery-architectures_comparison}.

\subsection{Reliability}
Reliability in \ac{SRIDS} architectures denotes the capacity to sustain uninterrupted service provision under node failures, network partitions, or software faults. Reliability is crucial: in the Health Guardian scenario of Section \ref{sec:basic_concepts}, where sub-millisecond anomaly detection, context enrichment, \ac{LLM} inference, and alert delivery must proceed across mobile edge nodes, any service outage directly endangers patient safety. Centralized designs (e.g., Napster, \ac{SLP}, \ac{DNS}/\ac{ONS}, Kubernetes, Consul, Detection Mesh) employ active–standby replication and transaction logs to mask failures but retain a residual single point of failure. Distributed architectures (Chord, Kademlia, \ac{CAN}, Skip Graph, Cycloid) scatter and replicate (key, value) pairs with stabilization protocols to recover orphaned records, trading maintenance overhead for resilience. Decentralized peer-to-peer schemes (Gnutella, Freenet,\ac{SD‐AMC}, Tree-based, Social-based) flood or gossip service descriptors among all peers, eliminating any coordinator dependency at the expense of higher control-plane traffic. Hybrid architectures (Jini, BitTorrent, \ac{CoAP}-based, eDonkey, Istio) fuse centralized policy enforcement with peer-centric dissemination, reconciling consistency. Overall, fully decentralized peer-to-peer architectures deliver the highest reliability by obviating any central coordinator.  

In \textit{Service Registration \& Indexing}, reliability hinges on preserving the metadata catalog under failures. Centralized registries (Napster servers, \ac{SLP} agents, \ac{DNS}/\ac{ONS}, Kubernetes API server, Consul catalog, Detection Mesh) rely on clustered backends and cache-invalidation to prevent index loss but must resolve split-brain when primaries crash. \ac{DHT}-based registries hash service \acp{ID}, replicate each entry across successor or leaf sets, and invoke periodic stabilization to recover lost data, ensuring no single outage erases registrations. Decentralized peer-to-peer indexing (Gnutella, Freenet, \ac{SD‐AMC}) embeds indices in every peer and propagates updates via controlled flooding or gossip, preserving metadata availability under arbitrary failures but delaying convergence. Hybrid schemes validate descriptors centrally and then push incremental updates into peer-to-peer overlays or sidecar caches, mixing authoritative consistency with distributed redundancy. Among these, fully decentralized peer-to-peer indexing is superior in reliability, as it eliminates any single point of failure.  

In \textit{Service Discovery \!\&\! Selection}, reliability depends on resilient lookup paths and fault-tolerant matching. Centralized brokers (Napster, \ac{DNS}/\ac{ONS} resolvers, Kubernetes \ac{DNS}, Consul) yield deterministic, low-latency responses but risk a total blackout if the coordinator fails. Structured distributed overlays route queries in $\mathcal{O}(\log N)$\footnote{$N$ is the total number of nodes.} hops and use alternate pointers to bypass failed nodes, sustaining high success rates under churn but incurring multi-hop latency. Decentralized flooding or random-walk discovery (Gnutella, Gia, Freenet, \ac{SD‐AMC}) guarantees that at least one provider responds if the overlay remains connected, achieving the highest availability under failures at the cost of excessive control traffic. Hybrid discovery (Jini, BitTorrent tracker \!+\! \ac{DHT}, \ac{CoAP} fog queries, eDonkey directory \!+\! peer-to-peer, Istio sidecar caches \!+\! control plane) combines rapid local caches with central fallbacks, balancing speed and consistency. Purely decentralized discovery maximizes reliability by removing all central dependencies.  

In \textit{Service Routing \& Communication}, reliability is measured by the ability to maintain end-to-end delivery and adapt paths under link or node failures. Centralized routing (\ac{DNS} resolution, Kubernetes kube-proxy, Consul mesh, Detection Mesh hierarchy) offers structured paths but hinges on healthy controllers and fresh mappings - failures here disrupt flows. Distributed overlays (Chord, Cycloid, Kademlia) integrate self-healing neighbor maintenance and multi-path forwarding to reroute around failed peers, preserving connectivity despite churn. Decentralized peer-to-peer routing (Gnutella flooding, Freenet, \ac{SD‐AMC} context-aware forwarding) dynamically reroutes at each hop around failures, ensuring reachability even if multiple nodes fail. Hybrid \ac{SRIDS} architectures (Istio Envoy sidecars, \ac{CoAP} gateways, eDonkey mediators) coordinate proxies for health checks and policy enforcement, enabling immediate failover. Nonetheless, fully decentralized peer-to-peer routing attains the highest reliability by obviating any centralized or hierarchical coordination point.

\subsection{Scalability}  
Scalability in \ac{SRIDS} architectures denotes the capacity to accommodate growing numbers of services, users, and data traffic without degrading performance or response time. Scalability is crucial: in the Health Guardian scenario, where millions of sensor streams, anomaly detections, context enrichments, and \ac{LLM} inferences must proceed in real time, any bottleneck breaks continuous monitoring. Centralized designs (Napster, \ac{SLP}, \ac{DNS}/\ac{ONS}, Kubernetes, Consul, Detection Mesh) exhibit poor scalability due to the single registry or broker becoming overloaded under surges. Distributed overlays (Chord, Kademlia, Pastry, \ac{CAN}, Cycloid) partition metadata and route lookups in $\mathcal{O}(\log N)$ hops, sustaining throughput as $N$ grows. Decentralized peer-to-peer schemes (Gnutella, Freenet, \ac{SD‐AMC}) scale linearly in storage and forwarding capacity but incur control-plane explosion from flooding or gossip. Hybrid solutions (Jini, BitTorrent, \ac{CoAP}-based, eDonkey) non-centralize high-volume tasks while retaining central coordination for policy, improving over pure centralization yet still facing coordination overhead. Overall, distributed overlays offer the most scalable foundation by evenly distributing load and bounding lookup costs.

In \textit{Service Registration \& Indexing}, scalability depends on how efficiently new services and instances are registered without creating hotspots. Centralized registries use replicas and sharding but ultimately bottleneck at the master index under high registration rates. Decentralized peer-to-peer indexing floods or gossip advertisements, which scale storage but trigger excessive control traffic and inconsistent convergence. Hybrid approaches push validated updates from a central registry into peer-to-peer caches, smoothing peaks but still requiring central throughput. In contrast, distributed indexing assigns each $(\mathrm{service \ac{ID}}, \mathrm{metadata})$ to a node in a structured overlay, replicates it across successor or leaf sets for resilience, and handles joins/leaves via stabilization routines. Since each node manages only $\mathcal{O}(\log N)$ pointers and entries, distributed architectures scale registration and indexing most effectively under large-scale service churn.

In \textit{Service Discovery \& Selection}, scalability hinges on handling massive query and lookup volumes with bound latency. Centralized brokers serve queries in constant time but collapse under heavy load. Decentralized flooding or random walks guarantee availability yet generate $\mathcal{O}(N)$ messages per lookup, overwhelming the network as $N$ grows. Hybrid discovery leverages local cache hits and central fallbacks, reducing average load but still relying on central capacity. Structured distributed overlays route queries in $\mathcal{O}(\log N)$ hops, employ alternate‐path pointers to bypass failures, and maintain balanced query distribution among nodes. This yields predictable, low-overhead discovery and selection at scale. To this end, distributed discovery and selection architectures outperform other paradigms in sustaining high lookup rates without incurring linear message overhead.

In \textit{Service Routing \& Communication}, scalability requires that request forwarding and endpoint resolution remain efficient as the system expands. Centralized routing (\ac{DNS} resolution, kube‐proxy, Consul mesh, Detection Mesh) offers fixed paths but saturates control nodes and mapping caches. Decentralized peer-to-peer routing (Gnutella flooding, Freenet hill‐climbing, \ac{SD‐AMC}) adapts dynamically but generates $\mathcal{O}(N)$ traffic under broad queries. Hybrid architectures colocate sidecar proxies to perform local load balancing and health checks, improving over pure centralization but adding coordination layers. Distributed overlay routing (Chord, Pastry, Cycloid, Kademlia) embeds self‐healing neighbor tables and multi-path forwarding in a structured topology, ensuring each node handles only $\mathcal{O}(\log N)$ neighbors and messages. This architecture maintains low-latency, high-throughput communication as $N$ grows, making distributed overlays the most scalable choice for service routing and communication.

\subsection{Automaticity and Adaptability}
Automaticity and adaptability denote the autonomous reconfiguration and self-tuning of \ac{SRIDS} modules in response to dynamic network conditions, service migrations, and shifting user intents. These criteria are crucial: in the Health Guardian scenario, patient mobility from home to ambulance triggers on-the-fly re-binding of anomaly detectors, context enrichers, and \ac{LLM} inference engines without human intervention. Centralized \ac{SRIDS} (Napster, \ac{SLP}, \ac{DNS}/\ac{ONS}, Kubernetes, Consul, Detection Mesh) exhibit limited automaticity, control-plane updates and policy changes often require explicit coordinator actions, hindering rapid adaptation. Distributed \ac{DHT} overlays (Chord, Kademlia, Pastry, \ac{CAN}, Cycloid) offer built-in stabilization and replication but adhere to fixed refresh schedules that may lag under abrupt topology changes. Decentralized peer-to-peer schemes (Gnutella, Freenet, \ac{SD‐AMC}) deliver maximal autonomy: peers self-organize, gossip, and flood updates without a central authority, achieving high adaptability at the expense of $\mathcal{O}(N)$ control traffic and inconsistent convergence times. Hybrid architectures (Jini, BitTorrent, \ac{CoAP}-based, eDonkey, Istio) fuse non-centralized diffusion with centralized governance, reconciling automated local adaptation with coordinated policy enforcement. Overall, fully decentralized \ac{SRIDS} architectures attain the highest degree of automaticity and adaptability by enabling each node to independently react to, and propagate changes, thereby ensuring rapid and localized responsiveness across the system and eliminating any mutual control points.

In \textit{Service Registration \& Indexing}, automaticity and adaptability govern how swiftly service metadata propagate, reconfigure, and converge under dynamic loads. Centralized registries (Kubernetes API server, \ac{SLP} directory, \ac{DNS}/\ac{ONS}) rely on manual or semi-automated sharding and cache invalidation, limiting responsiveness when services scale or relocate. \ac{DHT}-based registries autonomously hash, replicate, and stabilize each $(\mathrm{serviceID},\mathrm{metadata})$ across $\mathcal{O}(\log N)$ nodes, adapting to joins and failures without operator input, yet fixed stabilization intervals can delay updates under high churn. Decentralized peer-to-peer indexing (Gnutella, Freenet, \ac{SD‐AMC}) fully automates registration and deregistration: every peer maintains local indices and propagates changes via gossip or flooding, instantly reflecting service migrations but incurring linear message overhead. Hybrid schemes validate descriptors centrally, then push incremental updates into peer-to-peer overlay or sidecar caches, combining rapid adaptation with controlled consistency. Of these, decentralized peer-to-peer indexing is superior in automaticity and adaptability, enabling zero-touch registration across dynamic overlays.  

In \textit{Service Discovery \& Selection}, adaptability dictates how lookup flows adjust to varying query loads, network partitions, and instance churn. Centralized brokers (Napster, \ac{DNS}/\ac{ONS} resolvers, Kubernetes \ac{DNS}, Consul) provide deterministic, low-latency responses but mostly require manual scaling or replication policies to handle surges, impeding real-time responsiveness. Structured distributed overlay route queries in $\mathcal{O}(\log N)$ hops with alternate-path pointers that bypass failed nodes, maintaining availability under churn, yet bound by static routing-table refresh schedules. Decentralized flooding or random-walk discovery (Gnutella, Gia, Freenet, \ac{SD‐AMC}) inherently adapts the search radius according to local heuristics, delivering on-the-fly reachability at the cost of unpredictable latency. Hybrid discovery (Jini federation, BitTorrent + \ac{DHT}, \ac{CoAP} fog queries, eDonkey directory + peer-to-peer, Istio sidecars + control plane) integrates local caches and authoritative fallbacks, dynamically selecting the optimal resolution path per policy and network state. Nonetheless, fully decentralized discovery maximizes adaptability by autonomously tuning search scopes and exploiting local peer connectivity without centralized dependencies.  

In \textit{Service Routing \& Communication}, automaticity and adaptability measure the ability to self-adjust forwarding paths and load-balancing policies under shifting link states and instance availability. Centralized routing (\ac{DNS} resolution, Kubernetes kube-proxy, Consul mesh, Detection Mesh) depends on static control-plane directives, requiring manual updates when endpoints change. Distributed overlays (Chord, Pastry, Cycloid, Kademlia) implement self-healing neighbor maintenance and multi-path forwarding to reroute around failures, yet predefined neighbor-refresh protocols can lag during rapid churn. Decentralized peer-to-peer routing (Gnutella flooding, Freenet hill-climbing, \ac{SD‐AMC} context-aware forwarding) autonomously recalculates routes per hop based on local state, ensuring seamless adaptation but incurring high message overhead and variable convergence times. Hybrid routing and communications (Istio Envoy sidecars, \ac{CoAP} gateways, eDonkey mediators) utilize proxies that perform real-time health checks, evict unhealthy endpoints, and fetch updated routing rules from a control plane, balancing local agility with global governance. However, distributed and decentralized routing achieves the highest automaticity and adaptability by enabling each node to instantly reconfigure its forwarding logic without any centralized intervention.

\subsection{Determinism}
Determinism in \ac{SRIDS} architectures denotes the guarantee of bounded, predictable delays for service provision, as well as repeatable mappings from queries to service instances. In the Health Guardian scenario, where sub‐millisecond arrhythmia alerts and context‐aware \ac{LLM} inferences must execute reliably, any nondeterministic latency could compromise patient safety. Centralized systems (Napster, \ac{SLP}, \ac{DNS}/\ac{ONS}, Kubernetes, Consul, Detection Mesh) enforce determinism via a single authoritative registry and precomputed routing policies, yielding fixed lookup paths and response times. Distributed \ac{DHT} overlays (Chord, Kademlia, Pastry, \ac{CAN}, Cycloid) provide deterministic $\mathcal{O}(\log N)$ lookup sequences and hash‐based service‐to‐node mappings, but their stabilization protocols under churn introduce transient timing variability. Decentralized peer-to-peer schemes (Gnutella, Freenet, \ac{SD‐AMC}) rely on flooding or gossip, producing variable convergence times and inconsistent query latencies. Hybrid architectures fuse central control for policy enforcement with overlay‐based publication, achieving partial determinism through centralized fallbacks yet still incurring overlay‐induced timing fluctuations. Overall, centralized architectures remain superior in terms of determinism by enforcing single‐path, policy‐driven workflows.  

In \textit{Service Registration \& Indexing}, determinism demands that each service descriptor be stored at a fixed location and retrievable within known time bounds. Centralized registries (e.g., Kubernetes API server, \ac{DNS} directory, Consul catalog) guarantee deterministic registration delays and lookup times by funneling all updates through a master index and serving queries from a uniform repository. \ac{DHT}‐based distributed indexing ensures that each $(\mathrm{service \ac{ID}}, \mathrm{metadata})$ pair maps deterministically to the same overlay node in $\mathcal{O}(\log N)$ hops, but stabilization intervals and replica reallocation introduce nondeterministic update latencies. Decentralized peer-to-peer indexing (Gnutella, Freenet, \ac{SD‐AMC}) lacks any fixed mapping, as advertisements propagate via probabilistic gossip or flooding, resulting in variable convergence. Hybrid schemes centralize initial validation yet push updates into overlay caches, yielding partial determinism limited by asynchronous propagation. Thus, centralized indexing is superior to deterministic registration and retrieval.  
 
For \textit{Service Discovery \& Selection}, determinism implies that identical queries traverse fixed paths and return the same ordered results within known latency bounds. Centralized brokers (Napster, \ac{DNS}/\ac{ONS} resolvers, Kubernetes \ac{DNS}, Consul) deliver deterministic discovery by consulting a static mapping table and returning consistent selections, achieving bounded response times under uniform load. Structured \ac{DHT} overlays route lookups in deterministic $\mathcal{O}(\log N)$ hops via finger or routing tables, yet churn‐driven stabilization can transiently perturb routing tables, yielding variable latencies. Decentralized discovery (Gnutella, Gia, Freenet, \ac{SD‐AMC}) employs random walks or flooding, causing unpredictable message overhead and response timings. Hybrids use local cache hits for low‐latency resolution and fallback to a central coordinator for authoritative answers, thus offering semi‐deterministic behavior, but only when central paths are invoked. Accordingly, centralized discovery and selection excel at determinism.  

In \textit{Service Routing \& Communication}, determinism requires effective forwarding rules and path selections to ensure stable end‐to‐end delays. Centralized routing frameworks (\ac{DNS} resolution, Kubernetes kube‐proxy, Consul mesh, Detection Mesh hierarchy) precompute routes and update them synchronously via the control plane, yielding predictable packet paths and latency. Distributed overlay routing (Chord, Pastry, Cycloid, Kademlia) uses self‐healing neighbor tables and multi‐path routing, generally delivering deterministic $\mathcal{O}(\log N)$‐hop paths but experiencing transient perturbations during neighbor‐refresh cycles. Decentralized peer-to-peer routing (Gnutella flooding, Freenet hill‐climbing, \ac{SD‐AMC} context‐aware forwarding) adapts paths on‐the‐fly, leading to variable delays. Hybrid architectures (Istio Envoy sidecars, \ac{CoAP} gateways, eDonkey mediators) combine central policy distribution with local proxy routing, offering semi‐deterministic behavior contingent on control‐plane synchronization. Therefore, centralized routing architectures are superior to strict determinism.

\subsection{Efficiency}  
Efficiency in \ac{SRIDS} denotes the minimization of control‐plane and data‐plane overhead - computation, communication, memory, and message complexity - while sustaining rapid service provision under large-scale, dynamic loads. In the Health Guardian scenario, millions of sensor streams, anomaly detectors, and \ac{LLM} inferences must be orchestrated in real time without exhausting edge-node resources or violating sub-millisecond latency budgets. Centralized \ac{SRIDS} (Napster, \ac{SLP}, \ac{DNS}/\ac{ONS}, Kubernetes, Consul) leverage optimized database queries and caches to achieve low per-request costs but collapse into bottlenecks as $N$ and request rates grow. Distributed overlays (Chord, Kademlia, Pastry, \ac{CAN}, Cycloid) guarantee $\mathcal{O}(\log N)$ lookup and update costs, evenly distribute load, and maintain throughput at scale, at the expense of periodic stabilization traffic. Decentralized peer-to-peer schemes (Gnutella, Freenet, \ac{SD‐AMC}) offer full peer autonomy and fault tolerance but incur $\mathcal{O}(N)$ messaging for flooding or gossip, leading to high control-plane overhead. Hybrid architectures (Jini, BitTorrent, \ac{CoAP}-based, eDonkey, Istio) offload bulk advertisement and query diffusion to peers while retaining central registries for policy, balancing overhead and adaptability. Overall, distributed \ac{DHT} overlays deliver the highest efficiency by bounding per-operation complexity and avoiding both single-point and broadcast storms.

In \textit{Service Registration \& Indexing}, efficiency demands that service descriptors be stored, replicated, and retrieved with minimal messaging and processing costs. Centralized registries write to a master index and serve lookups in constant time, but suffer queuing delays once the arrival rate exceeds the server’s capacity. Decentralized peer-to-peer indexing floods or gossips each advertisement to all $N$ peers, causing $\mathcal{O}(N)$ messages per update. Hybrid schemes validate centrally and then multicast to a subset of super-peers, reducing peak load but still requiring synchronization. \ac{DHT}-based distributed registration hashes each $(\mathrm{service \ac{ID}}$ to $\mathcal{O}(\log N)$ nodes and replicates along successor or leaf sets, achieving $\mathcal{O}(\log N)$ messages per registration and balanced indexing load. Thus, distributed \ac{DHT} architectures are superior to registration and indexing efficiency.  

In \textit{Service Discovery \& Selection}, efficiency means resolving lookup requests with bounded hops and low control-plane cost. Centralized brokers require $\mathcal{O}(1)$ message cost but saturate under heavy queries. Flooding-based discovery in decentralized peer-to-peer yields high availability but at $\mathcal{O}(N)$ message cost per lookup. Hybrid discovery combines local caches and central fallback, moderating load but still relying on central coordination. Structured distributed overlays route queries in $\mathcal{O}(\log N)$ hops via finger tables or routing buckets, distributing query load evenly and incurring negligible per-node processing. By capping both messages and per-lookup delays on a logarithmic scale, distributed architectures outperform other paradigms in discovery and selection efficiency.  

In \textit{Service Routing \& Communication}, efficiency requires path computation and forwarding for service queries with minimal per-packet overhead, even under churn. Centralized routing policies yield predictable paths but impose high coordination costs for failover and congestion management. Decentralized flooding or context-aware forwarding (Gnutella, Freenet) achieves resilience but generates redundant traffic. Hybrid routing and communications rely on Envoy sidecars or \ac{CoAP} proxies - locally enforcing policies yet require control-plane updates that add latency. Distributed overlay routing (Chord, Pastry, Cycloid, Kademlia) maintains $\mathcal{O}(\log N)$ neighbor tables, computes next hops via prefix or XOR metrics, and self-heals with limited background traffic. This structured model ensures efficient, low-overhead routing at scale, making distributed overlays the most efficient approach for \ac{SRIDS} routing.

\subsection{Sustainability} 
Sustainability in \ac{SRIDS} architectures refers to the minimization of total energy expenditure across the \mbox{modules} responsible for service provision while preserving required \mbox{performance}. In the Health Guardian use case, where \mbox{wearable} sensors, anomaly detectors, context enrichers, and \ac{LLM} inferences operate on battery‐constrained edge devices, excessive energy draw shortens operational lifetime and risks service discontinuities. Centralized designs incur relentless consumption from always‐on directory servers and databases, yielding poor sustainability. Distributed \ac{DHT} overlays spread load but pay steady energy costs for periodic stabilization and replica maintenance. Fully decentralized peer-to-peer schemes eliminate central points yet incur $\mathcal{O}(N)$ flooding or gossip overhead, draining network energy under large $N$. Hybrid architectures offload bulk advertisement and query tasks to overlay peers, while central modules enforce policy and global consistency, curbing redundant transmissions. \red{As a result of balancing distributed work with controlled coordination, hybrid SRIDS architectures achieve the highest level of sustainability.}

In \textit{Service Registration \& Indexing}, sustainable operation demands that each registration incurs minimal messaging and storage power. Centralized registries consume constant power to maintain master indices, irrespective of registration rates. \ac{DHT}‐based registries hash each $(\mathrm{service \ac{ID}}$ to $\mathcal{O}(\log N)$ nodes, reducing per‐node load but generating regular stabilization traffic that wastes energy. Decentralized peer-to-peer indexing floods or gossips each advertisement to all $N$ peers, imposing $\mathcal{O}(N)$ message overhead per update. Hybrid schemes validate entries centrally, then multicast updates only to designated overlay nodes or super‐peers, slashing redundant transmissions and aligning energy use with actual registration needs. Hence, hybrid architectures are superior in sustainable registration and indexing.

Sustainable \textit{Service Discovery \& Selection} requires limiting redundant lookups and avoiding unnecessary probing. Centralized brokers serve queries with constant‐time database access but maintain persistent directory state, drawing high energy even under light loads. Structured distributed overlays route lookups in $\mathcal{O}(\log N)$ hops, distributing query load yet incurring background refresh overheads. Decentralized flooding (Gnutella, Freenet) or random‐walk discovery generates $\mathcal{O}(N)$ messages, wasting network energy when the services change infrequently. Hybrid discovery leverages local caches for most lookups and escalates to central fallbacks only on misses, dynamically adjusting the propagation scope to demand. This adaptive caching and controlled flooding minimize active querying, making hybrid designs the most sustainable for discovery and selection.

In \textit{Service Routing \& Communication}, sustainability hinges on reducing control‐plane chatter and optimizing data‐path energy. Centralized routing models (Kubernetes, \ac{DNS}) maintain always‐active controllers and proxies, incurring high baseline power. Distributed overlay routing (\ac{CAN}, Pastry, Kademlia) employs proximity‐aware multi‐hop forwarding with $\mathcal{O}(\log N)$ neighbor tables, cutting per‐hop transmission energy but paying periodic neighbor‐probe costs. Decentralized peer-to-peer forwarding (\ac{SD‐AMC}) relies on probabilistic or flooding‐based path discovery, generating redundant traffic. Hybrid architectures (Istio, \ac{CoAP}) colocate sidecar proxies that enforce policies and cache routes locally, reducing inter‐domain signaling and avoiding full‐mesh probes. By confining control updates to topology changes and exploiting in‐node caches for steady‐state routing, hybrid architectures deliver the highest communication sustainability.

\subsection{Semantic-Awareness} 
Semantic‐awareness in \ac{SRIDS} architectures denotes the ability to interpret, annotate, and reason over the conceptual meaning of service descriptions, user intents, and contextual metadata - moving beyond syntactic or keyword matches to ontology-driven classification and inference. In the Health Guardian exemplar, semantic tags (“arrhythmia detection,” “fall risk assessment,” “patient mobility context”) enable the orchestrator to bind appropriate anomaly‐detectors and \ac{LLM}‐inference engines without manual schema alignment. Across architectural classes, purely centralized systems maintain a single registry amenable to ontology extension but suffer from rigid schema evolution and high coupling. Distributed \ac{SRIDS} architectures scatter metadata without preserving semantic relationships, precluding inference across related service types. Fully decentralized peer-to-peer schemes exchange unstructured advertisements via flooding, lacking any unified vocabulary to support concept matching. Hybrid architectures combine a central ontology or policy engine for semantic annotation with peer‐to‐peer propagation at scale, reconciling global consistency with local autonomy. Consequently, hybrid \ac{SRIDS} emerge as the superior paradigm for embedding semantic‐awareness at scale. Nonetheless, none of the reviewed works consider semantic‐awareness as a critical requirement for 6G use cases.

In \textit{Service Registration \& Indexing}, semantic‐awareness demands that each service descriptor be annotated with ontology terms and context attributes, then stored and disseminated so that concept hierarchies and relationships remain accessible. Centralized registries can integrate \ac{OWL} or \ac{RDF} stores to tag profiles with semantic classes, yielding precise, queryable graphs, but they centralize schema management and incur versioning bottlenecks. \ac{DHT}‐based approaches map only hashed identifiers, oblivious to concept granularity, and cannot support multi-attribute semantic indexing absent costly overlays. Decentralized peer-to-peer indexing floods bare descriptors without shared vocabularies, rendering semantic tags inconsistent and mismatched. Hybrid schemes delegate ontology enforcement to a central controller - validating semantic schemas, resolving conflicts, and assigning context tags - then propagating enriched indices to selected peers for local lookups. This dual mechanism curtails redundant transmissions while preserving semantic coherence, making hybrid architectures the most effective for semantic‐aware registration and indexing.

For \textit{Service Discovery \& Selection}, semantic‐awareness enables users to express intent in conceptual terms (e.g., “high-sensitivity fall detector for elderly care”) and receive ranked results based on inferred service capabilities and context compatibility. Centralized brokers can deploy semantic matching engines (e.g., SPARQL over \ac{RDF} triples) to return conceptually related services, yet they centralize all inference logic and may become performance chokepoints. Structured \ac{DHT} lookups guarantee $\mathcal{O}(\log N)$ hops but offer only exact‐match routing on hashed keys, failing to exploit ontological hierarchies or similarity metrics. Decentralized flooding (Gnutella, Freenet) broadcasts untyped queries, yielding many irrelevant hits when concepts diverge from simple keywords. Hybrid discovery combines local semantic caches at edge peers - populated via controlled propagation - with central ontology services for complex inference, dynamically selecting the nearest node holding conceptually matched entries. This blend maximizes semantic recall and precision with bounded overhead, positioning hybrid \ac{SRIDS} as the superior choice for semantic‐aware discovery and selection. 

In \textit{Service Routing \& Communication}, semantic‐awareness implies that forwarding decisions consider service semantics and user context, prioritizing paths to nodes offering semantically relevant capabilities (e.g., routing anomaly alerts to cardiology-optimized edge servers). Centralized routing models (Kubernetes, \ac{DNS}‐based) can embed semantic policies in control-plane rules but require manual policy updates and global coordination for each new concept. Distributed overlays lack any semantic layer, routing solely based on identifier proximity. Decentralized peer-to-peer schemes (\ac{SD‐AMC}) adapt routes based on resource metrics but ignore high-level semantics, leading to suboptimal path choices. Hybrid architectures (Istio, \ac{CoAP}‐based) colocate proxies that interpret semantic tags - select routes depending on policy-driven intent hierarchies and local context - and fall back on central controllers for complex semantic resolution. By uniting local semantic filtering with global policy enforcement, hybrid \ac{SRIDS} deliver context-driven routing and communication that align network paths with user meanings, making them the most semantically aware approach.

\subsection{Security, Privacy, and Trust} 
Security, privacy, and trust in \ac{SRIDS} architectures encompass the guarantees of confidentiality and integrity of service metadata and user queries, the protection of sensitive context (e.g., patient vitals in the Health Guardian use case), and the assurance that participating nodes and services behave as advertised. In the Health Guardian scenario, unauthorized disclosure of physiological streams or forged anomaly alerts would compromise patient safety and erode user confidence. Across architectures, purely centralized designs enforce strong access control, authentication, and encryption in a single registry, yet introduce a single point of compromise. Distributed \ac{DHT} overlays eliminate that central bottleneck but lack a global trust anchor, making them vulnerable to Sybil, poisoning, and impersonation attacks without a heavy cryptographic overlay. Fully decentralized peer-to-peer schemes maximize autonomy but rely on reputation or ad‐hoc trust models, exposing metadata to all peers and offering minimal privacy. Hybrid architectures pair centralized identity and policy enforcement with non-centralized propagation, yielding both robust authentication and attack resilience. Accordingly, hybrid \ac{SRIDS} architectures provide the ideal balance of security, privacy, and trust.
 
In \textit{Service Registration \& Indexing}, security and privacy hinge on authenticating providers, encrypting descriptors, and preventing unauthorized registry tampering. Centralized indexing applies access control lists and end-to-end encryption to protect descriptor stores, but suffers from a single compromise point. Distributed registration spreads storage yet exposes metadata to $\mathcal{O}(\log N)$ nodes, risking leakage and requiring identity‐based encryption or verifiable credentials to forestall poisoning. Decentralized peer-to-peer indexing broadcasts plain descriptors network‐wide, offering no inherent confidentiality or integrity guarantees and relying on reputation systems that scale poorly. Hybrid schemes authenticate and authorize each entry centrally before propagating it to selected overlay peers over encrypted channels, thus minimizing metadata exposure while retaining non-centralization benefits. Hybrid \ac{SRIDS} thus emerges as the superior approach to secure, privacy‐preserving registration \& indexing.
 
For \textit{Service Discovery \& Selection}, preserving user query privacy and validating service authenticity are paramount. Centralized brokers enforce authentication and can anonymize queries, but a breached broker undermines the entire discovery process and reveals user preferences. Distributed lookups avoid central storage but require each lookup to traverse multiple nodes, exposing query patterns unless costly onion‐routing is used. Decentralized flood‐based discovery (Gnutella, Freenet) disperses queries broadly, infecting all peers with metadata and offering no trust validation - attackers can inject spoofed responses freely. Hybrid discovery confines sensitive matching to central semantic engines while permitting local cache hits via authenticated sidecars, reducing query surface and preserving privacy by encrypting fallback requests. By combining centralized verification with controlled peer caching, hybrid architectures deliver the strongest guarantees of authenticity, confidentiality, and trust in discovery and selection.  

In \textit{Service Routing \& Communication}, end-to-end encryption, policy-based forwarding, and node‐level attestation prevent eavesdropping, tampering, and route hijacking. Centralized control‐plane models (\ac{DNS}/\ac{ONS}) embed security policies centrally but are prime denial-of-service targets and leak topology to attackers. Distributed overlay routing disperses path computation yet lacks a unified trust fabric, leaving routing tables vulnerable to poisoning and man-in-the-middle attacks. Decentralized peer-to-peer forwarding (\ac{SD‐AMC}) grants full autonomy but omits mutual authentication and encryption by default. Hybrid architectures (Istio, \ac{CoAP}‐based) colocate sidecar proxies that enforce \ac{mTLS}, certificate‐based trust, and policy enforcement locally, while a central authority revokes credentials. With proxy‐anchored encryption and hierarchical trust, hybrid schemes offer the highest security, privacy, and trust in \ac{SRIDS} routing and communication.  
 

%% file: sections/sec6.tex
\begin{figure*}[t!]
    \centering
    \includegraphics[width=6.95in]{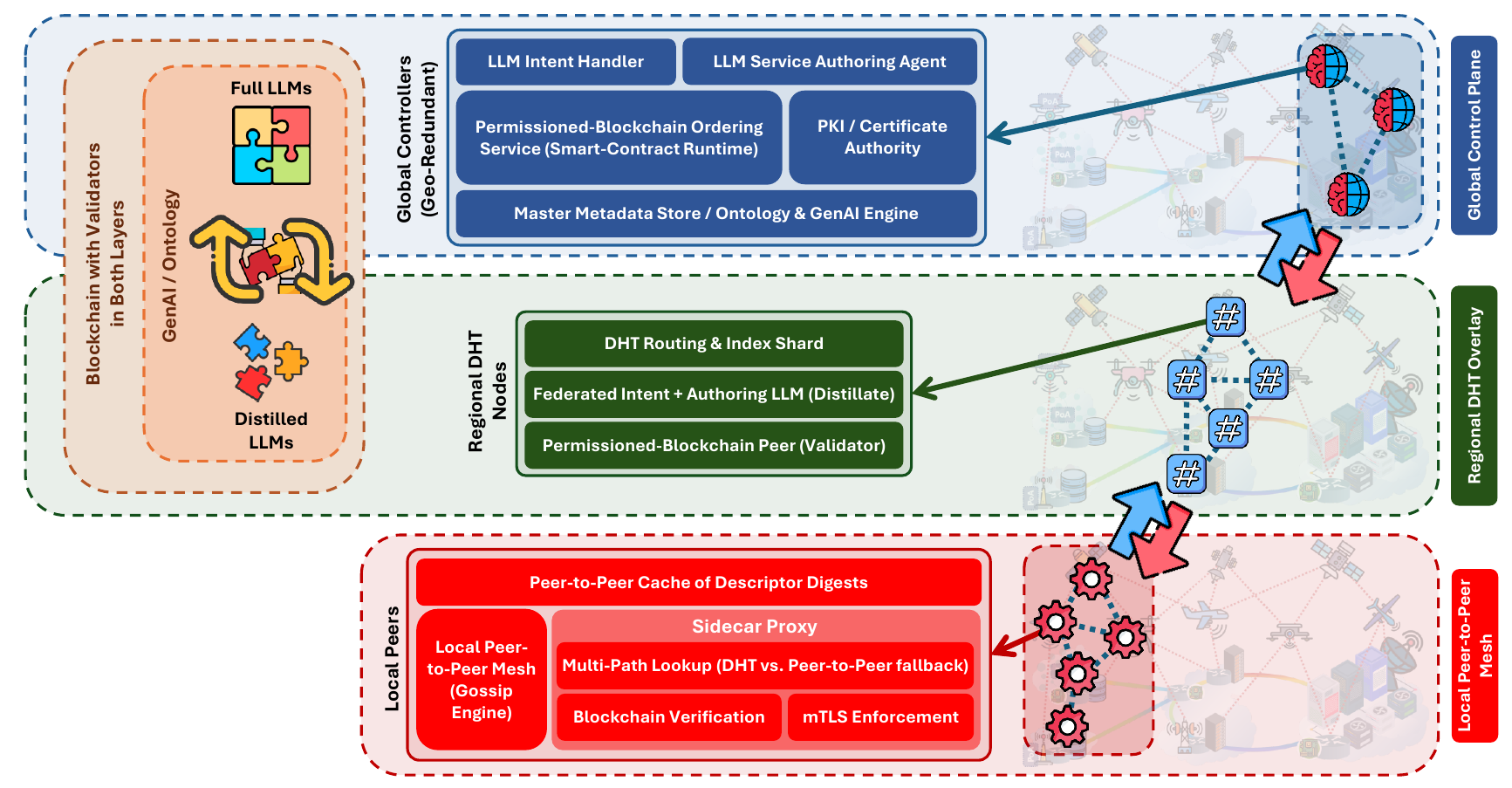}
    \vspace{-5pt}
    \caption{The modules of the proposed \ac{SRIDS} architecture aimed at facilitating service provision in future 6G systems.}
    \label{fig:architecture_components}
    \vspace{-8pt}
\end{figure*}

\section{Future Directions}
\label{sec:future-directions}
Considering the strengths and weaknesses of the studies in the literature, as well as the design objectives, the following subsections propose an architecture for \ac{SRIDS} in the 6G era and outline the open challenges that must be addressed for its practical implementation.

\subsection{Proposed Architecture}
The proposed architecture is a three-tiered, hybrid control-plane augmented by two cross-cutting planes (\ac{GenAI} and Blockchain) that consider all \ac{SRIDS} functionalities, depicted in Fig.~\ref{fig:architecture_components}. \blue{The architecture utilizes 6G as a distributed computing substrate that unifies networking and computation across the edge-cloud continuum.} At its foundation lies a geographically distributed continuum of computing and networking resources, partitioned into edge, regional, and core tiers. \blue{Within this hierarchy, the 6G orchestration fabric provides control to manage computing resources in a single system. Edge nodes are thus treated not merely as network relays but as active computing entities capable of hosting services.}

At the apex of our design sits a geo-redundant fleet of three to five \textit{Global Controllers}. Drawing on centralized coordination paradigms (Section \ref{ss_centralized}) like Kubernetes and hybrid federation models such as Jini, each controller integrates five tightly coupled modules: (i) an \textit{\ac{LLM} Intent Handler} that ingests and disambiguates natural-language requests; (ii) an \textit{\ac{LLM} Service Authoring Agent} that automatically generates, refines and version-controls service descriptors, policy snippets and deployment manifests; (iii) a \textit{\ac{PKI}/\ac{CA}} module (inspired by Istio) to issue, rotate and revoke \ac{mTLS} credentials across the fabric; (iv) a \textit{Permissioned-Blockchain Ordering Service \& Smart-Contract Runtime} (e.g. Hyperledger Fabric orderers or proof-of-authority Ethereum) that totally orders all transactions by service providers ($\mathrm{registerService}$) and users ($\mathrm{bindIntent}$), enforces on-chain policy checks, and anchors audit proofs; and (v) a \textit{Master Metadata Store / Ontology \& \ac{GenAI} Engine} - a graph-based \ac{OWL}/\ac{RDF} repository with schema-versioning, access policies and an embedded \ac{GenAI} inference layer for semantic classification, search ranking and dynamic ontology extension. All five modules participate in a \ac{BFT} consensus protocol, which enables agreement among controllers even in the presence of faulty or malicious nodes and ensures consistent coordination across the geo-redundant fleet.

Below the core tier, a fleet of regional computing nodes each hosts three co-resident distributed modules. First, a \textit{\ac{DHT} Routing \& Index Shard} (e.g., Kademlia, Chord, or Pastry as surveyed in Section \ref{ss_distributed}) is responsible for a contiguous key–range of the global $\mathrm{service ID} \to \mathrm{metadata}$ map, replicating entries across successor or $k$-bucket sets to ensure resilience and providing $\mathcal{O}(\log N)$ lookups. Second, a \textit{Federated Intent + Service Authoring \ac{LLM} Distillate} - a medium-sized (on the order of hundreds of megabytes) transformer model - performs local intent parsing, semantic ranking of candidate services, and on-the-fly refinement of service descriptors; it participates in a lightweight federated training cycle driven by the \ac{GenAI} plane to stay aligned with the core ontology. Third, a \textit{Permissioned-Blockchain Peer (validator)} commits all $\mathrm{registerService}$ and $\mathrm{bindIntent}$ transactions on a \ac{BFT} or proof-of-authority ledger (for example, Hyperledger Fabric peer or private Ethereum node), executes smart contracts to enforce schema and policy compliance, and anchors immutable audit proofs. These three modules cooperate via periodic \ac{DHT} gossip to edge caches, \ac{LLM} model and policy deltas from the core, and on-chain state broadcasts.

The \textit{Blockchain} plane captures all register/bind transactions on a permissioned ledger (informed by BlockONS and \ac{IOTA}-Tangle designs from Sections \ref{ss_centralized} and \ref{ss_distributed}). The \textit{\ac{GenAI}/Ontology} plane centralizes global schema and model training at the Core and propagates distilled \ac{LLM} models and context embeddings downward (as advocated by \ac{CoAP}-fog and Jini federation works in Section \ref{ss_hybrid}).

At the device or in-network edge, each node hosts three tightly coupled modules to enable fully decentralized service lookup and binding (Section \ref{ss_decentralized}). First, a \textit{Local Peer-to-Peer Mesh Gossip Engine} floods compact descriptor digests among immediate neighbors, rapidly disseminating $\mathrm{service ID} \to \mathrm{metadata}$ pointers without any central coordinator. Second, a \textit{Peer-to-Peer Cache of Descriptor Digests} maintains a local key–value map from $\mathrm{service ID}$ hashes to the regional \ac{DHT}‐shard endpoints (and optional policy tags), enabling sub‐millisecond cache hits for common lookups. Third, a \textit{Sidecar Proxy} intercepts all user requests, enforces \ac{mTLS} credentials issued by the \ac{PKI}/\ac{CA}, verifies on-chain transaction proofs, and implements a multi-path resolution strategy to select the most appropriate instance of the requested service: it first consults the local peer-to-peer cache, then falls back, via gRPC or \ac{UDP}, to the regional \ac{DHT} overlay, or even re-injects the request into the gossip mesh when needed. Moreover, two logically orthogonal planes span the core and regional tiers.

By combining these elements, we satisfy the ten critical \ac{SRIDS} design objectives. Reliability is achieved through multi-tier replication (\ac{DHT} successor sets, peer-to-peer gossip caching) and \ac{BFT} consensus at the core, removing single points of failure. Scalability derives from the logarithmic load distribution of \ac{DHT} lookups and the linear storage scaling of the peer-to-peer mesh, with only bounded, localized gossip traffic. Automaticity and adaptability emerge from embedded \ac{GenAI} agents, both in the core (for global descriptor authoring) and at regionals (for local intent ranking and policy adaptation), and from self-healing protocols (gossip flooding, \ac{DHT} stabilization) that respond autonomously to topology changes. \red{For critical control-plane operations, determinism is assured by the consensus protocols of the central controllers, and for data-plane lookups, determinism is ensured by the bounded hop counts of the structured overlays.} Efficiency and sustainability are realized by our multi-path resolution strategy: the sidecar proxy first probes the local peer-to-peer cache for sub-millisecond hits, resorts to the regional \ac{DHT} only on cache misses, and invokes gossip only as a last fallback, thereby minimizing protocol overhead and energy use. Semantic-awareness pervades all tiers via the graph-based ontology and \ac{LLM} inference layer that tags and ranks services according to user intent and contextual metadata. Finally, security, privacy, and trust rest on end-to-end \ac{mTLS} enforced by the \ac{PKI}/\ac{CA}, on-chain proof verification in the blockchain plane, and certificate-driven identity management, ensuring every $\mathrm{registerService}$ and $\mathrm{bindIntent}$ transaction is authenticated, authorized, and auditable across the continuum. \blue{The present architecture represents a harmonized conceptual model that promotes mutual compatibility among components while recognizing that fine-grained optimization remains a promising avenue for future research.}

\subsection{Health Guardian Workflow}
\subsubsection{Service Provider}
Upon electing to publish the Health Guardian service through the proposed architecture, the provider first submits a natural‐language intent, describing the end‐to‐end functional pipeline, its executable components, as well as quality‐of‐service targets and contextual requirements, to one of the geo‐redundant Global Controllers in the core tier. The \ac{LLM} Intent Handler ingests this invocation, resolves the ambiguities, and emits a structured intent schema. This schema is handed over to the \ac{LLM} Service Authoring Agent, which automatically generates a versioned service descriptor. The descriptor includes (1) the modular components annotated with their resource footprints, interface contracts, and inter‐component dependencies; (2) high‐level policy snippets that constrain execution latency and retention requirements; and (3) deployment manifests that target heterogeneous nodes across the edge-cloud continuum. Concurrently, the \ac{PKI}/\ac{CA} module provisions \ac{mTLS} certificates for Health Guardian endpoints. A $\mathrm{registerService}$ transaction, encapsulating the service identifier, semantic tags, and the provider’s digitally signed credentials, is submitted to the Permissioned‐Blockchain Ordering Service (Smart-Contract Runtime). Here, a smart contract enforces compliance with the on‐chain schema and policy. Upon consensus under the \ac{BFT} protocol, the transaction is ordered, anchored as an immutable audit proof, and the resulting block reference is returned to the provider. Once committed, the Master Metadata Store/Ontology \!\&\! \ac{GenAI} Engine integrates the Health Guardian descriptor into its graph‐based \ac{OWL}/\ac{RDF} repository.

The semantic engine classifies the components under existing ontological classes, assigns unique $\mathrm{service IDs}$, and propagates distilled embeddings to the regional tier via the \ac{GenAI} plane. In each region, the \ac{DHT} Routing \& Index Shard receives a contiguous slice of the global $\mathrm{service ID} \to \mathrm{metadata}$ map and replicates the new entries across successor nodes. Simultaneously, the Federated Intent + Service Authoring \ac{LLM} (Distillate) consumes core‐level ontology updates and participates in a lightweight federated training cycle, refining local ranking models so that region‐specific variants of Health Guardian remain aligned with the core’s semantic definitions. The regional Permissioned-Blockchain Peers commit matching $\mathrm{registerService}$ events on the local ledger, execute the same compliance smart contracts, and anchor audit proofs for cross-region traceability. At this critical juncture, the orchestrator is invoked to compute placements for each service of the Health Guardian service. The orchestration layer determines optimal locations for each service instance based on their specific resource requirements, current availability, and operational policies. Once the services are instantiated, the orchestrator gathers their actual runtime locations - node identifiers, \ac{IP} addresses, ports, and any associated interfaces. This information is then fed back into the Master Metadata Store, serving as a live index for future service discovery and selection.

At the network edge, the local Peer-to-Peer Mesh (Gossip Engine) floods compact descriptor digests of the Health Guardian $\mathrm{service IDs}$ to immediate neighbors, enabling sub‐millisecond awareness of available components. Each edge node's Peer-to-Peer Cache of Descriptor Digests maps the $\mathrm{service IDs}$ to one or more regional \ac{DHT} endpoints, annotated with policy tags indicating, for example, the maximum permitted data sampling rate or privacy‐level constraints. \red{In the case of the Health Guardian components, which are instantiated, the Sidecar Proxy intercepts all requests from users.} It enforces the \ac{mTLS} credentials issued by the \ac{PKI}/\ac{CA}, verifies the on‐chain proof of registration, and implements a multi-path resolution strategy: querying its local cache first, falling back to the regional \ac{DHT} via gRPC on a cache miss, or re-injecting into the gossip mesh if necessary.

\subsubsection{User}
Upon initiating a request through the proposed architecture, the user must articulate their functional requirements and quality-of-service targets using a standardized interface. This request is typically expressed in natural language, reflecting the user’s intent to engage with the Health Guardian service. After receiving the request at an edge node, the Sidecar Proxy in the Local Peer initially verifies the \ac{mTLS} credentials issued by the \ac{PKI}/\ac{CA}. This step is essential for establishing a trusted connection between the user and the Health Guardian service. Subsequently, the request is forwarded to the corresponding Regional \ac{DHT} Node for processing by the Federated Intent and Service Authoring \ac{LLM} (Distillate). The \ac{LLM} translates it into a structured intent schema so that the system can effectively query the available services. This structured representation includes key functional and quality-of-service criteria essential for the next stages of the workflow. The intent schema is then returned to the local edge node and forwarded to the proxy to retrieve the necessary on-chain proof of registration from the blockchain ledger, which validates the integrity of the service instance the user wishes to connect to. Then, the Peer-to-Peer Cache of Descriptor Digests is referenced, which maintains a mapping from $\mathrm{service IDs}$ to the corresponding regional \ac{DHT} endpoints, along with associated policy tags. If the desired service is present in the cache, the user’s request can be routed directly to the endpoint using the cached information, facilitating sub-millisecond response times under normal operational conditions.

For requests that cannot be satisfied through direct access to local caches, the Sidecar Proxy implements a multi-path resolution strategy. It first queries the regional \ac{DHT} using gRPC to retrieve service metadata. If this query does not yield the required service instance, the request undergoes a fallback mechanism where it is reinjected into the peer-to-peer gossip mesh. This last resort ensures that even under varying service loads or dynamic availability conditions, the user still has a chance to connect with the Health Guardian service through widely disseminated query protocols. Once a service match is found, whether through local cache queries, \ac{DHT} lookups, or through the gossip network, the Sidecar Proxy aggregates the service’s endpoint information, including node identifiers, \ac{IP} addresses, and ports. In the selection phase, the Sidecar Proxy assesses the gathered options in relation to established criteria, including quality-of-service requirements, as well as overarching system objectives such as energy consumption minimization and load balancing. This data is seamlessly returned to the user, who can now establish a direct communication channel with the most suitable Health Guardian components, effectively optimizing service performance based on the current operational context.

% \begin{figure*}[t!]
%     \centering
%     \includegraphics[scale=0.31]{figures/research_challenges.eps}
%     \vspace{-2pt}
%     \caption{An overview of the challenges and directions that should be taken into account in the SRIDS of futuristic dynamic networks.}
%     \label{fig:open_challenges}
%     \vspace{-5pt}
% \end{figure*}

\subsection{Potential Research Directions}
Based on the proposed \ac{SRIDS} architecture, potential open research directions can effectively be categorized as follows. Each category emphasizes methods that enhance specific components of the proposed architecture, thus facilitating a targeted discourse on future explorations.

\subsubsection{Adaptive Machine Learning Techniques}
Adaptive \ac{ML} techniques are fundamental to enhancing the capabilities of the \ac{SRIDS} architecture, particularly in the context of user-centric services. Research can delve into methods such as continual and lifelong learning, which equip algorithms to seamlessly adapt to evolving data and user requirements without the detriment of prior knowledge loss \cite{10.1145/3735633}. One of the potential techniques in this domain is CLFace, a continual learning framework that employs feature-level distillation to reduce drift between feature maps of student and teacher models across multiple stages, incorporating geometry-preserving distillation schemes and contrastive knowledge distillation to enhance discriminative power \cite{10943531}. Furthermore, federated learning can be utilized in reverse order to train regional \acp{LLM} and subsequently use these models to create global versions, all while maintaining user privacy. The most cutting-edge approach is the Federated Freeze A (FFA)-\ac{LoRA} technique, which addresses the instability in privacy-preserving federated learning by fixing randomly initialized non-zero matrices and only fine-tuning zero-initialized matrices, providing more consistent performance with better computational efficiency \cite{sun2024improving}. Another advanced method is the PP-BFL framework, a decentralized federated learning approach based on blockchain that introduces node reputation mechanisms and training accuracy verification through smart contracts \cite{10456734}. These approaches are particularly suitable for the \ac{LLM} Intent Handler and the \ac{LLM} Service Authoring Agent modules, as they can improve knowledge extraction and classification, thereby significantly contributing to the efficiency and effectiveness of service provision \cite{HOU2025121368}.

\subsubsection{Predictive Analytics and Caching Strategies}
The integration of predictive analytics and caching strategies is paramount to improving the response times and resource management capabilities of the proposed \ac{SRIDS} architecture. Studies can focus on predictive caching methodologies that anticipate user requests based on historical usage patterns, thereby minimizing latency in service access \cite{10379478}. The most advanced technique in this area is the KM-SVD++ approach, which combines Kalman filtering models for predicting vehicle locations with SVD++ algorithms for cache deployment and replacement decisions, achieving high hit rates and low latency in vehicular edge computing environments \cite{10664290}. This approach, in conjunction with the development of adaptive peer clustering techniques, can enhance the \ac{DHT} Routing \& Index Shard and Peer-to-Peer Cache of Descriptor Digests modules by ensuring that services are dynamically grouped based on contextual requirements and usage trends \cite{10707504}. Furthermore, FADE--a federated deep reinforcement learning approach--can be implemented, allowing peer nodes to collaboratively develop shared predictive models using first-round training settings as starting inputs for local training \cite{10923461}. Additionally, the CODR method can be integrated, which utilizes spatial-temporal graph neural networks for demand prediction combined with \ac{TD3} algorithms for optimal offloading schemes \cite{10034418}. These innovations can lead to substantial improvements in service responsiveness and overall user experience in dynamic environments like those envisaged for 6G \cite{10227281}.

\subsubsection{Smart Security Enhancements}
In the proposed \ac{SRIDS} architecture, security is a fundamental component that warrants thorough investigation by researchers. Specifically, the examination of AI-driven methods to enhance security protocols is of paramount importance. The deployment of advanced anomaly detection algorithms, powered by \ac{ML}, can facilitate real-time monitoring and identification of potential threats that may arise during service provision \cite{10705391}. For instance, the integration of blockchain technology within the \ac{PKI}/\ac{CA} and Permissioned-Blockchain Ordering Service (smart contract runtime) modules establishes a robust framework for user authentication and service integrity maintenance. This framework can be further augmented through the incorporation of autoencoder models. In this approach, \ac{ML} models are trained utilizing \ac{GPS} coordinates and activity data to detect deviations from typical behavior patterns, with blockchain technology ensuring decentralized identity verification \cite{10677323}. The synthesis of these methodologies creates a dual-layered security approach that considerably enhances the resilience of the infrastructure against unauthorized access and data breaches, a factor that is particularly critical in sensitive applications such as those anticipated in 6G networks \cite{10186353}.

As the architecture necessitates robust security measures, advanced cryptographic techniques will play a critical role, especially in scenarios where sensitive data is transmitted. Research can address the development of quantum-safe cryptographic methods that protect against future threats posed by quantum computing advancements, particularly benefiting the \ac{PKI}/\ac{CA} module and enhancing data integrity across the architecture \cite{BASERI2024103883}. The most advanced quantum-safe approach is the hybrid key exchange protocol, which incorporates both post-quantum cryptography and quantum key distribution in a modular and information-theoretic secure architecture, providing both forward and post-compromise security \cite{grams_202300304}. The CRYSTALS-Kyber for general encryption and CRYSTALS-Dilithium, FALCON, and SPHINCS+ for digital signatures represent cutting-edge NIST-standardized quantum-safe cryptography algorithms that outperform classical algorithms under various conditions \cite{tadepalli2024impact}. Additionally, implementing trusted enclave-based computation techniques can further bolster data privacy during operations handled by the Permissioned-Blockchain Peers, allowing for secure data processing without exposing sensitive information \cite{10628168}. The dual-factor quantum-safe blockchain architecture represents the most advanced approach, ensuring latency minimization while fortifying against quantum threats through a modular design \cite{10628167}. Incorporating homomorphic encryption could provide a balance between security and functionality when computations involve encrypted data \cite{10574838}.

\subsubsection{Addressing Network Challenges}
Addressing network challenges such as data divergence, replication latency, and synchronization delays is essential for maintaining the reliability of the \ac{SRIDS} architecture. Open research paths can include the examination of data consistency mechanisms that ensure coherent state management in both the Master Metadata Store/Ontology \& \ac{GenAI} Engine and \ac{DHT} Routing \& Index Shard components \cite{10634358}. The most advanced technique is the PPaxos (Pull Paxos) protocol, which employs a pull-based replication method where follower nodes proactively pull data from other nodes, featuring an adaptive load-aware pull strategy that allows dynamic adjustment of pull frequency and relay node selection \cite{10763844}. This approach achieves approximately 29.6\% improvement in throughput and 34.88\% reduction in latency under bandwidth-constrained edge environments \cite{10763844}. Exploring advanced replication strategies that minimize lag, along with enabling real-time synchronization techniques, can significantly enhance the overall robustness and reliability of the architecture \cite{10469511}. The Blockchain version of the Dwork-Lynch-Stockmeyer consensus protocol represents a breakthrough in \ac{BFT}, recognized for its efficiency and successfully integrated into Hyperledger Fabric as a practical alternative to traditional crash fault-tolerant protocols \cite{10.1145/3688225.3688237}. Techniques from distributed consensus protocols can also be employed to optimize service updates, ensuring that the entire system remains aligned with real-time information \cite{Venkatesan2024}.

\subsubsection{Global and Local Optimization Techniques}
The exploration of global and local optimization techniques is vital for achieving efficient service instance selection within the \ac{SRIDS} architecture. Research can investigate \ac{ML} algorithms that facilitate real-time selection of optimal service instances while balancing local resource constraints and global performance goals, which will enhance the decision-making capabilities of the Sidecar Proxy module \cite{10828850}. The most advanced approach is the quantum-inspired optimization algorithms, which offer superior performance in constrained environments by providing improved accuracy, reduced latency, and enhanced resource utilization compared to traditional optimization techniques \cite{10840586}. The differential evolution algorithm has shown superior performance for complex control systems, demonstrating excellence in dynamic performance metrics including overshoot, stabilization error, settling time, and energy consumption \cite{10828850}. Developing frameworks for automatic service configuration adjustments based on predictive analytics will further enhance the performance of 6G use cases, ensuring they meet stringent quality parameters expected in future environments \cite{10499767}.